\documentclass{fundam}

\publyear{25}
\papernumber{2}
\volume{195}
\issue{1-4}
\theDOI{10.46298/fi.12245}

\versionForARXIV

\usepackage{MathSpad,makeidx,ifthen,latexsym}
\usepackage{amssymb,graphicx} 
\def\mathit#1{#1}

\newcommand{\flagTikZ}{0}
\newcommand{\omittik}[1]{%
  \ifthenelse{\flagTikZ=1}{#1}{}}
\IfFileExists{tikzflag.tex}{\renewcommand{\flagTikZ}{1}}
 
\omittik{
\usepackage{tikz}
\usetikzlibrary{arrows,automata}
}

\usepackage{hyperref}

\newcommand{\rightsymdivision}{{\setminus}{\hspace{-1.2mm}}{\setminus}}
\newcommand{\leftsymdivision}{/{\hspace{-1.2mm}}/}

\DeclareSymbolFont{cmsygroup}{OMS}{cmsy}{m}{n}
\DeclareSymbolFont{cmrgroup}{OT1}{cmr}{m}{n}
\DeclareMathSymbol{\sim}{\mathrel}{cmsygroup}{"18}
\DeclareMathSymbol{=}{\mathrel}{cmrgroup}{"3D}
\DeclareMathSymbol{=}{\mathrel}{cmrgroup}{"3D}
\DeclareMathSymbol{[}{\mathopen}{cmrgroup}{"5B}
\DeclareMathSymbol{]}{\mathclose}{cmrgroup}{"5D}
\DeclareMathSymbol{(}{\mathopen}{cmrgroup}{"28}
\DeclareMathSymbol{)}{\mathclose}{cmrgroup}{"29}

\begin{document}
\title{The Index and Core of a Relation\\With Applications to the
Axiomatics of Relation Algebra}
\author{Roland Backhouse \\
 University of Nottingham, UK
\and
Ed Voermans \\
Independent Researcher}
\runninghead{R. Backhouse, E. Voermans}{The Index and Core of a Relation}

\maketitle
\begin{abstract}We introduce the general notions of an index and a core of a relation.  We postulate  a
limited form of the axiom of choice ---specifically that all partial equivalence relations have an index---
and explore the consequences of adding the axiom to standard axiom systems for point-free reasoning. 
Examples of the theorems we prove are that a core/index  of a difunction is a bijection, and that the so-called
``all or nothing'' axiom used to facilitate pointwise reasoning is derivable from our axiom of choice. \end{abstract}



\section{Introduction}\label{Indices:Introduction}

We introduce the general notions of an ``index'' and a ``core'' of a relation.  As suggested by the terminology, 
the practical significance of both notions is to substantially reduce the size of 
a (possibly very large) binary relation in such a way that the relation can nevertheless easily be
recovered.  Example \ref{core.example} illustrates the notions.
\begin{Example}\label{core.example}{\rm \ \ \ Fig.\ \ref{fig:core} depicts a relation (on the left) and two instances of  cores of
the relation   (in the middle and on the right).  All are depicted as bipartite graphs.  The relation $R$  is a
relation on blue  and red nodes.  The middle  figure depicts a core  as a relation on squares of blue nodes
and squares of  red nodes, each square being an equivalence class of the left per domain of $R$  (on the left)
or of the right per domain of $R$  (on the right).   The rightmost figure depicts a core as a relation on
representatives of the equivalence classes: the relation depicted by the thick green edges.  The rightmost
figure also depicts an index of the relation; the middle does not: although the relations depicted in the
middle and rightmost figures are isomorphic, they have different types.
\QED
\begin{figure}[h]
\centering \includegraphics{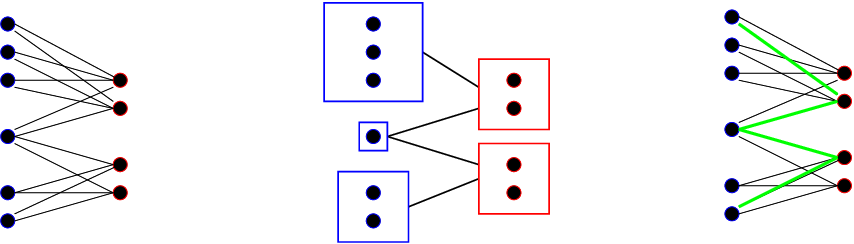}
 
\caption{A Relation, a  Core and an Index.}\label{fig:core}
\end{figure}
}%
\end{Example}

\vspace*{-4ex}

Although the notion of the ``core'' of a relation is more general than the notion of an  ``index'',  
a significant disadvantage   is that a ``core'' typically has a type that is different from the 
relation itself;  in contrast, an ``index'' of a relation is a ``core'' that has the same type as the relation. 
This is useful for practical purposes,  particularly in the context of heterogeneous relations, because it avoids the
necessity to introduce type judgements.  For this reason, our focus us on the notion of an index.

 The paper is divided into three  parts.  The first part, consisting of 
Sections \ref{Point-freeAxiomatisation}  and \ref{Domains},  sets  up the framework on which our calculations are
based.  Section \ref{Point-freeAxiomatisation} summarises the axiom system and the notation we use;
section \ref{Domains} lists a number of derived concepts and their properties.  These properties are given
without proof.

In Section \ref{Indices General}, we formalise  the notions of a ``core'' and an
``index'' of a relation in the context of point-free relation algebra.  We establish a large collection of
properties of these notions which form a basis for the third  part  of the paper.  (Because the notions are
new, almost all the properties are new.   An example of a property that some readers may recognise,
albeit expressed differently, is that a difunction has an index that is a bijection.)  

Section  \ref{Indexes of  Difunctions and Pers}  specialises the notion of an index to partial equivalence
relations and difunctions.  Section \ref{Indices and Pers}  concludes by the introduction of a restricted form of the axiom of
choice: we postulate that every partial equivalence relation has an index.  This is the same as saying that
it is possible to choose a representative element of every equivalence class of a partial equivalence
relation. Section \ref{From Pers To Relations}  then shows that every relation has an index.  
Section  \ref{per.difun.char}   is about applying our axiom of choice to the derivation of well-known
characterisations of pers and difunctions.  

 Section \ref{Formulations of Points} examines the consequences of adding our axiom of choice to point-free
relation algebra in order to facilitate pointwise reasoning.   We show that so doing has surprising and
remarkable consequences.  One such consequence is that we can derive the so-called ``all-or-nothing''
rule; this is a rule introduced by Gl\"uck \cite{Glueck17} also as a means of facilitating pointwise reasoning. 
(See \cite{BACKHOUSE2022100730} for examples of how the rule is used in reasoning about graphs.)  The
main theorem in Section \ref{Formulations of Points} is that, with the addition of our axiom of choice and an
extensionality axiom, the type
\setms{0.15em}$A{\sim}B$  of relations is isomorphic to the powerset $2^{A{\MPtimes}B}$ (the set of subsets of the
cartesian product of $A$ and $B$).  Section \ref{IndependenceOfTheAxioms}  presents several models of
point-free relation algebra clarifying the relation of the additional axioms to each other.

Section \ref{IndicesAndCoresConclusion} concludes the paper with a discussion of the significance of the
notions we have introduced and a pointer to the potential value for practical applications.



\section{Axioms of Point-free Relation Algebra}\label{Point-freeAxiomatisation}

In traditional, pointwise reasoning about relations, it is not the relations themselves that are the focus of
interest.   Rather, a relation $R$  of type $A{\sim}B$ is defined to be a subset of the cartesian product 
 $A{\MPtimes}B$ and the focus of interest is the boolean membership 
property $(a,b){\in}R$ where $a$ and $b$ are elements of type $A$ and $B$, respectively.  Equality of relations $R$ and $S$
is defined in terms of membership (typically in terms of ``if and only if''), leading to a lack  of
concision (and frequently precision).  In point-free relation algebra, the membership relation plays no
role, and reasoning is truly about properties of relations.  

In this section, we give a brief summary of the axioms of point-free relation algebra.  For full details
of the axioms, see \cite{BACKHOUSE2022100730}.    For a summary of our notational conventions, see the
appendix (Section \ref{IDCAppendix}).

\subsection{Summary}\label{Indices:The AxiomSystemSummary}

Point-free relation algebra comprises three layers with interfaces between the layers plus additional axioms
peculiar to relations.  The axiom system is typed.  For types $A$ and $B$,  $A{\sim}B$ denotes a set;  the elements of the set are called
\emph{(heterogeneous) relations of type}  $A{\sim}B$.  Elements of type $A{\sim}A$, for some  type  $A$, are called \emph{homogeneous
relations}.  

The first layer axiomatises the properties of a partially ordered  set.  We postulate that, for each pair of
types $A$ and $B$,  $A{\sim}B$ forms  a complete, universally distributive lattice.     In anticipation of Section~\ref{Formulations of Points}, where
we add axioms that require $A{\sim}B$ to be a powerset, we use the symbol ``${\subseteq}$'' for the ordering relation, and  ``${\cup}$''
and ``${\cap}$'' for the supremum and infimum operators.   We assume that this notation is familiar to the
reader, allowing us to skip a more detailed account of its  properties.   However, we use   ${\MPplatbottom}$
for the least element of the ordering  (rather than the conventional $\emptyset$)  and ${\MPplattop}$ for the greatest element.  In keeping with the
conventional practice of overloading the  symbol ``$\emptyset$'', both these symbols are overloaded.  The symbols
``${\MPplatbottom}$'' and ``${\MPplattop}$'' are pronounced ``bottom'' and ``top'', respectively.   (Strictly we should write something like
${}_{A}{\MPplatbottom}_{B}$ and ${}_{A}{\MPplattop}_{B}$  for the bottom and top elements of type $A{\sim}B$.  Of course, care needs to be taken when
overloading operators in this way but it is usually the case that elementary type considerations allow the
appropriate type to be deduced.)  

It is important to note that there is no axiom stating that a relation is a set, and there is no corresponding
notion of membership.  (In, for example,  \cite{REL92a} and \cite{Vo99}, we used the symbols ``${\sqsubseteq}$'', ``${\sqcup}$'' and ``${\sqcap}$''
and the name ``spec calculus'' rather than ``relation algebra''  in order to avoid
misunderstanding.)   The lack of a notion of membership distinguishes point-free relation algebra from
pointwise algebra.

The second layer adds a composition operator.  If $R$ is a relation of type $A{\sim}B$ and $S$ is a relation of type
$B{\sim}C$, the composition of $R$ and $S$ is a relation of type $A{\sim}C$ which we denote by $R{\MPcomp}S$.     Composition is
associative and, for each type $A$, there is an identity relation which we denote by $\mathbb{I}_{A}$.  We often overload
the notation for the identity relation, writing just $\mathbb{I}$.  Occasionally, for greater clarity, we do supply the
type information.

The interface between the first and second layers defines a relation algebra to be an instance of a \emph{regular
algebra} \cite{B2004a} (also  called a \emph{standard Kleene algebra}, or \textbf{S}-\emph{algebra} \cite{CON71b}).   For this paper, 
the most important aspect of this interface is the existence and properties of the factor operators.  These
are introduced in Section \ref{Barbosa:Basic Structures}.  Also, ${\MPplatbottom}$ is a zero of composition: for all $R$,
 ${\MPplatbottom}{\MPcomp}R\ms{1}{=}\ms{1}{\MPplatbottom}\ms{1}{=}\ms{1}R{\MPcomp}{\MPplatbottom}$.  

The completeness axiom in the first layer  allows the reflexive-transitive closure $R^{*}$ of
each element  $R$  of type  $A{\sim}A$, for some type $A$,  to be defined.  For practical applications, this is possibly the
 most important aspect of  regular algebra but such applications are not considered in this paper.  For this
paper, completeness is only relevant when we add axioms to the algebra that model pointwise reasoning.
We do require, however, the existence of $R{\cup}S$ and $R{\cap}S$, for all pairs of relations $R$ and $S$ of the same
type, and the usual  properties of set union and intersection.

The third layer is the introduction of a \emph{converse} operator.   If $R$ is a relation of type $A{\sim}B$, the converse of
$R$, which we denote by $R^{\MPrev}$ (pronounced $R$ ``wok'') is a relation of type $B{\sim}A$. 
The interface with the first  layer is that converse is a poset isomorphism (in particular,  ${\MPplatbottom}^{\MPrev}\ms{1}{=}\ms{1}{\MPplatbottom}$ and
${\MPplattop}^{\MPrev}\ms{1}{=}\ms{1}{\MPplattop}$), and the interface with the
second layer is formed by the two rules   $\mathbb{I}^{\MPrev}\ms{1}{=}\ms{1}\mathbb{I}$ and, for all relations $R$ and $S$ of appropriate type, 
 $(R{\MPcomp}S)^{\MPrev}\ms{2}{=}\ms{2}S^{\MPrev}\ms{1}{\MPcomp}\ms{1}R^{\MPrev}$. 

Additional axioms characterise properties peculiar to relations.
The modularity rule (aka Dede\-kind's rule \cite{Riguet48}) is that, for all relations $R$, $S$ and $T$,\begin{equation}\label{dedekind.rule}
R{\MPcomp}S\ms{1}{\cap}\ms{1}T\ms{3}{\subseteq}\ms{3}R\ms{1}{\MPcomp}\ms{1}(S\ms{3}{\cap}\ms{3}R^{\MPrev}\ms{1}{\MPcomp}\ms{1}T)~~.
\end{equation}The dual property, obtained by exploiting properties of the converse operator, is, for all relations $R$, $S$ and $T$,\begin{equation}\label{dedekind.rule.conv}
S{\MPcomp}R\ms{1}{\cap}\ms{1}T\ms{3}{\subseteq}\ms{3}(S\ms{3}{\cap}\ms{3}T\ms{1}{\MPcomp}\ms{1}R^{\MPrev})\ms{1}{\MPcomp}\ms{1}R~~.
\end{equation}The modularity rule is 
necessary to the derivation of some of the properties we state without proof (for example, the
properties of the domain operators given in section \ref{domain.ops}).  Another rule is  the \emph{cone rule}:\begin{equation}\label{cone.rule}
{\left\langle\forall{}R\ms{4}{:}{:}\ms{4}{\MPplattop}{\MPcomp}R{\MPcomp}{\MPplattop}\ms{2}{=}\ms{2}{\MPplattop}\ms{3}{\vee}\ms{3}R\ms{1}{=}\ms{1}{\MPplatbottom}\right\rangle}~~.
\end{equation}The cone rule limits consideration to ``unary'' relation algebras:  constructing the cartesian product of two
relation algebras to form a relation algebra (whereby the operators are defined pointwise) does not yield
an algebra satisfying the cone rule.

\subsection{Factors}\label{Barbosa:Basic Structures}

If $R$ is a relation of type $A{\sim}B$ and $S$ is a relation of type $A{\sim}C$, the relation $R{\setminus}S$ of type $B{\sim}C$ is defined by
the Galois connection, for all $T$ (of type $B{\sim}C$),\begin{equation}\label{under}
T\ms{2}{\subseteq}\ms{2}R{\setminus}S\ms{5}{\equiv}\ms{5}R{\MPcomp}T\ms{2}{\subseteq}\ms{2}S~~.
\end{equation}Similarly,  if $R$ is a relation of type $A{\sim}B$ and $S$ is a relation of type $C{\sim}B$, the relation $R{/}S$ of type $A{\sim}C$ is defined by
the Galois connection, for all $T$,\begin{equation}\label{over}
T\ms{2}{\subseteq}\ms{2}R{/}S\ms{5}{\equiv}\ms{5}T{\MPcomp}S\ms{2}{\subseteq}\ms{2}R~~.
\end{equation}The existence of factors is a property of a regular algebra; in relation algebra, factors are also known as 
``residuals''.    Factors have the \emph{cancellation} properties:\begin{equation}\label{factor.cancellation}
T\ms{1}{\MPcomp}\ms{1}T{\setminus}U\ms{4}{\subseteq}\ms{4}U\ms{9}{\wedge}\ms{9}R{/}S\ms{1}{\MPcomp}\ms{1}S\ms{4}{\subseteq}\ms{4}R~~.
\end{equation}The relations $R{\setminus}R$ (of type $B{\sim}B$ if $R$ has type $A{\sim}B$) and $R{/}R$ (of type $A{\sim}A$ if $R$ has type $A{\sim}B$) play a
central role in what follows.  As is easily verified, both are \emph{preorders}.  That is, both are \emph{transitive}:\begin{equation}\label{RunderRtransitive}
R{\setminus}R\ms{1}{\MPcomp}\ms{1}R{\setminus}R\ms{4}{\subseteq}\ms{4}R{\setminus}R\ms{9}{\wedge}\ms{9}R{/}R\ms{1}{\MPcomp}\ms{1}R{/}R\ms{4}{\subseteq}\ms{4}R{/}R
\end{equation}and both are \emph{reflexive}:\begin{equation}\label{RunderRreflexive}
\mathbb{I}\ms{3}{\subseteq}\ms{3}R{\setminus}R\ms{8}{\wedge}\ms{8}\mathbb{I}\ms{3}{\subseteq}\ms{3}R{/}R~~.
\end{equation}(The notation ``$\mathbb{I}$'' is overloaded in the above equation.  In the left conjunct, it
denotes the identity relation of type $B{\sim}B$ and, in the right conjunct, it
denotes the identity relation of type $A{\sim}A$, assuming $R$ has type $A{\sim}B$.) We also have the cancellation
property, for all $R$,\begin{equation}\label{factor.eq}
R\ms{1}{\MPcomp}\ms{1}R{\setminus}R\ms{6}{=}\ms{6}R\ms{6}{=}\ms{6}R{/}R\ms{1}{\MPcomp}\ms{1}R~~.
\end{equation}Factors enjoy a rich theory which underlies many of our calculations.  However, for space reasons, we
omit further details here.

\section{Some Definitions}\label{Domains}

In point-free relation algebra, ``coreflexives'' of a given type   represent sets of elements of that type. 
A \emph{coreflexive of type} $A$ is a relation $p$ such that $p\ms{1}{\subseteq}\ms{1}\mathbb{I}_{A}$.  Frequently used properties are that, for all
coreflexives $p$,  \begin{displaymath}p\ms{2}{=}\ms{2}p^{\MPrev}\ms{2}{=}\ms{2}p{\MPcomp}p\end{displaymath}and, for all coreflexives $p$ and $q$, \begin{displaymath}p{\MPcomp}q\ms{3}{=}\ms{3}p\ms{1}{\cap}\ms{1}q\ms{3}{=}\ms{3}q{\MPcomp}p~~.\end{displaymath}(The proof of these properties relies on the modularity rule.)  In the literature, coreflexives have several different names, usually depending on the application area in
question.  Examples are ``monotype'', ``pid'' (short for ``partial identity'')  and ``test''.

\subsection{The Domain Operators}\label{domain.ops}

The ``domain operators'' (see eg. \cite{BAC92b})  play a dominant and unavoidable role.  We
exploit their properties frequently in calculations, so much so that we assume great familiarity with
them.  
\begin{Definition}[Domain Operators]\label{lr.squares}{\rm \ \ \ Given  relation $R$ of type $A{\sim}B$,  
the \emph{left domain} $R{\MPldom{}}$  of   $R$ is a relation of type $A$ defined by the equation \begin{displaymath}R{\MPldom{}}\ms{6}{=}\ms{6}\mathbb{I}_{A}\ms{3}{\cap}\ms{3}R\ms{1}{\MPcomp}\ms{1}R^{\MPrev}\end{displaymath}and the \emph{right  domain} $R{\MPrdom{}}$  of    $R$ is a relation of type $B$  is defined by the equation \begin{displaymath}R{\MPrdom{}}\ms{6}{=}\ms{6}\mathbb{I}_{B}\ms{3}{\cap}\ms{3}R^{\MPrev}\ms{1}{\MPcomp}\ms{1}R~~.\end{displaymath}\vspace{-7mm}
}
\QED
\end{Definition}

The name ``domain operator'' is chosen because of the fundamental properties: for all $R$ and all
coreflexives $p$, \begin{equation}\label{rdom.is.least}
R\ms{1}{=}\ms{1}R{\MPcomp}p\ms{6}{\equiv}\ms{6}R{\MPrdom{}}\ms{2}{=}\ms{2}R{\MPrdom{}}\ms{1}{\MPcomp}\ms{1}p
\end{equation}and \begin{equation}\label{ldom.is.least}
R\ms{1}{=}\ms{1}p{\MPcomp}R\ms{6}{\equiv}\ms{6}R{\MPldom{}}\ms{2}{=}\ms{2}p\ms{1}{\MPcomp}\ms{1}R{\MPldom{}}~~.
\end{equation}A simple, often-used consequence of (\ref{rdom.is.least}) and (\ref{ldom.is.least}) is the property:\begin{equation}\label{ldom.and.rdom}
R{\MPldom{}}\ms{1}{\MPcomp}\ms{1}R\ms{4}{=}\ms{4}R\ms{4}{=}\ms{4}R\ms{1}{\MPcomp}\ms{1}R{\MPrdom{}}~~.
\end{equation}In words, $R{\MPrdom{}}$ is the least coreflexive $p$ such that restricting the ``domain'' of $R$ on the right has no effect on
$R$.    It is in  this sense that $R{\MPldom{}}$ and $R{\MPrdom{}}$ represent the set of points on the left and on the right on which the
relation $R$ is ``defined'', i.e.\ its left and right ``domains''.

By instantiating $p$ to ${\MPplatbottom}$ in (\ref{rdom.is.least}) and (\ref{ldom.is.least}) we get \begin{equation}\label{monos10.14'}
(R{\MPldom{}}\ms{1}{=}\ms{1}{\MPplatbottom})\ms{4}{=}\ms{4}(R\ms{1}{=}\ms{1}{\MPplatbottom})\ms{4}{=}\ms{4}(R{\MPrdom{}}\ms{1}{=}\ms{1}{\MPplatbottom})~~.
\end{equation}Additional properties used frequently below are as follows.
\begin{Theorem}\label{monos10.12}{\rm \ \ \ For all relations $R$ and coreflexives $p$,  \begin{equation}\label{monos10.9}
R{\MPrdom{}}\ms{1}{\subseteq}\ms{1}p\ms{3}{\equiv}\ms{3}R\ms{1}{\subseteq}\ms{1}{\MPplattop}{\MPcomp}p\mbox{~~and~~}R{\MPldom{}}\ms{1}{\subseteq}\ms{1}p\ms{3}{\equiv}\ms{3}R\ms{1}{\subseteq}\ms{1}p{\MPcomp}{\MPplattop}\mbox{~~,}
\end{equation}\begin{equation}\label{monos10.10}
R{\MPrdom{}}\ms{1}{\subseteq}\ms{1}p\ms{3}{\equiv}\ms{3}R\ms{1}{\subseteq}\ms{1}R{\MPcomp}p\mbox{~~and~~}R{\MPldom{}}\ms{1}{\subseteq}\ms{1}p\ms{3}{\equiv}\ms{3}R\ms{1}{=}\ms{1}p{\MPcomp}R~~.
\end{equation}
}
\QED
\end{Theorem}

\begin{Theorem}\label{domains}{\rm \ \ \ For all relations $R$ and  $S$,
\begin{description}
\item[(a)]${\MPplattop}\ms{1}{\MPcomp}\ms{1}R{\MPrdom{}}\ms{2}{=}\ms{2}{\MPplattop}{\MPcomp}R$ ~and~$R{\MPldom{}}\ms{1}{\MPcomp}\ms{1}{\MPplattop}\ms{2}{=}\ms{2}R{\MPcomp}{\MPplattop}$ ~~, 
\item[(b)]$(R^{\MPrev}){\MPrdom{}}\ms{2}{=}\ms{2}R{\MPldom{}}$ ~and~$(R^{\MPrev}){\MPldom{}}\ms{2}{=}\ms{2}R{\MPrdom{}}$~~, and 
\item[(c)]$(R{\MPcomp}S){\MPrdom{}}\ms{2}{=}\ms{2}(R{\MPrdom{}}\ms{1}{\MPcomp}\ms{1}S){\MPrdom{}}$ ~and~$(R{\MPcomp}S){\MPldom{}}\ms{2}{=}\ms{2}(R\ms{1}{\MPcomp}\ms{1}S{\MPldom{}}){\MPldom{}}$~~. \QED
 
\end{description}
}
\end{Theorem}

We also use the fact that the domain operators are monotonic (as is evident from Definition \ref{lr.squares}).

\subsection{Pers and Per Domains}\label{Per Domains}

Given relations $R$ of type $A{\sim}B$  and $S$ of type $A{\sim}C$,  the symmetric
\emph{right-division}  is the  relation $R\rightsymdivision{}S$  of type $B{\sim}C$  defined in terms of \emph{right} factors as  \begin{equation}\label{double.under.def}
R\rightsymdivision{}S\ms{5}{=}\ms{5}R{\setminus}S\ms{2}{\cap}\ms{2}(S{\setminus}R)^{\MPrev}~~.
\end{equation}Dually, given relations $R$ of type $B{\sim}A$ and $S$ of type $C{\sim}A$,  the
symmetric \emph{left-division} is the  relation $R\leftsymdivision{}S$  of type $B{\sim}C$ defined in terms of left factors as \begin{equation}\label{double.over.def}
R\leftsymdivision{}S\ms{5}{=}\ms{5}R{/}S\ms{3}{\cap}\ms{3}(S{/}R)^{\MPrev}~~.
\end{equation}The relation $R\rightsymdivision{}R$ is an equivalence relation\footnote{~This is a well-known fact:  the relation $R\rightsymdivision{}R$ is the
 symmetric closure of the preorder  $R{\setminus}R$.  The easy proof is left to the reader.}.  
 Voermans \cite{Vo99} calls it the ``greatest 
right domain'' of $R$. Riguet \cite{Riguet48} calls $R\rightsymdivision{}R$ the ``noyau''  of $R$ (but defines it using nested
complements).   Others  (see   \cite{Ol17} for references) call it the  ``kernel'' of $R$.  

As remarked elsewhere \cite{Ol17},  the \emph{symmetric left-division} inherits a number of (left) cancellation
properties from the properties of factorisation in terms of which it is defined.  For our purposes, the only
cancellation property we use is the following (inherited from  (\ref{factor.eq})).  For all $R$,\begin{equation}\label{sym.rightfactor.canc}
R\ms{1}{\MPcomp}\ms{1}R\rightsymdivision{}R\ms{4}{=}\ms{4}R\ms{4}{=}\ms{4}R\leftsymdivision{}R\ms{1}{\MPcomp}\ms{1}R~~.
\end{equation}In this section the focus is on the left and right
``per domains'' introduced by Voermans \cite{Vo99}.
\begin{Definition}[Right and Left Per Domains]\label{perdoms}{\rm \ \ \ The \emph{right per domain} of relation $R$, denoted
$R{\MPperrdom{}}$,   is defined by the equation\begin{equation}\label{per.rightdomain}
R{\MPperrdom{}}\ms{4}{=}\ms{4}R{\MPrdom{}}\ms{1}{\MPcomp}\ms{1}R\rightsymdivision{}R~~.
\end{equation}Dually, the \emph{left per domain} of relation $R$, denoted $R{\MPperldom{}}$, is defined  by the equation\begin{equation}\label{per.leftdomain}
R{\MPperldom{}}\ms{4}{=}\ms{4}R\leftsymdivision{}R\ms{1}{\MPcomp}\ms{1}R{\MPldom{}}~~.
\end{equation}
}
\QED
\end{Definition}

The left and right per domains are ``pers'' where ``per'' is an abbreviation of ``partial equivalence
relation''.  
\begin{Definition}[Partial Equivalence Relation (per)]\label{def:per}{\rm \ \ \ A relation  is a \emph{partial equivalence
relation} iff  it is symmetric and transitive. That is, $R$ is a partial equivalence relation iff \begin{displaymath}R\ms{1}{=}\ms{1}R^{\MPrev}\ms{5}{\wedge}\ms{5}R{\MPcomp}R\ms{1}{\subseteq}\ms{1}R~~.\end{displaymath}Henceforth we abbreviate partial equivalence relation to \emph{per}.
}
\QED
\end{Definition}

That $R{\MPperldom{}}$ and $R{\MPperrdom{}}$ are indeed pers is a direct consequence of the symmetry and transitivity of $R\rightsymdivision{}R$.  

The left and right per domains are called ``domains'' because,  like the coreflexive domains, we have the
properties: for all relations $R$ and pers $P$,\begin{equation}\label{per.rdom.is.least}
R\ms{1}{=}\ms{1}R{\MPcomp}P\ms{6}{\equiv}\ms{6}R{\MPperrdom{}}\ms{2}{=}\ms{2}R{\MPperrdom{}}\ms{1}{\MPcomp}\ms{1}P\mbox{~~, and}
\end{equation}\begin{equation}\label{per.ldom.is.least}
R\ms{1}{=}\ms{1}P{\MPcomp}R\ms{6}{\equiv}\ms{6}R{\MPperldom{}}\ms{2}{=}\ms{2}P\ms{1}{\MPcomp}\ms{1}R{\MPperldom{}}~~.
\end{equation}As with the coreflexive domains,  we also have:\begin{equation}\label{per.leftandrightdomain.eq}
R{\MPperldom{}}\ms{1}{\MPcomp}\ms{1}R\ms{4}{=}\ms{4}R\ms{4}{=}\ms{4}R\ms{1}{\MPcomp}\ms{1}R{\MPperrdom{}}~~.
\end{equation}(The second of these equalities is an  immediate consequence  of  (\ref{sym.rightfactor.canc}) and the
properties of (coreflexive) domains; the first is symmetric.)  
Indeed, $R{\MPperldom{}}$ and $R{\MPperrdom{}}$ are the ``least'' pers that satisfy the  equalities (\ref{per.leftandrightdomain.eq}).   See
\cite{Vo99} for details of the ordering relation on pers.

The right per domain  $R{\MPperrdom{}}$ can be defined equivalently by the equation \begin{equation}\label{per.rightdomain.equiv}
R{\MPperrdom{}}\ms{4}{=}\ms{4}R\rightsymdivision{}R\ms{1}{\MPcomp}\ms{1}R{\MPrdom{}}~~.
\end{equation}Moreover, \begin{equation}\label{per.rightdomain.doms}
(R{\MPperrdom{}}){\MPldom{}}\ms{5}{=}\ms{5}R{\MPrdom{}}\ms{5}{=}\ms{5}(R{\MPperrdom{}}){\MPrdom{}}~~.
\end{equation} (See \cite{RCB2020} for the proofs of these properties.)  Symmetrical properties hold of $R{\MPperldom{}}$.

The following lemma  extends   \cite[Corollaire, p.134]{Riguet48}  from equivalence relations to pers.
\begin{Lemma}\label{per.perdom}{\rm \ \ \ For all  relations $R$, the following statements are all equivalent.
\begin{description}
\item[(i)]$R$ is a per (i.e.\ $R\ms{1}{=}\ms{1}R^{\MPrev}\ms{2}{\wedge}\ms{2}R{\MPcomp}R\ms{1}{\subseteq}\ms{1}R$)~~,  
\item[(ii)]$R\ms{2}{=}\ms{2}R^{\MPrev}\ms{1}{\MPcomp}\ms{1}R$~~, 
\item[(iii)]$R\ms{1}{=}\ms{1}R{\MPperldom{}}$~~, 
\item[(iv)]$R\ms{1}{=}\ms{1}R{\MPperrdom{}}$~~. \QED
 
\end{description}
}%
\end{Lemma}%

For further properties of pers and per domains, see \cite{Vo99}.

\subsection{Functionality}\label{Functionality}

A relation $R$ of type $A{\sim}B$
is said to be  \emph{left}-\emph{functional} iff $R\ms{1}{\MPcomp}\ms{1}R^{\MPrev}\ms{2}{=}\ms{2}R{\MPldom{}}$.    Equivalently,   $R$ is \emph{left-functional}  iff $R\ms{1}{\MPcomp}\ms{1}R^{\MPrev}\ms{2}{\subseteq}\ms{2}\mathbb{I}_{A}$.
It is said to be  \emph{right-functional}  iff  $R^{\MPrev}\ms{1}{\MPcomp}\ms{1}R\ms{2}{=}\ms{2}R{\MPrdom{}}$ (equivalently, \linebreak $R^{\MPrev}\ms{1}{\MPcomp}\ms{1}R\ms{2}{\subseteq}\ms{2}\mathbb{I}_{B}$).   A relation $R$
is said to be a \emph{bijection} iff it is both left- and right-functional.  

Rather than left-  and right-functional, the more common terminology is ``functional'' and ``injective''
but publications differ on which of left- or right-functional is ``functional'' or ``injective''.  We choose to
abbreviate ``left-functional'' to \emph{functional} and to use the term \emph{injective} instead of right-functional. 
Typically, we use  $f$ and $g$ to denote functional relations, and Greek letters to denote bijections (although
the latter is not always the case).  Other authors make the opposite choice.


\subsection{Difunctions}\label{bd:Difunctions}

Formally,  relation $R$ is \emph{difunctional} iff \begin{equation}\label{difunctional.def}
R\ms{1}{\MPcomp}\ms{1}R^{\MPrev}\ms{1}{\MPcomp}\ms{1}R\ms{3}{\subseteq}\ms{3}R~~.
\end{equation}As for pers,  there are several equivalent definitions of ``difunctional''.    We begin with the point-free
definitions:
\begin{Theorem}\label{difunctional.strongdef}{\rm \ \ \ For all $R$, the following statements are all equivalent.
\begin{description}
\item[(i)~~]$R$ is difunctional (i.e.\  $R\ms{1}{\MPcomp}\ms{1}R^{\MPrev}\ms{1}{\MPcomp}\ms{1}R\ms{3}{\subseteq}\ms{3}R$)~~, 
\item[(ii)~]$R\ms{3}{=}\ms{3}R\ms{1}{\MPcomp}\ms{1}R^{\MPrev}\ms{1}{\MPcomp}\ms{1}R$~~, 
\item[(iii)]$R{\MPrdom{}}\ms{1}{\MPcomp}\ms{1}R{\setminus}R\ms{3}{=}\ms{3}R^{\MPrev}\ms{1}{\MPcomp}\ms{1}R$~~, 
\item[(iv)~]$R{\MPperrdom{}}\ms{3}{=}\ms{3}R^{\MPrev}\ms{1}{\MPcomp}\ms{1}R$~~, 
\item[(v)~~]$R{/}R\ms{1}{\MPcomp}\ms{1}R{\MPldom{}}\ms{3}{=}\ms{3}R\ms{1}{\MPcomp}\ms{1}R^{\MPrev}$~~, 
\item[(vi)~]$R{\MPperldom{}}\ms{3}{=}\ms{3}R\ms{1}{\MPcomp}\ms{1}R^{\MPrev}$~~, 
\item[(vii)]$R\ms{3}{=}\ms{3}R\ms{1}{\cap}\ms{1}(R{\setminus}R{/}R)^{\MPrev}$~~. \QED
\end{description}
}
\end{Theorem}

The equivalence of \ref{difunctional.strongdef}(i) and \ref{difunctional.strongdef}(ii) is well-known and due to
Riguet \cite{Riguet48,Riguet50}; the equivalence of \ref{difunctional.strongdef}(i),  (iv) and (vi)  is due to Voermans 
\cite{Vo99}.  The equivalence of \ref{difunctional.strongdef}(i),  (iii) and (v) is formally stronger: a
consequence is that, if $R$ is difunctional, \begin{equation}\label{difunctional.strongdef.2}
R{\MPperrdom{}}\ms{3}{=}\ms{3}R{\MPrdom{}}\ms{1}{\MPcomp}\ms{1}R{\setminus}R\ms{7}{\wedge}\ms{7}R{\MPperldom{}}\ms{3}{=}\ms{3}R{/}R\ms{1}{\MPcomp}\ms{1}R{\MPldom{}}~~.
\end{equation}(Cf.  (\ref{per.rightdomain}).)  

Riguet \cite{Riguet51} calls the right side of \ref{difunctional.strongdef}(vii) the ``diff\a'{e}rence'' of relation $R$,
although he defines it using nested complements rather than factors.   The name ``diff\a'{e}rence'' is
motivated by his  definition.  Since we avoid the  use of complements  (and for other reasons) we prefer 
the term  ``diagonal'';  see \cite{VB2022,VB2023b}  for further discussion.  

The equivalence of (i) and (vii) is
a straightforward but beautiful application of the Galois connections defining intersection, converse and
factors.  

Definition (\ref{difunctional.def}) is the most  useful when it is required to establish that a 
particular  relation is difunctional, whereas properties \ref{difunctional.strongdef}(ii)-(vii) are more useful
when it is required to  exploit the fact that a particular relation is difunctional.

The combination of Theorem \ref{difunctional.strongdef} (in particular \ref{difunctional.strongdef}(ii) and 
\ref{difunctional.strongdef}(iv) with Lemma \ref{per.perdom})   allows one to prove that a per is a symmetric
difunction.  (We leave the easy calculation to the reader.)  This property is sometimes used to specialise
properties of difunctions to properties of pers.

\subsection{Squares and  Rectangles}\label{Difun:Rectangles}

 We now introduce the notions  of a ``rectangle'' and a ``square''; rectangles  are typically heterogeneous
whilst squares are, by definition, homogeneous relations.  Squares are rectangles;  properties of squares
are typically  obtained by specialising properties of rectangles.  (Riguet \cite{Riguet48} uses the terms
``rectangle'' and ``carr\a'{e}''.)  
\begin{Definition}[Rectangle and Square]\label{rectangle}\rm \ \ \ A relation $R$ is a \emph{rectangle} iff  $R\ms{1}{=}\ms{1}R{\MPcomp}{\MPplattop}{\MPcomp}R$.  A relation $R$
is a \emph{square} iff $R$ is a symmetric rectangle.
\QED
\end{Definition}

It is easily shown that a rectangle is a difunction and a square is a per.  

\begin{Lemma}\label{rTOPs}{\rm \ \ \ For all relations $R$ and $S$,  $R{\MPcomp}{\MPplattop}{\MPcomp}S$ is a rectangle.  
It follows that $R{\MPcomp}T{\MPcomp}S$ is a rectangle  if $T$ is a rectangle.
}%
\end{Lemma}%
\begin{proof}
Because the proof is based on the cone rule, a case analysis is necessary.  In the case that either $R$
or $S$ is the empty relation, the lemma clearly holds (because $R{\MPcomp}{\MPplattop}{\MPcomp}S$ is the empty relation, and the empty
relation  is a rectangle).  Suppose now that both $R$ and $S$ are non-empty.  Then
\begin{mpdisplay}{0.15em}{6.5mm}{0mm}{2}
	$R{\MPcomp}{\MPplattop}{\MPcomp}S{\MPcomp}{\MPplattop}{\MPcomp}R{\MPcomp}{\MPplattop}{\MPcomp}S$\push\-\\
	$=$	\>	\>$\{$	\>\+\+\+cone rule: (\ref{cone.rule}) (applied twice), assumption:    $R\ms{1}{\neq}\ms{1}{\MPplatbottom}$ and $S\ms{1}{\neq}\ms{1}{\MPplatbottom}$\-\-$~~~ \}$\pop\\
	$R{\MPcomp}{\MPplattop}{\MPcomp}S~~.$
\end{mpdisplay}
If $T$ is a rectangle,  $R{\MPcomp}T{\MPcomp}S\ms{1}{=}\ms{1}R{\MPcomp}T{\MPcomp}{\MPplattop}{\MPcomp}T{\MPcomp}S$; thus $R{\MPcomp}T{\MPcomp}S$ is a rectangle.
\end{proof}

\subsection{Isomorphic Relations}\label{Isomorphic Relations}

The (yet-to-be-defined) cores and indexes of a given relation are not unique; in common mathematical
jargon, they are unique ``up to isomorphism''.  In order to make this precise, we need to define the
notion of isomorphic relation and establish a number of properties.
\begin{Definition}\label{rel.iso}{\rm \ \ \ Suppose $R$ and $S$ are two relations (not necessarily of the same type).  Then we say
that $R$ and $S$ are \emph{isomorphic} and write $R\ms{1}{\cong}\ms{1}S$ iff
\begin{mpdisplay}{0.15em}{6.5mm}{0mm}{2}
	\push$\langle\exists\ms{1}\phi{,}\psi$\=	\>\+\\
	$:$	\>\+$\phi\ms{1}{\MPcomp}\ms{1}\phi^{\MPrev}\ms{2}{=}\ms{2}R{\MPldom{}}\ms{5}{\wedge}\ms{5}\phi^{\MPrev}\ms{1}{\MPcomp}\ms{1}\phi\ms{2}{=}\ms{2}S{\MPldom{}}\ms{5}{\wedge}\ms{5}\psi\ms{1}{\MPcomp}\ms{1}\psi^{\MPrev}\ms{2}{=}\ms{2}R{\MPrdom{}}\ms{5}{\wedge}\ms{5}\psi^{\MPrev}\ms{1}{\MPcomp}\ms{1}\psi\ms{2}{=}\ms{2}S{\MPrdom{}}$\-\\
	$:$	\>\+$R\ms{2}{=}\ms{2}\phi\ms{1}{\MPcomp}\ms{1}S\ms{1}{\MPcomp}\ms{1}\psi^{\MPrev}$\-\-\\
	$\rangle$\pop$~~.$
\end{mpdisplay} 
}
\QED
\end{Definition}
The relation between $R$ and $S$ in definition \ref{rel.iso} can be strengthened to the conjunction \begin{equation}\label{rel.iso.both}
R\ms{2}{=}\ms{2}\phi\ms{1}{\MPcomp}\ms{1}S\ms{1}{\MPcomp}\ms{1}\psi^{\MPrev}\ms{6}{\wedge}\ms{6}\phi^{\MPrev}\ms{1}{\MPcomp}\ms{1}R\ms{1}{\MPcomp}\ms{1}\psi\ms{2}{=}\ms{2}S~~.
\end{equation}Alternatively, the leftmost conjunct can be replaced  by the rightmost conjunct.  This is a consequence of
the following lemma. 
\begin{Lemma}\label{rel.iso.equiv}{\rm \ \ \ For all $\phi$, $\psi$, $R$ and $S$,
\begin{mpdisplay}{0.15em}{6.5mm}{0mm}{2}
	\push$~~~\ms{7}$\=\+$(R\ms{2}{=}\ms{2}\phi\ms{1}{\MPcomp}\ms{1}S\ms{1}{\MPcomp}\ms{1}\psi^{\MPrev}\ms{3}{\equiv}\ms{3}\phi^{\MPrev}\ms{1}{\MPcomp}\ms{1}R\ms{1}{\MPcomp}\ms{1}\psi\ms{2}{=}\ms{2}S)$\push\-\\
	$\Leftarrow$	\>\+\pop$\phi\ms{1}{\MPcomp}\ms{1}\phi^{\MPrev}\ms{2}{=}\ms{2}R{\MPldom{}}\ms{5}{\wedge}\ms{5}\phi^{\MPrev}\ms{1}{\MPcomp}\ms{1}\phi\ms{2}{=}\ms{2}S{\MPldom{}}\ms{5}{\wedge}\ms{5}\psi\ms{1}{\MPcomp}\ms{1}\psi^{\MPrev}\ms{2}{=}\ms{2}R{\MPrdom{}}\ms{5}{\wedge}\ms{5}\psi^{\MPrev}\ms{1}{\MPcomp}\ms{1}\psi\ms{2}{=}\ms{2}S{\MPrdom{}}~~.$\-\pop
\end{mpdisplay}
}%
\QED
\end{Lemma}%
We often choose one or other of the conjuncts in  (\ref{rel.iso.both}), whichever being most convenient at the
time.
\begin{Lemma}\label{rel.iso.eq}{\rm \ \ \ The relation ${\cong}$ is reflexive,  transitive and symmetric.  That is, ${\cong}$ is an equivalence
relation.
}%
\QED
\end{Lemma}%

The task of  proving that two relations are isomorphic involves constructing $\phi$ and $\psi$ that satisfy the
conditions of the existential quantification in Definition \ref{rel.iso}; we call the constructed values 
\emph{witnesses} to  the  isomorphism.

Note that the requirement on $\phi$   in Definition \ref{rel.iso}  is that it is  both functional and injective; thus it
is required to ``witness''   a (1--1)   correspondence   between the points in the left domain of $R$  and the points
in the left  domain of $S$.   Similarly, the requirement on $\psi$ is that it ``witnesses''  a (1--1)   correspondence  
between the  points in the right  domain of $R$  and the points in the right   domain of $S$.  Formally,  $R{\MPldom{}}$ and 
$S{\MPldom{}}$ are isomorphic as  ``witnessed''  by $\phi$ and $R{\MPrdom{}}$ and $S{\MPrdom{}}$ are isomorphic as ``witnessed'' by $\psi$:
\begin{Lemma}\label{dom.iso}{\rm \ \ \ Suppose $R$ and $S$ are relations such that $R\ms{1}{\cong}\ms{1}S$.  Then  $R{\MPldom{}}\ms{1}{\cong}\ms{1}S{\MPldom{}}$ and $R{\MPrdom{}}\ms{1}{\cong}\ms{1}S{\MPrdom{}}$.
Specifically, if $\phi$ and $\psi$ witness the isomorphism  $R\ms{1}{\cong}\ms{1}S$,  \begin{displaymath}R{\MPldom{}}\ms{2}{=}\ms{2}\phi\ms{1}{\MPcomp}\ms{1}S{\MPldom{}}\ms{1}{\MPcomp}\ms{1}\phi^{\MPrev}\ms{8}{\wedge}\ms{8}R{\MPrdom{}}\ms{2}{=}\ms{2}\psi\ms{1}{\MPcomp}\ms{1}S{\MPrdom{}}\ms{1}{\MPcomp}\ms{1}\psi^{\MPrev}~~.\end{displaymath}
}%
\end{Lemma}%
\begin{proof}
Suppose $\phi$ and $\psi$ are such that \begin{displaymath}\phi\ms{1}{\MPcomp}\ms{1}\phi^{\MPrev}\ms{2}{=}\ms{2}R{\MPldom{}}\ms{5}{\wedge}\ms{5}\phi^{\MPrev}\ms{1}{\MPcomp}\ms{1}\phi\ms{2}{=}\ms{2}S{\MPldom{}}\ms{5}{\wedge}\ms{5}\psi\ms{1}{\MPcomp}\ms{1}\psi^{\MPrev}\ms{2}{=}\ms{2}R{\MPrdom{}}\ms{5}{\wedge}\ms{5}\psi^{\MPrev}\ms{1}{\MPcomp}\ms{1}\psi\ms{2}{=}\ms{2}S{\MPrdom{}}~~.\end{displaymath}Then
\begin{mpdisplay}{0.15em}{6.5mm}{0mm}{2}
	$R{\MPldom{}}$\push\-\\
	$=$	\>	\>$\{$	\>\+\+\+$R{\MPldom{}}$ is a coreflexive\-\-$~~~ \}$\pop\\
	$R{\MPldom{}}\ms{1}{\MPcomp}\ms{1}R{\MPldom{}}$\push\-\\
	$=$	\>	\>$\{$	\>\+\+\+assumption\-\-$~~~ \}$\pop\\
	$\phi\ms{1}{\MPcomp}\ms{1}\phi^{\MPrev}\ms{1}{\MPcomp}\ms{1}\phi\ms{1}{\MPcomp}\ms{1}\phi^{\MPrev}$\push\-\\
	$=$	\>	\>$\{$	\>\+\+\+assumption\-\-$~~~ \}$\pop\\
	$\phi\ms{1}{\MPcomp}\ms{1}S{\MPldom{}}\ms{1}{\MPcomp}\ms{1}\phi^{\MPrev}~~.$
\end{mpdisplay}
That is   $R{\MPldom{}}\ms{2}{=}\ms{2}\phi\ms{1}{\MPcomp}\ms{1}S{\MPldom{}}\ms{1}{\MPcomp}\ms{1}\phi^{\MPrev}$.  Similarly,  $R{\MPrdom{}}\ms{2}{=}\ms{2}\psi\ms{1}{\MPcomp}\ms{1}S{\MPrdom{}}\ms{1}{\MPcomp}\ms{1}\psi^{\MPrev}$.  But also (because the domain operators are closure
operators),\begin{displaymath}\phi\ms{1}{\MPcomp}\ms{1}\phi^{\MPrev}\ms{2}{=}\ms{2}(R{\MPldom{}}){\MPldom{}}\ms{3}{\wedge}\ms{3}\phi^{\MPrev}\ms{1}{\MPcomp}\ms{1}\phi\ms{2}{=}\ms{2}(S{\MPldom{}}){\MPldom{}}\ms{3}{\wedge}\ms{3}\psi\ms{1}{\MPcomp}\ms{1}\psi^{\MPrev}\ms{2}{=}\ms{2}(R{\MPrdom{}}){\MPrdom{}}\ms{3}{\wedge}\ms{3}\psi^{\MPrev}\ms{1}{\MPcomp}\ms{1}\psi\ms{2}{=}\ms{2}(S{\MPrdom{}}){\MPrdom{}}~~.\end{displaymath}Applying Definition \ref{rel.iso} with $R{,}S{,}\phi{,}\psi\ms{2}{:=}\ms{2}R{\MPldom{}}\ms{1}{,}\ms{1}S{\MPldom{}}\ms{1}{,}\ms{1}\phi\ms{1}{,}\ms{1}\phi$ and $R{,}S{,}\phi{,}\psi\ms{2}{:=}\ms{2}R{\MPrdom{}}\ms{1}{,}\ms{1}S{\MPrdom{}}\ms{1}{,}\ms{1}\psi\ms{1}{,}\ms{1}\psi$, the lemma is proved.
\end{proof}

The property of the left and right domains stated in Lemma  \ref{dom.iso} is also valid for the left and 
right per domains:
\begin{Lemma}\label{perdom.iso}{\rm \ \ \ Suppose $R$ and $S$ are relations such that $R\ms{1}{\cong}\ms{1}S$.  Then  $R{\MPperldom{}}\ms{1}{\cong}\ms{1}S{\MPperldom{}}$ and $R{\MPperrdom{}}\ms{1}{\cong}\ms{1}S{\MPperrdom{}}$. 
Specifically, if $\phi$ and $\psi$ witness the isomorphism  $R\ms{1}{\cong}\ms{1}S$,  \begin{displaymath}R{\MPperldom{}}\ms{2}{=}\ms{2}\phi\ms{1}{\MPcomp}\ms{1}S{\MPperldom{}}\ms{1}{\MPcomp}\ms{1}\phi^{\MPrev}\ms{8}{\wedge}\ms{8}R{\MPperrdom{}}\ms{2}{=}\ms{2}\psi\ms{1}{\MPcomp}\ms{1}S{\MPperrdom{}}\ms{1}{\MPcomp}\ms{1}\psi^{\MPrev}~~.\end{displaymath}
}%
\end{Lemma}%
\begin{proof} 
Suppose $\phi$ and $\psi$ witness the isomorphism  $R\ms{1}{\cong}\ms{1}S$.  We show that the pair $(\psi,\psi)$ witnesses the
isomorphism $R{\MPperrdom{}}\ms{1}{\cong}\ms{1}S{\MPperrdom{}}$.  By assumption,  $\psi\ms{1}{\MPcomp}\ms{1}\psi^{\MPrev}\ms{2}{=}\ms{2}R{\MPrdom{}}$,  $\psi^{\MPrev}\ms{1}{\MPcomp}\ms{1}\psi\ms{2}{=}\ms{2}S{\MPrdom{}}$. Moreover, for all   
$R$,  $$(R{\MPperrdom{}}){\MPrdom{}}\ms{2}{=}\ms{2}(R{\MPperrdom{}}){\MPldom{}}\ms{2}{=}\ms{2}R{\MPrdom{}};$$ thus
 $\psi\ms{1}{\MPcomp}\ms{1}\psi^{\MPrev}\ms{2}{=}\ms{2}(R{\MPperrdom{}}){\MPrdom{}}$ and $\psi^{\MPrev}\ms{1}{\MPcomp}\ms{1}\psi\ms{2}{=}\ms{2}(S{\MPperrdom{}}){\MPrdom{}}$.  So it remains to show that $R{\MPperrdom{}}\ms{4}{=}\ms{4}\psi\ms{1}{\MPcomp}\ms{1}S{\MPperrdom{}}\ms{1}{\MPcomp}\ms{1}\psi^{\MPrev}$.   Now
\begin{mpdisplay}{0.15em}{6.5mm}{0mm}{2}
	$R{\MPperrdom{}}\ms{4}{=}\ms{4}\psi\ms{1}{\MPcomp}\ms{1}S{\MPperrdom{}}\ms{1}{\MPcomp}\ms{1}\psi^{\MPrev}$\push\-\\
	$\Leftarrow$	\>	\>$\{$	\>\+\+\+transitivity\-\-$~~~ \}$\pop\\
	$R{\MPperrdom{}}\ms{4}{=}\ms{4}R{\MPperrdom{}}\ms{1}{\MPcomp}\ms{1}\psi\ms{1}{\MPcomp}\ms{1}S{\MPperrdom{}}\ms{1}{\MPcomp}\ms{1}\psi^{\MPrev}\ms{4}{=}\ms{4}\psi\ms{1}{\MPcomp}\ms{1}S{\MPperrdom{}}\ms{1}{\MPcomp}\ms{1}\psi^{\MPrev}~~.$
\end{mpdisplay}
The calculation thus splits into two steps: the proof of the leftmost equality and the proof of the rightmost
equality.  The leftmost equality proceeds as follows.
\begin{mpdisplay}{0.15em}{6.5mm}{0mm}{2}
	$R{\MPperrdom{}}\ms{4}{=}\ms{4}R{\MPperrdom{}}\ms{1}{\MPcomp}\ms{1}\psi\ms{1}{\MPcomp}\ms{1}S{\MPperrdom{}}\ms{1}{\MPcomp}\ms{1}\psi^{\MPrev}$\push\-\\
	$=$	\>	\>$\{$	\>\+\+\+(\ref{per.rdom.is.least}),  $\psi\ms{1}{\MPcomp}\ms{1}S{\MPperrdom{}}\ms{1}{\MPcomp}\ms{1}\psi^{\MPrev}$ is a per (see below)\-\-$~~~ \}$\pop\\
	$R\ms{4}{=}\ms{4}R\ms{1}{\MPcomp}\ms{1}\psi\ms{1}{\MPcomp}\ms{1}S{\MPperrdom{}}\ms{1}{\MPcomp}\ms{1}\psi^{\MPrev}~~.$
\end{mpdisplay}
Continuing with the right hand side:
\begin{mpdisplay}{0.15em}{6.5mm}{0mm}{2}
	$R\ms{1}{\MPcomp}\ms{1}\psi\ms{1}{\MPcomp}\ms{1}S{\MPperrdom{}}\ms{1}{\MPcomp}\ms{1}\psi^{\MPrev}$\push\-\\
	$=$	\>	\>$\{$	\>\+\+\+$R\ms{2}{=}\ms{2}\phi\ms{1}{\MPcomp}\ms{1}S\ms{1}{\MPcomp}\ms{1}\psi^{\MPrev}$\-\-$~~~ \}$\pop\\
	$\phi\ms{1}{\MPcomp}\ms{1}S\ms{1}{\MPcomp}\ms{1}\psi^{\MPrev}\ms{1}{\MPcomp}\ms{1}\psi\ms{1}{\MPcomp}\ms{1}S{\MPperrdom{}}\ms{1}{\MPcomp}\ms{1}\psi^{\MPrev}$\push\-\\
	$=$	\>	\>$\{$	\>\+\+\+$\psi^{\MPrev}\ms{1}{\MPcomp}\ms{1}\psi\ms{2}{=}\ms{2}S{\MPrdom{}}$, domains: (\ref{ldom.and.rdom}) and (\ref{per.leftandrightdomain.eq})  \-\-$~~~ \}$\pop\\
	$\phi\ms{1}{\MPcomp}\ms{1}S\ms{1}{\MPcomp}\ms{1}\psi^{\MPrev}$\push\-\\
	$=$	\>	\>$\{$	\>\+\+\+see Lemma \ref{rel.iso.eq}\-\-$~~~ \}$\pop\\
	$R~~.$
\end{mpdisplay}
Combining the two calculations, we have established that \begin{displaymath}R{\MPperrdom{}}\ms{4}{=}\ms{4}R{\MPperrdom{}}\ms{1}{\MPcomp}\ms{1}\psi\ms{1}{\MPcomp}\ms{1}S{\MPperrdom{}}\ms{1}{\MPcomp}\ms{1}\psi^{\MPrev}~~.\end{displaymath}Now, for the rightmost equality, we have:
\begin{mpdisplay}{0.15em}{6.5mm}{0mm}{2}
	$R{\MPperrdom{}}\ms{1}{\MPcomp}\ms{1}\psi\ms{1}{\MPcomp}\ms{1}S{\MPperrdom{}}\ms{1}{\MPcomp}\ms{1}\psi^{\MPrev}\ms{4}{=}\ms{4}\psi\ms{1}{\MPcomp}\ms{1}S{\MPperrdom{}}\ms{1}{\MPcomp}\ms{1}\psi^{\MPrev}$\push\-\\
	$=$	\>	\>$\{$	\>\+\+\+$(R{\MPperrdom{}}){\MPldom{}}\ms{2}{=}\ms{2}R{\MPrdom{}}$, domains: (\ref{ldom.and.rdom})\-\-$~~~ \}$\pop\\
	$R{\MPrdom{}}\ms{1}{\MPcomp}\ms{1}R{\MPperrdom{}}\ms{1}{\MPcomp}\ms{1}\psi\ms{1}{\MPcomp}\ms{1}S{\MPperrdom{}}\ms{1}{\MPcomp}\ms{1}\psi^{\MPrev}\ms{4}{=}\ms{4}\psi\ms{1}{\MPcomp}\ms{1}S{\MPperrdom{}}\ms{1}{\MPcomp}\ms{1}\psi^{\MPrev}$\push\-\\
	$=$	\>	\>$\{$	\>\+\+\+$R{\MPrdom{}}\ms{2}{=}\ms{2}\psi\ms{1}{\MPcomp}\ms{1}\psi^{\MPrev}$\-\-$~~~ \}$\pop\\
	$\psi\ms{1}{\MPcomp}\ms{1}\psi^{\MPrev}\ms{1}{\MPcomp}\ms{1}R{\MPperrdom{}}\ms{1}{\MPcomp}\ms{1}\psi\ms{1}{\MPcomp}\ms{1}S{\MPperrdom{}}\ms{1}{\MPcomp}\ms{1}\psi^{\MPrev}\ms{4}{=}\ms{4}\psi\ms{1}{\MPcomp}\ms{1}S{\MPperrdom{}}\ms{1}{\MPcomp}\ms{1}\psi^{\MPrev}$\push\-\\
	$\Leftarrow$	\>	\>$\{$	\>\+\+\+Leibniz\-\-$~~~ \}$\pop\\
	$\psi^{\MPrev}\ms{1}{\MPcomp}\ms{1}R{\MPperrdom{}}\ms{1}{\MPcomp}\ms{1}\psi\ms{1}{\MPcomp}\ms{1}S{\MPperrdom{}}\ms{4}{=}\ms{4}S{\MPperrdom{}}$\push\-\\
	$=$	\>	\>$\{$	\>\+\+\+converse (noting that $R{\MPperrdom{}}$ and $S{\MPperrdom{}}$ are symmetric)\-\-$~~~ \}$\pop\\
	$S{\MPperrdom{}}\ms{1}{\MPcomp}\ms{1}\psi^{\MPrev}\ms{1}{\MPcomp}\ms{1}R{\MPperrdom{}}\ms{1}{\MPcomp}\ms{1}\psi\ms{4}{=}\ms{4}S{\MPperrdom{}}$\push\-\\
	$=$	\>	\>$\{$	\>\+\+\+(\ref{per.rdom.is.least}),  $\psi^{\MPrev}\ms{1}{\MPcomp}\ms{1}R{\MPperrdom{}}\ms{1}{\MPcomp}\ms{1}\psi$ is a per (see below)\-\-$~~~ \}$\pop\\
	$S\ms{1}{\MPcomp}\ms{1}\psi^{\MPrev}\ms{1}{\MPcomp}\ms{1}R{\MPperrdom{}}\ms{1}{\MPcomp}\ms{1}\psi\ms{4}{=}\ms{4}S$\push\-\\
	$=$	\>	\>$\{$	\>\+\+\+as above, with $R{,}S{,}\psi\ms{2}{:=}\ms{2}S\ms{1}{,}\ms{1}R\ms{1}{,}\ms{1}\psi^{\MPrev}$\-\-$~~~ \}$\pop\\
	$\mathsf{true}~~.$
\end{mpdisplay}
Note that the usage of (\ref{per.rdom.is.least}) relies on the fact that both $\psi\ms{1}{\MPcomp}\ms{1}S{\MPperrdom{}}\ms{1}{\MPcomp}\ms{1}\psi^{\MPrev}$ and $\psi^{\MPrev}\ms{1}{\MPcomp}\ms{1}R{\MPperrdom{}}\ms{1}{\MPcomp}\ms{1}\psi$ are pers. 
The straightforward  proof is omitted.
\end{proof}

\begin{Lemma}\label{pid.iso.bijection}{\rm \ \ \ A relation $R$ is isomorphic to a coreflexive iff $R$ is a bijection.
}%
\end{Lemma}%
\begin{proof}
The proof is by mutual implication.  Suppose first that $R$ is a bijection.  That is,\begin{displaymath}R\ms{1}{\MPcomp}\ms{1}R^{\MPrev}\ms{2}{=}\ms{2}R{\MPldom{}}\ms{6}{\wedge}\ms{6}R^{\MPrev}\ms{1}{\MPcomp}\ms{1}R\ms{2}{=}\ms{2}R{\MPrdom{}}~~.\end{displaymath}We prove that $R$ is isomorphic to $R{\MPldom{}}$.  (Symmetrically, $R$ is isomorphic to $R{\MPrdom{}}$.)  For the witnesses we take 
$R{\MPldom{}}$ and $R$.  Instantiating Definition \ref{rel.iso},  we have to verify that \begin{displaymath}R{\MPldom{}}\ms{1}{\MPcomp}\ms{1}(R{\MPldom{}})^{\MPrev}\ms{2}{=}\ms{2}R{\MPldom{}}\ms{5}{\wedge}\ms{5}(R{\MPldom{}})^{\MPrev}\ms{1}{\MPcomp}\ms{1}R{\MPldom{}}\ms{2}{=}\ms{2}R{\MPldom{}}\ms{5}{\wedge}\ms{5}R\ms{1}{\MPcomp}\ms{1}R^{\MPrev}\ms{2}{=}\ms{2}(R{\MPldom{}}){\MPrdom{}}\ms{5}{\wedge}\ms{5}R^{\MPrev}\ms{1}{\MPcomp}\ms{1}R\ms{2}{=}\ms{2}R{\MPrdom{}}\end{displaymath}and
\begin{mpdisplay}{0.15em}{6.5mm}{0mm}{2}
	$R{\MPldom{}}\ms{2}{=}\ms{2}R{\MPldom{}}\ms{1}{\MPcomp}\ms{1}R\ms{1}{\MPcomp}\ms{1}R^{\MPrev}~~.$
\end{mpdisplay}
The verification is a straightforward application of properties of the left domain operator.

Now suppose that  coreflexive  $p$  is isomorphic to  $R$.  Suppose the witnesses are $\phi$ and $\psi$.  That is,\begin{equation}\label{discrete.ass1}
\phi\ms{1}{\MPcomp}\ms{1}\phi^{\MPrev}\ms{3}{=}\ms{3}p\ms{5}{\wedge}\ms{5}\phi^{\MPrev}\ms{1}{\MPcomp}\ms{1}\phi\ms{2}{=}\ms{2}R{\MPldom{}}\ms{5}{\wedge}\ms{5}\psi^{\MPrev}\ms{1}{\MPcomp}\ms{1}\psi\ms{2}{=}\ms{2}R{\MPrdom{}}
\end{equation}and \begin{equation}\label{discrete.ass2}
p\ms{2}{=}\ms{2}\phi\ms{1}{\MPcomp}\ms{1}R\ms{1}{\MPcomp}\ms{1}\psi^{\MPrev}~~.
\end{equation}Then
\begin{mpdisplay}{0.15em}{6.5mm}{0mm}{2}
	$R{\MPldom{}}$\push\-\\
	$=$	\>	\>$\{$	\>\+\+\+$\phi^{\MPrev}\ms{1}{\MPcomp}\ms{1}\phi\ms{2}{=}\ms{2}R{\MPldom{}}\ms{2}{=}\ms{2}R{\MPldom{}}\ms{1}{\MPcomp}\ms{1}R{\MPldom{}}$\-\-$~~~ \}$\pop\\
	$\phi^{\MPrev}\ms{1}{\MPcomp}\ms{1}\phi\ms{1}{\MPcomp}\ms{1}\phi^{\MPrev}\ms{1}{\MPcomp}\ms{1}\phi$\push\-\\
	$=$	\>	\>$\{$	\>\+\+\+$\phi\ms{1}{\MPcomp}\ms{1}\phi^{\MPrev}\ms{2}{=}\ms{2}p\ms{2}{=}\ms{2}p\ms{1}{\MPcomp}\ms{1}p^{\MPrev}$\-\-$~~~ \}$\pop\\
	$\phi^{\MPrev}\ms{1}{\MPcomp}\ms{1}p\ms{1}{\MPcomp}\ms{1}p^{\MPrev}\ms{1}{\MPcomp}\ms{1}\phi$\push\-\\
	$=$	\>	\>$\{$	\>\+\+\+(\ref{discrete.ass2})\-\-$~~~ \}$\pop\\
	$\phi^{\MPrev}\ms{1}{\MPcomp}\ms{1}\phi\ms{1}{\MPcomp}\ms{1}R\ms{1}{\MPcomp}\ms{1}\psi^{\MPrev}\ms{1}{\MPcomp}\ms{1}(\phi\ms{1}{\MPcomp}\ms{1}R\ms{1}{\MPcomp}\ms{1}\psi^{\MPrev})^{\MPrev}\ms{1}{\MPcomp}\ms{1}\phi$\push\-\\
	$=$	\>	\>$\{$	\>\+\+\+converse\-\-$~~~ \}$\pop\\
	$\phi^{\MPrev}\ms{1}{\MPcomp}\ms{1}\phi\ms{1}{\MPcomp}\ms{1}R\ms{1}{\MPcomp}\ms{1}\psi^{\MPrev}\ms{1}{\MPcomp}\ms{1}\psi\ms{1}{\MPcomp}\ms{1}R^{\MPrev}\ms{1}{\MPcomp}\ms{1}\phi^{\MPrev}\ms{1}{\MPcomp}\ms{1}\phi$\push\-\\
	$=$	\>	\>$\{$	\>\+\+\+(\ref{discrete.ass1})\-\-$~~~ \}$\pop\\
	$R{\MPldom{}}\ms{1}{\MPcomp}\ms{1}R\ms{1}{\MPcomp}\ms{1}R{\MPrdom{}}\ms{1}{\MPcomp}\ms{1}R^{\MPrev}\ms{1}{\MPcomp}\ms{1}R{\MPldom{}}$\push\-\\
	$=$	\>	\>$\{$	\>\+\+\+domains: (\ref{ldom.and.rdom})\-\-$~~~ \}$\pop\\
	$R\ms{1}{\MPcomp}\ms{1}R^{\MPrev}~~.$
\end{mpdisplay}
We conclude that $R{\MPldom{}}\ms{2}{=}\ms{2}R\ms{1}{\MPcomp}\ms{1}R^{\MPrev}$.   Symmetrically,  $R{\MPrdom{}}\ms{2}{=}\ms{2}R^{\MPrev}\ms{1}{\MPcomp}\ms{1}R$.   That is,  $R$ is a bijection.
\end{proof}

\begin{Theorem}\label{core.if.iso}{\rm \ \ \ Suppose $P$ is a per.  Then,\begin{displaymath}P{\MPldom{}}\ms{2}{=}\ms{2}P\ms{6}{\Leftarrow}\ms{6}P{\MPldom{}}\ms{2}{\cong}\ms{2}P~~.\end{displaymath}In particular,  for all $R$,\begin{displaymath}R{\MPldom{}}\ms{2}{=}\ms{2}R{\MPperldom{}}\ms{6}{\Leftarrow}\ms{6}R{\MPldom{}}\ms{2}{\cong}\ms{2}R{\MPperldom{}}~~.\end{displaymath}Symmetrically, for all $R$,\begin{displaymath}R{\MPrdom{}}\ms{2}{=}\ms{2}R{\MPperrdom{}}\ms{6}{\Leftarrow}\ms{6}R{\MPrdom{}}\ms{2}{\cong}\ms{2}R{\MPperrdom{}}~~.\end{displaymath}
}
\end{Theorem}
\begin{proof}
This is an instance of Lemma \ref{pid.iso.bijection}.  Specifically, assuming that $P{\MPldom{}}\ms{2}{\cong}\ms{2}P$, we may 
apply the instantiation $p{,}R\ms{2}{:=}\ms{2}P{\MPldom{}}\ms{1}{,}\ms{1}P$  in  Lemma \ref{pid.iso.bijection}  to deduce that $P$ is a bijection.  That
is,  $P\ms{1}{\MPcomp}\ms{1}P^{\MPrev}\ms{2}{=}\ms{2}P{\MPldom{}}$.     But $P$ is a per (i.e.\  $P\ms{2}{=}\ms{2}P\ms{1}{\MPcomp}\ms{1}P^{\MPrev}$).  So we conclude  that \begin{displaymath}P\ms{1}{=}\ms{1}P{\MPldom{}}~~.\end{displaymath}That, for all $R$,  $R{\MPldom{}}\ms{1}{=}\ms{1}R{\MPperldom{}}$  if $R{\MPldom{}}\ms{2}{\cong}\ms{2}R{\MPperldom{}}$  now follows by making the instantiation $P\ms{1}{:=}\ms{1}R{\MPperldom{}}$ and using the fact
that $(R{\MPperldom{}}){\MPldom{}}\ms{1}{=}\ms{1}R{\MPldom{}}$.    The symmetric property of the right domain operators follows by making the
instantiation $P\ms{1}{:=}\ms{1}R{\MPperrdom{}}$ and using the fact  that $(R{\MPperrdom{}}){\MPldom{}}\ms{1}{=}\ms{1}R{\MPrdom{}}$.
\end{proof}

\section{Indexes and Core Relations}\label{Indices General}

This section introduces the notions of ``index'' and ``core'' of a relation. An ``index'' is a special case of a
``core'' of a relation but, in general, it is more useful.  The properties of both notions are explored in depth.

\subsection{Indexes}\label{Indexes General}

Recall Fig.\ \ref{fig:core}.  We said that the middle and rightmost figures depict  ``core relations''.  
The property that is common to both is captured by the following definition. 
\begin{Definition}[Core Relation]\label{Core.gen}{\rm \ \ \ A relation $R$ is a \emph{core relation}  iff $R{\MPldom{}}\ms{1}{=}\ms{1}R{\MPperldom{}}$ and $R{\MPrdom{}}\ms{1}{=}\ms{1}R{\MPperrdom{}}$.
}
\QED
\end{Definition}


The rightmost figure of Fig.\ \ref{fig:core} is what we call an ``index'' of the relation depicted by the leftmost
figure.  The definition of an ``index'' of a relation is as follows.  
\begin{Definition}[Index]\label{gen.index}{\rm \ \ \ An \emph{index} of a relation $R$ is a relation $J$ that  has the following properties:
\begin{description}
\item[(a)]$J\ms{1}{\subseteq}\ms{1}R~~,$ 
\item[(b)]$R{\MPperldom{}}\ms{1}{\MPcomp}\ms{1}J\ms{1}{\MPcomp}\ms{1}R{\MPperrdom{}}\ms{3}{=}\ms{3}R~~,$ 
\item[(c)]$J{\MPldom{}}\ms{1}{\MPcomp}\ms{1}R{\MPperldom{}}\ms{1}{\MPcomp}\ms{1}J{\MPldom{}}\ms{3}{=}\ms{3}J{\MPldom{}}~~,$ 
\item[(d)]$J{\MPrdom{}}\ms{1}{\MPcomp}\ms{1}R{\MPperrdom{}}\ms{1}{\MPcomp}\ms{1}J{\MPrdom{}}\ms{3}{=}\ms{3}J{\MPrdom{}}~~.$\QED 
\end{description}
}
\end{Definition}

Note particularly requirement \ref{gen.index}(a).  A consequence of this requirement is that an index of a
relation has the same type as the relation.  This means that the relation depicted by the middle figure of
Fig.\ \ref{fig:core} is \emph{not} an index of the  relation depicted by the leftmost figure because the relations  have
different types.    

An obvious property is that a core relation is an index of itself:
\begin{Theorem}\label{core.is.own.index}{\rm \ \ \ Suppose $R$ is a core relation.  Then $R$ is an index of $R$.
}
\end{Theorem}
\begin{proof}
Straightforward application of Definitions \ref{Core.gen} and \ref{gen.index} together with the properties
of  (coreflexive and per) domains.
\end{proof}

In general, the existence of an index of an arbitrary relation is \emph{not} derivable in systems that axiomatise 
point-free relation algebra.   In Section \ref{Indices and Pers}  we add a limited form of the axiom of choice
that guarantees the existence of indexes of arbitrary pers; we also show that this then guarantees the 
existence of indexes for arbitrary relations.  For the moment, we establish a number of properties of
indexes assuming they exist.  For example, we show that all indexes of a given relation are isomorphic:
see Theorem \ref{index.iso}.  
\begin{Lemma}\label{index.per.is.pid}{\rm \ \ \ If $J$ is an index of the  relation $R$ then  \begin{displaymath}J{\MPperldom{}}\ms{2}{\subseteq}\ms{2}R{\MPperldom{}}\ms{6}{\wedge}\ms{6}J{\MPperrdom{}}\ms{2}{\subseteq}\ms{2}R{\MPperrdom{}}~~.\end{displaymath}It follows that \begin{displaymath}J{\MPldom{}}\ms{2}{=}\ms{2}J{\MPperldom{}}\ms{6}{\wedge}\ms{6}J{\MPrdom{}}\ms{2}{=}\ms{2}J{\MPperrdom{}}~~.\end{displaymath}That is,  an index is a core relation.
}%
\end{Lemma}%
\begin{proof}
We first prove that $J{\MPperldom{}}\ms{2}{\subseteq}\ms{2}R{\MPperldom{}}$.  
\begin{mpdisplay}{0.15em}{6.5mm}{0mm}{2}
	$R{\MPperldom{}}$\push\-\\
	$=$	\>	\>$\{$	\>\+\+\+definition\-\-$~~~ \}$\pop\\
	$R\leftsymdivision{}R\ms{2}{\MPcomp}\ms{2}R{\MPldom{}}$\push\-\\
	$\supseteq$	\>	\>$\{$	\>\+\+\+\ref{gen.index}(a) and monotonicity\-\-$~~~ \}$\pop\\
	$R\leftsymdivision{}R\ms{2}{\MPcomp}\ms{2}J{\MPldom{}}$\push\-\\
	$\supseteq$	\>	\>$\{$	\>\+\+\+see below\-\-$~~~ \}$\pop\\
	$J{\MPperldom{}}~~.$
\end{mpdisplay}
The last step in the above calculation proceeds as follows.
\begin{mpdisplay}{0.15em}{6.5mm}{0mm}{2}
	$J{\MPperldom{}}\ms{4}{\subseteq}\ms{4}R\leftsymdivision{}R\ms{2}{\MPcomp}\ms{2}J{\MPldom{}}$\push\-\\
	$\Leftarrow$	\>	\>$\{$	\>\+\+\+$(J{\MPperldom{}}){\MPrdom{}}\ms{2}{=}\ms{2}J{\MPldom{}}$ (so $J{\MPperldom{}}\ms{3}{=}\ms{3}J{\MPperldom{}}\ms{1}{\MPcomp}\ms{1}J{\MPldom{}}$)  and $J{\MPldom{}}\ms{1}{\MPcomp}\ms{1}J{\MPldom{}}\ms{2}{=}\ms{2}J{\MPldom{}}$\\
	monotonicity\-\-$~~~ \}$\pop\\
	$J{\MPperldom{}}\ms{4}{\subseteq}\ms{4}R\leftsymdivision{}R$\push\-\\
	$=$	\>	\>$\{$	\>\+\+\+definition of $R\leftsymdivision{}R$\-\-$~~~ \}$\pop\\
	$J{\MPperldom{}}\ms{4}{\subseteq}\ms{4}R{/}R\ms{2}{\cap}\ms{2}(R{/}R)^{\MPrev}$\push\-\\
	$=$	\>	\>$\{$	\>\+\+\+$J{\MPperldom{}}\ms{2}{=}\ms{2}(J{\MPperldom{}})^{\MPrev}$\-\-$~~~ \}$\pop\\
	$J{\MPperldom{}}\ms{4}{\subseteq}\ms{4}R{/}R$\push\-\\
	$=$	\>	\>$\{$	\>\+\+\+shunting \-\-$~~~ \}$\pop\\
	$J{\MPperldom{}}\ms{1}{\MPcomp}\ms{1}R\ms{5}{\subseteq}\ms{5}R~~.$
\end{mpdisplay}
We continue with the lefthand side of the above inclusion.
\begin{mpdisplay}{0.15em}{6.5mm}{0mm}{2}
	$J{\MPperldom{}}\ms{1}{\MPcomp}\ms{1}R$\push\-\\
	$=$	\>	\>$\{$	\>\+\+\+\ref{gen.index}(b)\-\-$~~~ \}$\pop\\
	$J{\MPperldom{}}\ms{1}{\MPcomp}\ms{1}R{\MPperldom{}}\ms{1}{\MPcomp}\ms{1}J\ms{1}{\MPcomp}\ms{1}R{\MPperrdom{}}$\push\-\\
	$=$	\>	\>$\{$	\>\+\+\+$(J{\MPperldom{}}){\MPrdom{}}\ms{1}{=}\ms{1}J{\MPldom{}}$ and domains: (\ref{ldom.and.rdom})\-\-$~~~ \}$\pop\\
	$J{\MPperldom{}}\ms{1}{\MPcomp}\ms{1}J{\MPldom{}}\ms{1}{\MPcomp}\ms{1}R{\MPperldom{}}\ms{1}{\MPcomp}\ms{1}J{\MPldom{}}\ms{1}{\MPcomp}\ms{1}J\ms{1}{\MPcomp}\ms{1}R{\MPperrdom{}}$\push\-\\
	$=$	\>	\>$\{$	\>\+\+\+\ref{gen.index}(c)\-\-$~~~ \}$\pop\\
	$J{\MPperldom{}}\ms{1}{\MPcomp}\ms{1}J{\MPldom{}}\ms{1}{\MPcomp}\ms{1}J\ms{1}{\MPcomp}\ms{1}R{\MPperrdom{}}$\push\-\\
	$=$	\>	\>$\{$	\>\+\+\+(corefexive and per) domains: (\ref{ldom.and.rdom}) and (\ref{per.leftandrightdomain.eq})\-\-$~~~ \}$\pop\\
	$J\ms{1}{\MPcomp}\ms{1}R{\MPperrdom{}}$\push\-\\
	$\subseteq$	\>	\>$\{$	\>\+\+\+\ref{gen.index}(a)\-\-$~~~ \}$\pop\\
	$R\ms{1}{\MPcomp}\ms{1}R{\MPperrdom{}}$\push\-\\
	$=$	\>	\>$\{$	\>\+\+\+per domains: (\ref{per.leftandrightdomain.eq})\-\-$~~~ \}$\pop\\
	$R~~.$
\end{mpdisplay}
We conclude that $J{\MPperldom{}}\ms{2}{\subseteq}\ms{2}R{\MPperldom{}}$.  The equation $J{\MPperldom{}}\ms{1}{=}\ms{1}J{\MPldom{}}$ uses anti-symmetry.
\begin{mpdisplay}{0.15em}{6.5mm}{0mm}{2}
	$J{\MPperldom{}}$\push\-\\
	$\supseteq$	\>	\>$\{$	\>\+\+\+per domains: (\ref{per.leftdomain}), and reflexivity of $J\leftsymdivision{}J$\-\-$~~~ \}$\pop\\
	$J{\MPldom{}}$\push\-\\
	$=$	\>	\>$\{$	\>\+\+\+\ref{gen.index}(c)\-\-$~~~ \}$\pop\\
	$J{\MPldom{}}\ms{1}{\MPcomp}\ms{1}R{\MPperldom{}}\ms{1}{\MPcomp}\ms{1}J{\MPldom{}}$\push\-\\
	$\supseteq$	\>	\>$\{$	\>\+\+\+$J{\MPperldom{}}\ms{2}{\subseteq}\ms{2}R{\MPperldom{}}$  (see above), composition of coreflexives is idempotent\-\-$~~~ \}$\pop\\
	$J{\MPldom{}}~~.$
\end{mpdisplay}
The other two properties are symmetrical.
\end{proof}

An immediate corollary of Lemma \ref{index.per.is.pid} is the following theorem.
\begin{Theorem}\label{index.closure}{\rm \ \ \ If $J$ is an index (of some relation)  then $J$ is an index of $J$. 
}
\end{Theorem}
\begin{proof}
Suppose $J$ is an index of $R$.  Then we have to prove the properties \ref{gen.index}(a), (b), (c) and (d)
with $R\ms{1}{:=}\ms{1}J.$  These are the properties:
\begin{description}
\item[(e)]$J\ms{1}{\subseteq}\ms{1}J~~,$ 
\item[(f)]$J{\MPperldom{}}\ms{1}{\MPcomp}\ms{1}J\ms{1}{\MPcomp}\ms{1}J{\MPperrdom{}}\ms{3}{=}\ms{3}J~~,$ 
\item[(g)]$J{\MPldom{}}\ms{1}{\MPcomp}\ms{1}J{\MPperldom{}}\ms{1}{\MPcomp}\ms{1}J{\MPldom{}}\ms{3}{=}\ms{3}J{\MPldom{}}~~,$ 
\item[(h)]$J{\MPrdom{}}\ms{1}{\MPcomp}\ms{1}J{\MPperrdom{}}\ms{1}{\MPcomp}\ms{1}J{\MPrdom{}}\ms{3}{=}\ms{3}J{\MPrdom{}}~~.$ 
 
\end{description}
Properties (e) and (f) are true of all relations $J$.   Properties (g) and (h) follow from  Lemma \ref{index.per.is.pid}
and the fact that composition of coreflexives is idempotent.
\end{proof}

The indexes of a relation are uniquely defined by their left and right domains.  See Corollary 
\ref{index.doms},  which is an immediate consequence of the following lemma.
\begin{Lemma}\label{index.dom.def}{\rm \ \ \ Suppose $J$ is an  index of the relation $R$.  Then\begin{displaymath}J\ms{4}{=}\ms{4}J{\MPldom{}}\ms{1}{\MPcomp}\ms{1}R\ms{1}{\MPcomp}\ms{1}J{\MPrdom{}}~~.\end{displaymath}
}%
\end{Lemma}%
\begin{shortproof}
\begin{mpdisplay}{0.15em}{6.5mm}{0mm}{2}
	$J$\push\-\\
	$=$	\>	\>$\{$	\>\+\+\+domains: (\ref{ldom.and.rdom})\-\-$~~~ \}$\pop\\
	$J{\MPldom{}}\ms{1}{\MPcomp}\ms{1}J\ms{1}{\MPcomp}\ms{1}J{\MPrdom{}}$\push\-\\
	$=$	\>	\>$\{$	\>\+\+\+\ref{gen.index}(c) and (d)\-\-$~~~ \}$\pop\\
	$J{\MPldom{}}\ms{1}{\MPcomp}\ms{1}R{\MPperldom{}}\ms{1}{\MPcomp}\ms{1}J{\MPldom{}}\ms{1}{\MPcomp}\ms{1}J\ms{1}{\MPcomp}\ms{1}J{\MPrdom{}}\ms{1}{\MPcomp}\ms{1}R{\MPperrdom{}}\ms{1}{\MPcomp}\ms{1}J{\MPrdom{}}$\push\-\\
	$=$	\>	\>$\{$	\>\+\+\+domains: (\ref{ldom.and.rdom})\-\-$~~~ \}$\pop\\
	$J{\MPldom{}}\ms{1}{\MPcomp}\ms{1}R{\MPperldom{}}\ms{1}{\MPcomp}\ms{1}J\ms{1}{\MPcomp}\ms{1}R{\MPperrdom{}}\ms{1}{\MPcomp}\ms{1}J{\MPrdom{}}$\push\-\\
	$=$	\>	\>$\{$	\>\+\+\+\ref{gen.index}(b)\-\-$~~~ \}$\pop\\
	$J{\MPldom{}}\ms{1}{\MPcomp}\ms{1}R\ms{1}{\MPcomp}\ms{1}J{\MPrdom{}}~~.$
\end{mpdisplay}
\end{shortproof}

\begin{Corollary}\label{index.doms}{\rm \ \ \ Suppose $J$ and $K$ are both indexes of the relation $R$.  Then\begin{displaymath}J\ms{1}{=}\ms{1}K\ms{8}{\equiv}\ms{8}J{\MPldom{}}\ms{1}{=}\ms{1}K{\MPldom{}}\ms{4}{\wedge}\ms{4}J{\MPrdom{}}\ms{1}{=}\ms{1}K{\MPrdom{}}~~.\end{displaymath}
}%
\end{Corollary}%
\begin{proof}
Implication is an immediate consequence of Leibniz's rule.  For the ``if'' part, we assume that 
$J{\MPldom{}}\ms{1}{=}\ms{1}K{\MPldom{}}$ and $J{\MPrdom{}}\ms{1}{=}\ms{1}K{\MPrdom{}}$.  Then
\begin{mpdisplay}{0.15em}{6.5mm}{0mm}{2}
	$J$\push\-\\
	$=$	\>	\>$\{$	\>\+\+\+$J$ is an index of $R$,  Lemma \ref{index.dom.def}\-\-$~~~ \}$\pop\\
	$J{\MPldom{}}\ms{1}{\MPcomp}\ms{1}R\ms{1}{\MPcomp}\ms{1}J{\MPrdom{}}$\push\-\\
	$=$	\>	\>$\{$	\>\+\+\+assumption:  $J{\MPldom{}}\ms{1}{=}\ms{1}K{\MPldom{}}\ms{4}{\wedge}\ms{4}J{\MPrdom{}}\ms{1}{=}\ms{1}K{\MPrdom{}}$\-\-$~~~ \}$\pop\\
	$K{\MPldom{}}\ms{1}{\MPcomp}\ms{1}R\ms{1}{\MPcomp}\ms{1}K{\MPrdom{}}$\push\-\\
	$=$	\>	\>$\{$	\>\+\+\+$K$  is an index of $R$,  Lemma \ref{index.dom.def} with $J\ms{1}{:=}\ms{1}K$\-\-$~~~ \}$\pop\\
	$K~~.$
\end{mpdisplay}
\end{proof}

The following lemma becomes relevant when we study indexes of difunctions.  (See Section 
\ref{Indices and Difunctions}.)
\begin{Lemma}\label{index.simp}{\rm \ \ \ Suppose  $J$ is an index of $R$.  Then\begin{displaymath}R\ms{1}{\MPcomp}\ms{1}J^{\MPrev}\ms{1}{\MPcomp}\ms{1}R\ms{5}{=}\ms{5}R\ms{1}{\MPcomp}\ms{1}R^{\MPrev}\ms{1}{\MPcomp}\ms{1}R~~.\end{displaymath}
}%
\end{Lemma}%
\begin{shortproof}
\begin{mpdisplay}{0.15em}{6.5mm}{0mm}{2}
	$R\ms{1}{\MPcomp}\ms{1}J^{\MPrev}\ms{1}{\MPcomp}\ms{1}R$\push\-\\
	$=$	\>	\>$\{$	\>\+\+\+per domains: (\ref{per.rightdomain}) and (\ref{per.leftdomain})\-\-$~~~ \}$\pop\\
	$R\ms{1}{\MPcomp}\ms{1}R{\MPperrdom{}}\ms{1}{\MPcomp}\ms{1}J^{\MPrev}\ms{1}{\MPcomp}\ms{1}R{\MPperldom{}}\ms{1}{\MPcomp}\ms{1}R$\push\-\\
	$=$	\>	\>$\{$	\>\+\+\+\ref{gen.index}(b) and converse\-\-$~~~ \}$\pop\\
	$R\ms{1}{\MPcomp}\ms{1}R^{\MPrev}\ms{1}{\MPcomp}\ms{1}R~~.$
\end{mpdisplay}
\end{shortproof}

We now formulate a couple  of lemmas that lead to Lemma \ref{R-perleft} which, in turn,  leads to
Theorem \ref{index-perdoms}.
\begin{Lemma}\label{R-perleftJR-perleft}{\rm \ \ \ Suppose  $J$ is an index of $R$.  Then 
$R{\MPperldom{}}\ms{1}{\MPcomp}\ms{1}J{\MPldom{}}\ms{1}{\MPcomp}\ms{1}R{\MPperldom{}}$  and   $R{\MPperrdom{}}\ms{1}{\MPcomp}\ms{1}J{\MPrdom{}}\ms{1}{\MPcomp}\ms{1}R{\MPperrdom{}}$ are pers.
}%
\end{Lemma}%
\begin{proof}
We prove that \begin{displaymath}R{\MPperldom{}}\ms{1}{\MPcomp}\ms{1}J{\MPldom{}}\ms{1}{\MPcomp}\ms{1}R{\MPperldom{}}\ms{5}{=}\ms{5}R{\MPperldom{}}\ms{1}{\MPcomp}\ms{1}J{\MPldom{}}\ms{1}{\MPcomp}\ms{1}R{\MPperldom{}}\ms{1}{\MPcomp}\ms{1}(R{\MPperldom{}}\ms{1}{\MPcomp}\ms{1}J{\MPldom{}}\ms{1}{\MPcomp}\ms{1}R{\MPperldom{}})^{\MPrev}~~.\end{displaymath}We have:
\begin{mpdisplay}{0.15em}{6.5mm}{0mm}{2}
	$R{\MPperldom{}}\ms{1}{\MPcomp}\ms{1}J{\MPldom{}}\ms{1}{\MPcomp}\ms{1}R{\MPperldom{}}\ms{1}{\MPcomp}\ms{1}(R{\MPperldom{}}\ms{1}{\MPcomp}\ms{1}J{\MPldom{}}\ms{1}{\MPcomp}\ms{1}R{\MPperldom{}})^{\MPrev}$\push\-\\
	$=$	\>	\>$\{$	\>\+\+\+$R{\MPperldom{}}$ is a per, $J{\MPldom{}}$ is a coreflexive,  converse\-\-$~~~ \}$\pop\\
	$R{\MPperldom{}}\ms{1}{\MPcomp}\ms{1}J{\MPldom{}}\ms{1}{\MPcomp}\ms{1}R{\MPperldom{}}\ms{1}{\MPcomp}\ms{1}J{\MPldom{}}\ms{1}{\MPcomp}\ms{1}R{\MPperldom{}}$\push\-\\
	$=$	\>	\>$\{$	\>\+\+\+\ref{gen.index}(c)\-\-$~~~ \}$\pop\\
	$R{\MPperldom{}}\ms{1}{\MPcomp}\ms{1}J{\MPldom{}}\ms{1}{\MPcomp}\ms{1}R{\MPperldom{}}~~.$
\end{mpdisplay}
\end{proof}

\begin{Lemma}\label{R-perleftleft}{\rm \ \ \ Suppose  $J$ is an index of $R$.  Then\begin{displaymath}(R{\MPperldom{}}\ms{1}{\MPcomp}\ms{1}J{\MPldom{}}\ms{1}{\MPcomp}\ms{1}R{\MPperldom{}}){\MPldom{}}\ms{3}{=}\ms{3}R{\MPldom{}}~~.\end{displaymath}Symmetrically,\begin{displaymath}(R{\MPperrdom{}}\ms{1}{\MPcomp}\ms{1}J{\MPrdom{}}\ms{1}{\MPcomp}\ms{1}R{\MPperrdom{}}){\MPrdom{}}\ms{3}{=}\ms{3}R{\MPrdom{}}~~.\end{displaymath}
}%
\end{Lemma}%
\begin{shortproof}
\begin{mpdisplay}{0.15em}{6.5mm}{0mm}{2}
	$(R{\MPperldom{}}\ms{1}{\MPcomp}\ms{1}J{\MPldom{}}\ms{1}{\MPcomp}\ms{1}R{\MPperldom{}}){\MPldom{}}$\push\-\\
	$=$	\>	\>$\{$	\>\+\+\+domains: Theorem \ref{domains}(c),  $(R{\MPperldom{}}){\MPldom{}}\ms{3}{=}\ms{3}R{\MPldom{}}$\-\-$~~~ \}$\pop\\
	$(R{\MPperldom{}}\ms{1}{\MPcomp}\ms{1}J{\MPldom{}}\ms{1}{\MPcomp}\ms{1}R{\MPldom{}}){\MPldom{}}$\push\-\\
	$=$	\>	\>$\{$	\>\+\+\+by \ref{gen.index}(a),  $J{\MPldom{}}\ms{1}{\subseteq}\ms{1}R{\MPldom{}}$, domains\-\-$~~~ \}$\pop\\
	$(R{\MPperldom{}}\ms{1}{\MPcomp}\ms{1}J){\MPldom{}}$\push\-\\
	$=$	\>	\>$\{$	\>\+\+\+by \ref{gen.index}(a),  $J{\MPrdom{}}\ms{1}{\subseteq}\ms{1}R{\MPrdom{}}$, domains\-\-$~~~ \}$\pop\\
	$(R{\MPperldom{}}\ms{1}{\MPcomp}\ms{1}J\ms{1}{\MPcomp}\ms{1}R{\MPrdom{}}){\MPldom{}}$\push\-\\
	$=$	\>	\>$\{$	\>\+\+\+domains: Theorem \ref{domains}(c),  $(R{\MPperrdom{}}){\MPldom{}}\ms{3}{=}\ms{3}R{\MPrdom{}}$\-\-$~~~ \}$\pop\\
	$(R{\MPperldom{}}\ms{1}{\MPcomp}\ms{1}J\ms{1}{\MPcomp}\ms{1}R{\MPperrdom{}}){\MPldom{}}$\push\-\\
	$=$	\>	\>$\{$	\>\+\+\+\ref{gen.index}(b)\-\-$~~~ \}$\pop\\
	$R{\MPldom{}}~~.$
\end{mpdisplay}
\end{shortproof}

\begin{Lemma}\label{R-perleft}{\rm \ \ \ Suppose  $J$ is an index of $R$.  Then
\begin{description}
\item[(a)]$R{\MPperldom{}}\ms{1}{\MPcomp}\ms{1}J{\MPldom{}}\ms{1}{\MPcomp}\ms{1}R{\MPperldom{}}\ms{3}{=}\ms{3}R{\MPperldom{}}~~,$ 
\item[(b)]$R{\MPperrdom{}}\ms{1}{\MPcomp}\ms{1}J{\MPrdom{}}\ms{1}{\MPcomp}\ms{1}R{\MPperrdom{}}\ms{3}{=}\ms{3}R{\MPperrdom{}}~~.$ 
 
\end{description}

}%
\end{Lemma}%
\begin{shortproof}
\begin{mpdisplay}{0.15em}{6.5mm}{0mm}{2}
	$R{\MPperldom{}}$\push\-\\
	$=$	\>	\>$\{$	\>\+\+\+$R{\MPperldom{}}$ is a per\-\-$~~~ \}$\pop\\
	$R{\MPperldom{}}\ms{1}{\MPcomp}\ms{1}R{\MPperldom{}}\ms{1}{\MPcomp}\ms{1}R{\MPperldom{}}$\push\-\\
	$\supseteq$	\>	\>$\{$	\>\+\+\+$R{\MPperldom{}}\ms{1}{\supseteq}\ms{1}R{\MPldom{}}$\-\-$~~~ \}$\pop\\
	$R{\MPperldom{}}\ms{1}{\MPcomp}\ms{1}R{\MPldom{}}\ms{1}{\MPcomp}\ms{1}R{\MPperldom{}}$\push\-\\
	$\supseteq$	\>	\>$\{$	\>\+\+\+$J$ is an index of $R$;  Definition \ref{gen.index}(a) and monotonicity\-\-$~~~ \}$\pop\\
	$R{\MPperldom{}}\ms{1}{\MPcomp}\ms{1}J{\MPldom{}}\ms{1}{\MPcomp}\ms{1}R{\MPperldom{}}$\push\-\\
	$=$	\>	\>$\{$	\>\+\+\+$R{\MPperldom{}}$ is a per\-\-$~~~ \}$\pop\\
	$R{\MPperldom{}}\ms{1}{\MPcomp}\ms{1}J{\MPldom{}}\ms{1}{\MPcomp}\ms{1}R{\MPperldom{}}\ms{1}{\MPcomp}\ms{1}R{\MPperldom{}}$\push\-\\
	$\supseteq$	\>	\>$\{$	\>\+\+\+Lemma \ref{R-perleftJR-perleft}:   $R{\MPperldom{}}\ms{1}{\MPcomp}\ms{1}J{\MPldom{}}\ms{1}{\MPcomp}\ms{1}R{\MPperldom{}}$ is a per\-\-$~~~ \}$\pop\\
	$(R{\MPperldom{}}\ms{1}{\MPcomp}\ms{1}J{\MPldom{}}\ms{1}{\MPcomp}\ms{1}R{\MPperldom{}}){\MPldom{}}\ms{2}{\MPcomp}\ms{2}R{\MPperldom{}}$\push\-\\
	$=$	\>	\>$\{$	\>\+\+\+Lemma \ref{R-perleftleft}\-\-$~~~ \}$\pop\\
	$R{\MPldom{}}\ms{2}{\MPcomp}\ms{2}R{\MPperldom{}}$\push\-\\
	$=$	\>	\>$\{$	\>\+\+\+$(R{\MPperldom{}}){\MPldom{}}\ms{3}{=}\ms{3}R{\MPldom{}}$\-\-$~~~ \}$\pop\\
	$R{\MPperldom{}}~~.$
\end{mpdisplay}
By anti-symmetry of the subset relation we have proved (a).  Property (b) is symmetrical.
\end{shortproof}

\begin{Theorem}\label{index-perdoms}{\rm \ \ \ Suppose  $J$ is an index of $R$.  Then $J{\MPldom{}}$ is an index of $R{\MPperldom{}}$ and $J{\MPrdom{}}$ is an index of $R{\MPperrdom{}}$.
}
\end{Theorem}
\begin{proof}
We prove that $J{\MPldom{}}$ is an index of $R{\MPperldom{}}$.  That $J{\MPrdom{}}$ is an index of $R{\MPperrdom{}}$ is symmetrical.

Instantiating Definition \ref{gen.index} with $R{,}J\ms{2}{:=}\ms{2}R{\MPperldom{}}\ms{1}{,}\ms{1}J{\MPldom{}}$,   our task is to prove the four properties:
\begin{description}
\item[(a)]$J{\MPldom{}}\ms{2}{\subseteq}\ms{2}R{\MPperldom{}}~~,$ 
\item[(b)]$(R{\MPperldom{}}){\MPperldom{}}\ms{1}{\MPcomp}\ms{1}(J{\MPldom{}}){\MPldom{}}\ms{1}{\MPcomp}\ms{1}(R{\MPperrdom{}}){\MPperldom{}}\ms{3}{=}\ms{3}R{\MPperldom{}}~~,$ 
\item[(c)]$(J{\MPldom{}}){\MPldom{}}\ms{1}{\MPcomp}\ms{1}(R{\MPperldom{}}){\MPperldom{}}\ms{1}{\MPcomp}\ms{1}(J{\MPldom{}}){\MPldom{}}\ms{3}{=}\ms{3}(J{\MPldom{}}){\MPldom{}}~~,$ 
\item[(d)]$(J{\MPldom{}}){\MPrdom{}}\ms{1}{\MPcomp}\ms{1}(R{\MPperldom{}}){\MPperrdom{}}\ms{1}{\MPcomp}\ms{1}(J{\MPldom{}}){\MPrdom{}}\ms{3}{=}\ms{3}(J{\MPldom{}}){\MPrdom{}}~~.$ 
 
\end{description}

The proof of property (a) is straightforward:
\begin{mpdisplay}{0.15em}{6.5mm}{0mm}{2}
	$J{\MPldom{}}\ms{2}{\subseteq}\ms{2}R{\MPperldom{}}$\push\-\\
	$\Leftarrow$	\>	\>$\{$	\>\+\+\+$R{\MPldom{}}\ms{1}{\subseteq}\ms{1}R{\MPperldom{}}$, transitivity\-\-$~~~ \}$\pop\\
	$J{\MPldom{}}\ms{2}{\subseteq}\ms{2}R{\MPldom{}}$\push\-\\
	$\Leftarrow$	\>	\>$\{$	\>\+\+\+monotonicity\-\-$~~~ \}$\pop\\
	$J\ms{1}{\subseteq}\ms{1}R$\push\-\\
	$=$	\>	\>$\{$	\>\+\+\+$J$ is an index of $R$,  \ref{gen.index}(a)\-\-$~~~ \}$\pop\\
	$\mathsf{true}~~.$
\end{mpdisplay}
Property (b) simplifies using the fact that $(R{\MPperldom{}}){\MPperldom{}}\ms{1}{=}\ms{1}R{\MPperldom{}}$,  $(R{\MPperrdom{}}){\MPperldom{}}\ms{1}{=}\ms{1}R{\MPperrdom{}}$ and    $J{\MPldom{}}\ms{1}{=}\ms{1}(J{\MPldom{}}){\MPldom{}}$ to:
\begin{description}
\item[(b')]$R{\MPperldom{}}\ms{1}{\MPcomp}\ms{1}J{\MPldom{}}\ms{1}{\MPcomp}\ms{1}R{\MPperrdom{}}\ms{3}{=}\ms{3}R{\MPperldom{}}~~,$ 
 
\end{description}
This is the first  of the two properties proved in Lemma \ref{R-perleft}.
Using the fact that $(R{\MPperldom{}}){\MPperldom{}}\ms{1}{=}\ms{1}R{\MPperldom{}}$  and    $J{\MPldom{}}\ms{1}{=}\ms{1}(J{\MPldom{}}){\MPldom{}}$,  property (c) is the same as property (c) of Definition 
\ref{gen.index};  similarly, using the fact that 
 $R{\MPperldom{}}\ms{1}{=}\ms{1}(R{\MPperldom{}}){\MPperrdom{}}$,  and $J{\MPldom{}}\ms{1}{=}\ms{1}(J{\MPldom{}}){\MPrdom{}}$, property (d) is also the same as property (c) of Definition 
\ref{gen.index}.
\end{proof}

We show later that the converse of Theorem \ref{index-perdoms} is a prescription for constructing an index of an
arbitrary relation.  See Theorem \ref{per.to.relation}.  
\begin{Theorem}\label{index.iso}{\rm \ \ \ If $R$ and $S$ are isomorphic relations then indexes of $R$ and $S$ are also
isomorphic.   In particular,   indexes  of a relation  $R$  are  isomorphic.
}
\end{Theorem}
\begin{proof}
Suppose $\phi$ and $\psi$ witness the isomorphism $R\ms{1}{\cong}\ms{1}S$ and $J$ is an index of $R$ and $K$ is an
index of $S$.   We verify that $\lambda$ and $\rho$ defined by \begin{displaymath}\lambda\ms{2}{=}\ms{2}J{\MPldom{}}\ms{1}{\MPcomp}\ms{1}R{\MPperldom{}}\ms{1}{\MPcomp}\ms{1}\phi\ms{1}{\MPcomp}\ms{1}S{\MPperldom{}}\ms{1}{\MPcomp}\ms{1}K{\MPldom{}}\ms{7}{\wedge}\ms{7}\rho\ms{2}{=}\ms{2}J{\MPrdom{}}\ms{1}{\MPcomp}\ms{1}R{\MPperrdom{}}\ms{1}{\MPcomp}\ms{1}\psi\ms{1}{\MPcomp}\ms{1}S{\MPperrdom{}}\ms{1}{\MPcomp}\ms{1}K{\MPrdom{}}\end{displaymath}witness the isomorphism $J\ms{1}{\cong}\ms{1}K$.  

The task is to verify that \begin{displaymath}J{\MPldom{}}\ms{2}{=}\ms{2}\lambda\ms{1}{\MPcomp}\ms{1}\lambda^{\MPrev}\ms{7}{\wedge}\ms{7}\lambda^{\MPrev}\ms{1}{\MPcomp}\ms{1}\lambda\ms{2}{=}\ms{2}K{\MPldom{}}\ms{7}{\wedge}\ms{7}\rho\ms{1}{\MPcomp}\ms{1}\rho^{\MPrev}\ms{2}{=}\ms{2}J{\MPrdom{}}\ms{7}{\wedge}\ms{7}\rho^{\MPrev}\ms{1}{\MPcomp}\ms{1}\rho\ms{2}{=}\ms{2}K{\MPrdom{}}\end{displaymath}and
\begin{mpdisplay}{0.15em}{6.5mm}{0mm}{2}
	$J\ms{3}{=}\ms{3}\lambda\ms{1}{\MPcomp}\ms{1}K\ms{1}{\MPcomp}\ms{1}\rho^{\MPrev}~~.$
\end{mpdisplay}
The four domain properties are all essentially the same so we only verify the first conjunct: 
\begin{mpdisplay}{0.15em}{6.5mm}{0mm}{2}
	$\lambda\ms{1}{\MPcomp}\ms{1}\lambda^{\MPrev}$\push\-\\
	$=$	\>	\>$\{$	\>\+\+\+definition, converse\-\-$~~~ \}$\pop\\
	$J{\MPldom{}}\ms{1}{\MPcomp}\ms{1}R{\MPperldom{}}\ms{1}{\MPcomp}\ms{1}\phi\ms{1}{\MPcomp}\ms{1}S{\MPperldom{}}\ms{1}{\MPcomp}\ms{1}K{\MPldom{}}\ms{1}{\MPcomp}\ms{1}S{\MPperldom{}}\ms{1}{\MPcomp}\ms{1}\phi^{\MPrev}\ms{1}{\MPcomp}\ms{1}R{\MPperldom{}}\ms{1}{\MPcomp}\ms{1}J{\MPldom{}}$\push\-\\
	$=$	\>	\>$\{$	\>\+\+\+$K$ is an index of $S$,  Lemma \ref{R-perleftJR-perleft} with $J{,}R\ms{1}{:=}\ms{1}K{,}S$\-\-$~~~ \}$\pop\\
	$J{\MPldom{}}\ms{1}{\MPcomp}\ms{1}R{\MPperldom{}}\ms{1}{\MPcomp}\ms{1}\phi\ms{1}{\MPcomp}\ms{1}S{\MPperldom{}}\ms{1}{\MPcomp}\ms{1}\phi^{\MPrev}\ms{1}{\MPcomp}\ms{1}R{\MPperldom{}}\ms{1}{\MPcomp}\ms{1}J{\MPldom{}}$\push\-\\
	$=$	\>	\>$\{$	\>\+\+\+Theorem \ref{index.iso}\-\-$~~~ \}$\pop\\
	$J{\MPldom{}}\ms{1}{\MPcomp}\ms{1}R{\MPperldom{}}\ms{1}{\MPcomp}\ms{1}R{\MPperldom{}}\ms{1}{\MPcomp}\ms{1}R{\MPperldom{}}\ms{1}{\MPcomp}\ms{1}J{\MPldom{}}$\push\-\\
	$=$	\>	\>$\{$	\>\+\+\+$R{\MPperldom{}}$ is a per,  $J$ is an index of $R$,  Definition \ref{gen.index}(c)\-\-$~~~ \}$\pop\\
	$J{\MPldom{}}~~.$
\end{mpdisplay}
Finally,
\begin{mpdisplay}{0.15em}{6.5mm}{0mm}{2}
	$\lambda\ms{1}{\MPcomp}\ms{1}K\ms{1}{\MPcomp}\ms{1}\rho^{\MPrev}$\push\-\\
	$=$	\>	\>$\{$	\>\+\+\+definition, converse\-\-$~~~ \}$\pop\\
	$J{\MPldom{}}\ms{1}{\MPcomp}\ms{1}R{\MPperldom{}}\ms{1}{\MPcomp}\ms{1}\phi\ms{1}{\MPcomp}\ms{1}S{\MPperldom{}}\ms{1}{\MPcomp}\ms{1}K{\MPldom{}}\ms{1}{\MPcomp}\ms{1}K\ms{1}{\MPcomp}\ms{1}K{\MPrdom{}}\ms{1}{\MPcomp}\ms{1}S{\MPperrdom{}}\ms{1}{\MPcomp}\ms{1}\psi^{\MPrev}\ms{1}{\MPcomp}\ms{1}R{\MPperrdom{}}\ms{1}{\MPcomp}\ms{1}J{\MPrdom{}}$\push\-\\
	$=$	\>	\>$\{$	\>\+\+\+domains: (\ref{ldom.and.rdom})\-\-$~~~ \}$\pop\\
	$J{\MPldom{}}\ms{1}{\MPcomp}\ms{1}R{\MPperldom{}}\ms{1}{\MPcomp}\ms{1}\phi\ms{1}{\MPcomp}\ms{1}S{\MPperldom{}}\ms{1}{\MPcomp}\ms{1}K\ms{1}{\MPcomp}\ms{1}S{\MPperrdom{}}\ms{1}{\MPcomp}\ms{1}\psi^{\MPrev}\ms{1}{\MPcomp}\ms{1}R{\MPperrdom{}}\ms{1}{\MPcomp}\ms{1}J{\MPrdom{}}$\push\-\\
	$=$	\>	\>$\{$	\>\+\+\+$K$ is an index of $S$,  Definition \ref{gen.index}(b)\-\-$~~~ \}$\pop\\
	$J{\MPldom{}}\ms{1}{\MPcomp}\ms{1}R{\MPperldom{}}\ms{1}{\MPcomp}\ms{1}\phi\ms{1}{\MPcomp}\ms{1}S\ms{1}{\MPcomp}\ms{1}\psi^{\MPrev}\ms{1}{\MPcomp}\ms{1}R{\MPperrdom{}}\ms{1}{\MPcomp}\ms{1}J{\MPrdom{}}$\push\-\\
	$=$	\>	\>$\{$	\>\+\+\+$R\ms{2}{=}\ms{2}\phi\ms{1}{\MPcomp}\ms{1}S\ms{1}{\MPcomp}\ms{1}\psi^{\MPrev}$\-\-$~~~ \}$\pop\\
	$J{\MPldom{}}\ms{1}{\MPcomp}\ms{1}R{\MPperldom{}}\ms{1}{\MPcomp}\ms{1}R\ms{1}{\MPcomp}\ms{1}R{\MPperrdom{}}\ms{1}{\MPcomp}\ms{1}J{\MPrdom{}}$\push\-\\
	$=$	\>	\>$\{$	\>\+\+\+per domains: (\ref{per.leftandrightdomain.eq})\-\-$~~~ \}$\pop\\
	$J{\MPldom{}}\ms{1}{\MPcomp}\ms{1}R\ms{1}{\MPcomp}\ms{1}J{\MPrdom{}}$\push\-\\
	$=$	\>	\>$\{$	\>\+\+\+$J$ is an index of $R$,  Definition \ref{gen.index}(b)\-\-$~~~ \}$\pop\\
	$J~~.$
\end{mpdisplay}
That the indexes of a relation $R$ are isomorphic follows because $R$ is isomorphic to itself (with witnesses $R{\MPldom{}}$
and $R{\MPrdom{}}$), i.e.\ the isomorphism relation is reflexive.
\end{proof}

The construction of the witnesses $\lambda$ and $\rho$ looks very much like the proverbial rabbit out of a hat!   In fact,
they were calculated using the type judgements formulated in Voermans' thesis \cite{Vo99}.  We hope at a
later date to exploit Voermans' calculus in order to make the process of constructing witnesses much
more methodical.  


\subsection{Core Relations}\label{Cores}

Indexes are a special case of what we call ``core'' relations.  (Recall Definition \ref{Core.gen}.)  This section is
about the properties of a ``core'' of a given relation $R$, first introduced in \cite{BO23}.
\begin{Definition}[Core]\label{core}{\rm \ \ \ Suppose $R$ is an arbitrary relation and suppose $C$ is a relation
such that \begin{displaymath}C\ms{4}{=}\ms{4}\lambda\ms{1}{\MPcomp}\ms{1}R\ms{1}{\MPcomp}\ms{1}\rho^{\MPrev}\end{displaymath}for some relations $\lambda$ and $\rho$ satisfying  \begin{displaymath}R{\MPperldom{}}\ms{3}{=}\ms{3}\lambda^{\MPrev}\ms{1}{\MPcomp}\ms{1}\lambda\ms{7}{\wedge}\ms{7}\lambda{\MPldom{}}\ms{2}{=}\ms{2}\lambda\ms{1}{\MPcomp}\ms{1}\lambda^{\MPrev}\ms{7}{\wedge}\ms{7}R{\MPperrdom{}}\ms{3}{=}\ms{3}\rho^{\MPrev}\ms{1}{\MPcomp}\ms{1}\rho\ms{7}{\wedge}\ms{7}\rho{\MPldom{}}\ms{2}{=}\ms{2}\rho\ms{1}{\MPcomp}\ms{1}\rho^{\MPrev}~~.\end{displaymath}Then $C$ is said to be a \emph{core of}  $R$ \emph{as witnessed by} $\lambda$ \emph{and} $\rho$.
}
\QED
\end{Definition}

(The terminology just introduced anticipates Theorem \ref{core.perdomain} which establishes
that a core of a relation is indeed a  core relation according to Definition \ref{Core.gen}.)

The existence of a core of a given relation $R$ has a constructive element:  it is necessary to construct the
``witnesses'' $\lambda$ and $\rho$.  In general, given a per $P$,  a functional relation $f$ with the property that $P$ equals 
$f^{\MPrev}\ms{1}{\MPcomp}\ms{1}f$ is called a ``splitting'' of $P$.  Constructing a core of relation $R$ thus involves  ``splitting '' the pers $R{\MPperldom{}}$
and $R{\MPperrdom{}}$ into  functional  relations  $\lambda$ and $\rho$.   As with indexes,  the existence of cores is not derivable in
point-free relation algebra.    However, just as for
indexes,  all cores of a given relation are isomorphic  in the sense of Definition \ref{rel.iso}.  
See   Section \ref{per.difun.char} for further discussion of the construction of cores of pers.  

Immediately obvious  is that an index of a relation  is a core of the relation:
\begin{Theorem}\label{core.existence}{\rm \ \ \ Suppose $R$ is an arbitrary relation and suppose $J$ is an index of $R$.  Then $J$ 
is a  core of $R$ as witnessed by $J{\MPldom{}}\ms{1}{\MPcomp}\ms{1}R{\MPperldom{}}$ and $J{\MPrdom{}}\ms{1}{\MPcomp}\ms{1}R{\MPperrdom{}}$.
}
\end{Theorem}
\begin{shortproof}
First,
\begin{mpdisplay}{0.15em}{6.5mm}{0mm}{2}
	$J$\push\-\\
	$=$	\>	\>$\{$	\>\+\+\+Lemma \ref{index.dom.def}\-\-$~~~ \}$\pop\\
	$J{\MPldom{}}\ms{1}{\MPcomp}\ms{1}R\ms{1}{\MPcomp}\ms{1}J{\MPrdom{}}$\push\-\\
	$=$	\>	\>$\{$	\>\+\+\+per domains: (\ref{per.leftandrightdomain.eq})\-\-$~~~ \}$\pop\\
	$J{\MPldom{}}\ms{1}{\MPcomp}\ms{1}R{\MPperldom{}}\ms{1}{\MPcomp}\ms{1}R\ms{1}{\MPcomp}\ms{1}R{\MPperrdom{}}\ms{1}{\MPcomp}\ms{1}J{\MPrdom{}}$\push\-\\
	$=$	\>	\>$\{$	\>\+\+\+converse, domains are coreflexive\-\-$~~~ \}$\pop\\
	$(J{\MPldom{}}\ms{1}{\MPcomp}\ms{1}R{\MPperldom{}})\ms{1}{\MPcomp}\ms{1}R\ms{1}{\MPcomp}\ms{1}(J{\MPrdom{}}\ms{1}{\MPcomp}\ms{1}R{\MPperrdom{}})^{\MPrev}~~.$
\end{mpdisplay}
This establishes the required property of $C$ in Definition \ref{core}, with $C\ms{1}{:=}\ms{1}J$.
(The parentheses in the last line of the calculation indicate the definitions of the splittings $\lambda$ and $\rho$.)  
Second,
\begin{mpdisplay}{0.15em}{6.5mm}{0mm}{2}
	$(J{\MPldom{}}\ms{1}{\MPcomp}\ms{1}R{\MPperldom{}})^{\MPrev}\ms{1}{\MPcomp}\ms{1}J{\MPldom{}}\ms{1}{\MPcomp}\ms{1}R{\MPperldom{}}$\push\-\\
	$=$	\>	\>$\{$	\>\+\+\+converse,  $(R{\MPperldom{}})^{\MPrev}\ms{1}{=}\ms{1}R{\MPperldom{}}$ and $(J{\MPldom{}})^{\MPrev}\ms{1}{\MPcomp}\ms{1}J{\MPldom{}}\ms{2}{=}\ms{2}J{\MPldom{}}$\-\-$~~~ \}$\pop\\
	$R{\MPperldom{}}\ms{1}{\MPcomp}\ms{1}J{\MPldom{}}\ms{1}{\MPcomp}\ms{1}R{\MPperrdom{}}$\push\-\\
	$=$	\>	\>$\{$	\>\+\+\+Lemma \ref{R-perleft}\-\-$~~~ \}$\pop\\
	$R{\MPperldom{}}~~.$
\end{mpdisplay}
Third,
\begin{mpdisplay}{0.15em}{6.5mm}{0mm}{2}
	$J{\MPldom{}}\ms{1}{\MPcomp}\ms{1}R{\MPperldom{}}\ms{1}{\MPcomp}\ms{1}(J{\MPldom{}}\ms{1}{\MPcomp}\ms{1}R{\MPperldom{}})^{\MPrev}$\push\-\\
	$=$	\>	\>$\{$	\>\+\+\+converse,  $(J{\MPldom{}})^{\MPrev}\ms{1}{=}\ms{1}J{\MPldom{}}$ and $R{\MPperldom{}}\ms{1}{\MPcomp}\ms{1}(R{\MPperldom{}})^{\MPrev}\ms{2}{=}\ms{2}R{\MPperldom{}}$\-\-$~~~ \}$\pop\\
	$J{\MPldom{}}\ms{1}{\MPcomp}\ms{1}R{\MPperldom{}}\ms{1}{\MPcomp}\ms{1}J{\MPldom{}}$\push\-\\
	$=$	\>	\>$\{$	\>\+\+\+$J$ is an index of $R$,  Definition \ref{gen.index}(c)\-\-$~~~ \}$\pop\\
	$J{\MPldom{}}$\push\-\\
	$=$	\>	\>$\{$	\>\+\+\+Theorem \ref{index-perdoms}; in particular, $J{\MPldom{}}\ms{1}{\subseteq}\ms{1}R{\MPldom{}}$\-\-$~~~ \}$\pop\\
	$(J{\MPldom{}}\ms{1}{\MPcomp}\ms{1}R{\MPldom{}}){\MPldom{}}$\push\-\\
	$=$	\>	\>$\{$	\>\+\+\+$(R{\MPperldom{}}){\MPldom{}}\ms{1}{=}\ms{1}R{\MPldom{}}$,   domains: Theorem \ref{domains}(c)\-\-$~~~ \}$\pop\\
	$(J{\MPldom{}}\ms{1}{\MPcomp}\ms{1}R{\MPperldom{}}){\MPldom{}}~~.$
\end{mpdisplay}
This establishes the required properties of $\lambda$ in Definition \ref{core} (with $\lambda\ms{2}{:=}\ms{2}J{\MPldom{}}\ms{1}{\MPcomp}\ms{1}R{\MPperldom{}}$).  
The properties of $\rho$ in Definition \ref{core} (with $\rho\ms{2}{:=}\ms{2}J{\MPrdom{}}\ms{1}{\MPcomp}\ms{1}R{\MPperrdom{}}$) are established similarly.
\end{shortproof}

\begin{figure}[h]
\centering \includegraphics{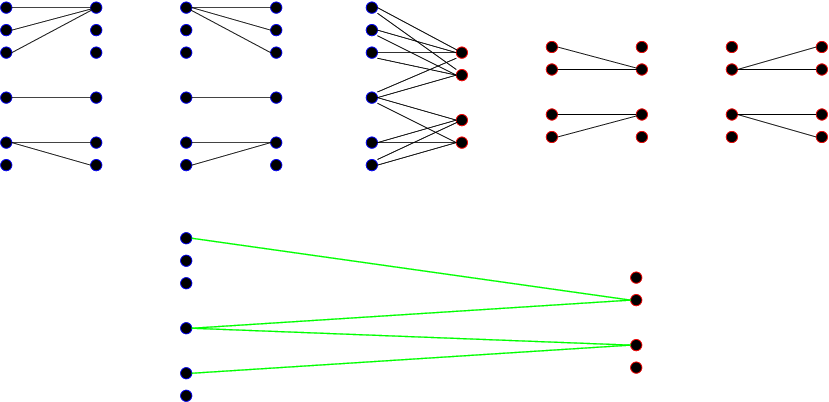}
 
\caption{Decomposition of a Relation into a  Core and Witnesses}\label{fig:coresplitting}
\end{figure}

{}  
Fig.\ \ref{fig:coresplitting}  illustrates Theorem \ref{core.existence} applied to the relation introduced in 
Fig.\ \ref{fig:core}.  The index $J$ is depicted by the green edges in the  lower bipartite  graph.  The decomposition
of the relation in the definition of a  core is illustrated by the row of bipartite graphs at the top;
the relations depicted are, in order, $\lambda^{\MPrev}$, $\lambda$,  $R$, $\rho^{\MPrev}$ and $\rho$.  The composition of the middle three  figures is the
index $J$. 

A number of properties of indexes are derived from the fact that indexes are cores.  The remainder of
this section catalogues such properties.

The name ``core'' in Definition \ref{core} anticipates Theorem \ref{core.perdomain} where we show
that the  relation $C$ is a core relation as defined by Definition \ref{Core.gen}.   Some preliminary lemmas are
needed first.

For later use, we calculate the left and right domains of the core of a relation.
\begin{Lemma}\label{core.domains}{\rm \ \ \ Suppose $R$, $\lambda$,  $\rho$  and $C$ are as in Definition \ref{core}.  Then\begin{displaymath}R{\MPldom{}}\ms{2}{=}\ms{2}\lambda{\MPrdom{}}\ms{5}{\wedge}\ms{5}C{\MPldom{}}\ms{2}{=}\ms{2}\lambda{\MPldom{}}\ms{5}{\wedge}\ms{5}R{\MPrdom{}}\ms{2}{=}\ms{2}\rho{\MPrdom{}}\ms{5}{\wedge}\ms{5}C{\MPrdom{}}\ms{2}{=}\ms{2}\rho{\MPldom{}}~~.\end{displaymath}
}%
\end{Lemma}%
\begin{shortproof}
We prove the middle  two equations.  First,
\begin{mpdisplay}{0.15em}{6.5mm}{0mm}{2}
	$R{\MPrdom{}}$\push\-\\
	$=$	\>	\>$\{$	\>\+\+\+(\ref{per.rightdomain.doms})\-\-$~~~ \}$\pop\\
	$(R{\MPperrdom{}}){\MPldom{}}$\push\-\\
	$=$	\>	\>$\{$	\>\+\+\+Definition \ref{core}\-\-$~~~ \}$\pop\\
	$(\rho^{\MPrev}\ms{1}{\MPcomp}\ms{1}\rho){\MPldom{}}$\push\-\\
	$=$	\>	\>$\{$	\>\+\+\+domains\-\-$~~~ \}$\pop\\
	$\rho{\MPrdom{}}~~.$
\end{mpdisplay}
The dual equation,  $R{\MPldom{}}\ms{2}{=}\ms{2}\lambda{\MPrdom{}}$,   is proved similarly.  Second,
\begin{mpdisplay}{0.15em}{6.5mm}{0mm}{2}
	$C{\MPldom{}}$\push\-\\
	$=$	\>	\>$\{$	\>\+\+\+Definition \ref{core}\-\-$~~~ \}$\pop\\
	$(\lambda\ms{1}{\MPcomp}\ms{1}R\ms{1}{\MPcomp}\ms{1}\rho^{\MPrev}){\MPldom{}}$\push\-\\
	$=$	\>	\>$\{$	\>\+\+\+$R{\MPrdom{}}\ms{2}{=}\ms{2}\rho{\MPrdom{}}$ (just proved)\-\-$~~~ \}$\pop\\
	$(\lambda\ms{1}{\MPcomp}\ms{1}R\ms{1}{\MPcomp}\ms{1}R{\MPrdom{}}){\MPldom{}}$\push\-\\
	$=$	\>	\>$\{$	\>\+\+\+domains: (\ref{ldom.and.rdom})\-\-$~~~ \}$\pop\\
	$(\lambda\ms{1}{\MPcomp}\ms{1}R{\MPldom{}}){\MPldom{}}$\push\-\\
	$=$	\>	\>$\{$	\>\+\+\+$R{\MPldom{}}\ms{2}{=}\ms{2}\lambda{\MPrdom{}}$ (see above)\-\-$~~~ \}$\pop\\
	$\lambda{\MPldom{}}~~.$
\end{mpdisplay}
The final equation is  also proved  similarly.
\end{shortproof}

\begin{Lemma}\label{core.iso.index}{\rm \ \ \ Suppose $R$, $\lambda$,  $\rho$  and $C$  are as in Definition \ref{core}. Suppose also that $J$ is an index of
$R$.  Then  $C\ms{1}{\cong}\ms{1}J$ as witnessed by $\lambda\ms{1}{\MPcomp}\ms{1}J{\MPldom{}}$ and $\rho\ms{1}{\MPcomp}\ms{1}J{\MPrdom{}}$.
}%
\end{Lemma}%
\begin{shortproof}
We construct the witnesses as follows.
\begin{mpdisplay}{0.15em}{6.5mm}{0mm}{2}
	$C$\push\-\\
	$=$	\>	\>$\{$	\>\+\+\+Definition \ref{core}\-\-$~~~ \}$\pop\\
	$\lambda\ms{1}{\MPcomp}\ms{1}R\ms{1}{\MPcomp}\ms{1}\rho^{\MPrev}$\push\-\\
	$=$	\>	\>$\{$	\>\+\+\+$J$ is an index of $R$,  Definition \ref{gen.index}(b)\-\-$~~~ \}$\pop\\
	$\lambda\ms{1}{\MPcomp}\ms{1}R{\MPperldom{}}\ms{1}{\MPcomp}\ms{1}J\ms{1}{\MPcomp}\ms{1}R{\MPperrdom{}}\ms{1}{\MPcomp}\ms{1}\rho^{\MPrev}$\push\-\\
	$=$	\>	\>$\{$	\>\+\+\+Definition \ref{core}\-\-$~~~ \}$\pop\\
	$\lambda\ms{1}{\MPcomp}\ms{1}\lambda^{\MPrev}\ms{1}{\MPcomp}\ms{1}\lambda\ms{1}{\MPcomp}\ms{1}J\ms{1}{\MPcomp}\ms{1}\rho^{\MPrev}\ms{1}{\MPcomp}\ms{1}\rho\ms{1}{\MPcomp}\ms{1}\rho^{\MPrev}$\push\-\\
	$=$	\>	\>$\{$	\>\+\+\+$\lambda$ and $\rho$ are functional,  \\
	so $\lambda{\MPldom{}}\ms{2}{=}\ms{2}\lambda\ms{1}{\MPcomp}\ms{1}\lambda^{\MPrev}$ and $\rho{\MPldom{}}\ms{2}{=}\ms{2}\rho\ms{1}{\MPcomp}\ms{1}\rho^{\MPrev}$\-\-$~~~ \}$\pop\\
	$\lambda\ms{1}{\MPcomp}\ms{1}J\ms{1}{\MPcomp}\ms{1}\rho^{\MPrev}$\push\-\\
	$=$	\>	\>$\{$	\>\+\+\+domains: (\ref{ldom.and.rdom}) and converse\-\-$~~~ \}$\pop\\
	$\lambda\ms{1}{\MPcomp}\ms{1}J{\MPldom{}}\ms{1}{\MPcomp}\ms{1}J\ms{1}{\MPcomp}\ms{1}(\rho\ms{1}{\MPcomp}\ms{1}J{\MPrdom{}})^{\MPrev}~~.$
\end{mpdisplay}
Comparing the last line with the definition of an isomorphism of relations (Definition \ref{rel.iso} with the
instantiation $R{,}S{,}\phi{,}\psi\ms{3}{:=}\ms{3}C\ms{2}{,}\ms{2}J\ms{2}{,}\ms{2}\lambda\ms{1}{\MPcomp}\ms{1}J{\MPldom{}}\ms{2}{,}\ms{2}\rho\ms{1}{\MPcomp}\ms{1}J{\MPrdom{}}$), we postulate 
$\lambda\ms{1}{\MPcomp}\ms{1}J{\MPldom{}}$ and $\rho\ms{1}{\MPcomp}\ms{1}J{\MPrdom{}}$ as witnesses to the isomorphism.

It remains to show that $\lambda\ms{1}{\MPcomp}\ms{1}J{\MPldom{}}$ and $\rho\ms{1}{\MPcomp}\ms{1}J{\MPrdom{}}$  are bijections  on  the appropriate domains.  First,
\begin{mpdisplay}{0.15em}{6.5mm}{0mm}{2}
	$(\rho\ms{1}{\MPcomp}\ms{1}J{\MPrdom{}})^{\MPrev}\ms{1}{\MPcomp}\ms{1}\rho\ms{1}{\MPcomp}\ms{1}J{\MPrdom{}}$\push\-\\
	$=$	\>	\>$\{$	\>\+\+\+converse\-\-$~~~ \}$\pop\\
	$J{\MPrdom{}}\ms{1}{\MPcomp}\ms{1}\rho^{\MPrev}\ms{1}{\MPcomp}\ms{1}\rho\ms{1}{\MPcomp}\ms{1}J{\MPrdom{}}$\push\-\\
	$=$	\>	\>$\{$	\>\+\+\+definition \ref{core}\-\-$~~~ \}$\pop\\
	$J{\MPrdom{}}\ms{1}{\MPcomp}\ms{1}R{\MPperrdom{}}\ms{1}{\MPcomp}\ms{1}J{\MPrdom{}}$\push\-\\
	$=$	\>	\>$\{$	\>\+\+\+$J$ is an index of $R$,  Definition \ref{gen.index}(d)\-\-$~~~ \}$\pop\\
	$J{\MPrdom{}}~~.$
\end{mpdisplay}
Symmetrically,  \begin{displaymath}(\lambda\ms{1}{\MPcomp}\ms{1}J{\MPldom{}})^{\MPrev}\ms{1}{\MPcomp}\ms{1}\lambda\ms{1}{\MPcomp}\ms{1}J{\MPldom{}}\ms{4}{=}\ms{4}J{\MPldom{}}~~.\end{displaymath}
Finally, 
\begin{mpdisplay}{0.15em}{6.5mm}{0mm}{2}
	$(\rho\ms{1}{\MPcomp}\ms{1}J{\MPrdom{}}){\MPldom{}}$\push\-\\
	$=$	\>	\>$\{$	\>\+\+\+$\rho$ is functional, and $\rho^{\MPrev}\ms{1}{\MPcomp}\ms{1}\rho\ms{2}{=}\ms{2}R{\MPperrdom{}}$,\\
	i.e.\   $\rho\ms{2}{=}\ms{2}\rho\ms{1}{\MPcomp}\ms{1}\rho^{\MPrev}\ms{1}{\MPcomp}\ms{1}\rho\ms{2}{=}\ms{2}\rho\ms{1}{\MPcomp}\ms{1}R{\MPperrdom{}}$\-\-$~~~ \}$\pop\\
	$(\rho\ms{1}{\MPcomp}\ms{1}R{\MPperrdom{}}\ms{1}{\MPcomp}\ms{1}J{\MPrdom{}}){\MPldom{}}$\push\-\\
	$=$	\>	\>$\{$	\>\+\+\+$J{\MPrdom{}}\ms{1}{\subseteq}\ms{1}R{\MPrdom{}}$  and $R{\MPrdom{}}\ms{1}{=}\ms{1}(R{\MPperrdom{}}){\MPrdom{}}$\-\-$~~~ \}$\pop\\
	$(\rho\ms{1}{\MPcomp}\ms{1}R{\MPperrdom{}}\ms{1}{\MPcomp}\ms{1}J{\MPrdom{}}\ms{1}{\MPcomp}\ms{1}(R{\MPperrdom{}}){\MPrdom{}}){\MPldom{}}$\push\-\\
	$=$	\>	\>$\{$	\>\+\+\+domains: Theorem \ref{domains}(b) and (c),  $R{\MPperrdom{}}\ms{1}{=}\ms{1}(R{\MPperrdom{}})^{\MPrev}$\-\-$~~~ \}$\pop\\
	$(\rho\ms{2}{\MPcomp}\ms{2}R{\MPperrdom{}}\ms{1}{\MPcomp}\ms{1}J{\MPrdom{}}\ms{1}{\MPcomp}\ms{1}R{\MPperrdom{}}){\MPldom{}}$\push\-\\
	$=$	\>	\>$\{$	\>\+\+\+domains: Theorem \ref{domains}(c)\-\-$~~~ \}$\pop\\
	$(\rho\ms{2}{\MPcomp}\ms{2}(R{\MPperrdom{}}\ms{1}{\MPcomp}\ms{1}J{\MPrdom{}}\ms{1}{\MPcomp}\ms{1}R{\MPperrdom{}}){\MPldom{}}){\MPldom{}}$\push\-\\
	$=$	\>	\>$\{$	\>\+\+\+Lemmas \ref{R-perleftJR-perleft} and  \ref{R-perleftleft}(b)\-\-$~~~ \}$\pop\\
	$(\rho\ms{2}{\MPcomp}\ms{2}R{\MPrdom{}}){\MPldom{}}$\push\-\\
	$=$	\>	\>$\{$	\>\+\+\+(\ref{per.rightdomain.doms}) and domains: Theorem \ref{domains}(c)\-\-$~~~ \}$\pop\\
	$(\rho\ms{2}{\MPcomp}\ms{2}R{\MPperrdom{}}){\MPldom{}}$\push\-\\
	$=$	\>	\>$\{$	\>\+\+\+$\rho\ms{2}{=}\ms{2}\rho\ms{1}{\MPcomp}\ms{1}R{\MPperrdom{}}$ (see first step)\-\-$~~~ \}$\pop\\
	$\rho{\MPldom{}}$\push\-\\
	$=$	\>	\>$\{$	\>\+\+\+Lemma \ref{core.domains}\-\-$~~~ \}$\pop\\
	$C{\MPrdom{}}~~.$
\end{mpdisplay}
Symmetrically,  $(\lambda\ms{1}{\MPcomp}\ms{1}J{\MPldom{}}){\MPldom{}}\ms{1}{=}\ms{1}C{\MPldom{}}$.  

Putting all the calculations together, we conclude that $\lambda\ms{1}{\MPcomp}\ms{1}J{\MPldom{}}$ and $\rho\ms{1}{\MPcomp}\ms{1}J{\MPrdom{}}$ are bijections; the left domain of
$\lambda\ms{1}{\MPcomp}\ms{1}J{\MPldom{}}$ is $C{\MPldom{}}$ and its right domain is $J{\MPldom{}}$;  the left domain of $\rho\ms{1}{\MPcomp}\ms{1}J{\MPrdom{}}$  is $C{\MPrdom{}}$ and its right domain is $J{\MPrdom{}}$.
\end{shortproof}

We now prove the theorem alluded to by the nomenclature of Definition \ref{core},  namely any core of a
given relation $R$  is a core relation in the sense of Definition \ref{Core.gen}.   
\begin{Theorem}\label{core.perdomain}{\rm \ \ \ Suppose $C$ is a core of $R$.  Then, if $R$ has an index,  \begin{equation}\label{core.perdomain.r}
C{\MPperrdom{}}\ms{3}{=}\ms{3}C{\MPrdom{}}\mbox{~~, and}
\end{equation}\begin{equation}\label{core.perdomain.l}
C{\MPperldom{}}\ms{3}{=}\ms{3}C{\MPldom{}}~~.
\end{equation}That is,  if $R$ has an index,  any core   $C$ of $R$  is a core relation.  (See definition \ref{Core.gen}.)
}
\end{Theorem}
\begin{proof}
Assume that $J$ is an index of $R$.  The proof is a combination of several preceding lemmas and 
theorems.
\begin{mpdisplay}{0.15em}{6.5mm}{0mm}{2}
	$C{\MPperldom{}}\ms{3}{=}\ms{3}C{\MPldom{}}$\push\-\\
	$\Leftarrow$	\>	\>$\{$	\>\+\+\+Theorem \ref{core.if.iso}\-\-$~~~ \}$\pop\\
	$C{\MPperldom{}}\ms{3}{\cong}\ms{3}C{\MPldom{}}$\push\-\\
	$\Leftarrow$	\>	\>$\{$	\>\+\+\+Leibniz\-\-$~~~ \}$\pop\\
	$J{\MPperldom{}}\ms{1}{=}\ms{1}J{\MPldom{}}\ms{4}{\wedge}\ms{4}C{\MPperldom{}}\ms{2}{\cong}\ms{2}J{\MPperldom{}}\ms{4}{\wedge}\ms{4}J{\MPldom{}}\ms{2}{\cong}\ms{2}C{\MPldom{}}$\push\-\\
	$\Leftarrow$	\>	\>$\{$	\>\+\+\+index $J$ is a core relation (Lemma \ref{index.per.is.pid})\-\-$~~~ \}$\pop\\
	$C{\MPperldom{}}\ms{2}{\cong}\ms{2}J{\MPperldom{}}\ms{4}{\wedge}\ms{4}J{\MPldom{}}\ms{2}{\cong}\ms{2}C{\MPldom{}}$\push\-\\
	$\Leftarrow$	\>	\>$\{$	\>\+\+\+Lemmas \ref{perdom.iso} and \ref{dom.iso}\-\-$~~~ \}$\pop\\
	$C\ms{2}{\cong}\ms{2}J$\push\-\\
	$=$	\>	\>$\{$	\>\+\+\+Lemma \ref{core.iso.index}\-\-$~~~ \}$\pop\\
	$\mathsf{true}~~.$
\end{mpdisplay}
\end{proof}

\noindent
\textbf{Note.}  Theorem \ref{core.perdomain} assumes that relation $R$ has an index $J$.  Likewise, a  corollary of
Lemma \ref{core.iso.index} is that, assuming  relation $R$ has an index, all cores of $R$ are isomorphic.  It is
straightforward to  prove  that all cores of $R$ are isomorphic without the assumption that $R$  has
 an index.  Similarly,  Theorem \ref{core.perdomain}   can be proved without this assumption but the proof is
quite long and complex.  See \cite{RCB2020} for full details. 

We argue later that this assumption has no practical significance: in Section \ref{From Pers To Relations} we
show that every relation $R$  has an index if both its per domains have an index.  This means that, for a
given $R$, it is necessary to calculate indices of $R{\MPperldom{}}$ and $R{\MPperrdom{}}$; however, in practice, this is not an issue.  
\textbf{End of Note}

\section{Indexes of  Difunctions and Pers}\label{Indexes of  Difunctions and Pers}

\subsection{Indexes of  Difunctions}\label{Indices and Difunctions}

We now specialise the notion of index to difunctions. 
\begin{Lemma}\label{difun.index.implies.difun}{\rm \ \ \ Suppose $J$ is an index of relation $R$ and $J$ is difunctional.  Then $R$
is difunctional.
}%
\end{Lemma}%
\begin{shortproof}
\begin{mpdisplay}{0.15em}{6.5mm}{0mm}{2}
	$R\ms{1}{\MPcomp}\ms{1}R^{\MPrev}\ms{1}{\MPcomp}\ms{1}R$\push\-\\
	$=$	\>	\>$\{$	\>\+\+\+$J$ is an index of $R$,  Lemma \ref{index.simp}\-\-$~~~ \}$\pop\\
	$R\ms{1}{\MPcomp}\ms{1}J^{\MPrev}\ms{1}{\MPcomp}\ms{1}R$\push\-\\
	$=$	\>	\>$\{$	\>\+\+\+$J$ is an index of $R$,  \ref{gen.index}(b)\-\-$~~~ \}$\pop\\
	$R{\MPperldom{}}\ms{1}{\MPcomp}\ms{1}J\ms{1}{\MPcomp}\ms{1}R{\MPperrdom{}}\ms{1}{\MPcomp}\ms{1}J^{\MPrev}\ms{1}{\MPcomp}\ms{1}R{\MPperldom{}}\ms{1}{\MPcomp}\ms{1}J\ms{1}{\MPcomp}\ms{1}R{\MPperrdom{}}$\push\-\\
	$=$	\>	\>$\{$	\>\+\+\+domains: (\ref{ldom.and.rdom}) and Theorem \ref{domains}(b)\-\-$~~~ \}$\pop\\
	$R{\MPperldom{}}\ms{1}{\MPcomp}\ms{1}J\ms{1}{\MPcomp}\ms{1}J{\MPrdom{}}\ms{1}{\MPcomp}\ms{1}R{\MPperrdom{}}\ms{1}{\MPcomp}\ms{1}J{\MPrdom{}}\ms{1}{\MPcomp}\ms{1}J^{\MPrev}\ms{1}{\MPcomp}\ms{1}J{\MPldom{}}\ms{1}{\MPcomp}\ms{1}R{\MPperldom{}}\ms{1}{\MPcomp}\ms{1}J{\MPldom{}}\ms{1}{\MPcomp}\ms{1}J\ms{1}{\MPcomp}\ms{1}R{\MPperrdom{}}$\push\-\\
	$=$	\>	\>$\{$	\>\+\+\+$J$ is an index of $R$,  \ref{gen.index}(d) and (c)\-\-$~~~ \}$\pop\\
	$R{\MPperldom{}}\ms{1}{\MPcomp}\ms{1}J\ms{1}{\MPcomp}\ms{1}J{\MPrdom{}}\ms{1}{\MPcomp}\ms{1}J^{\MPrev}\ms{1}{\MPcomp}\ms{1}J{\MPldom{}}\ms{1}{\MPcomp}\ms{1}J\ms{1}{\MPcomp}\ms{1}R{\MPperrdom{}}$\push\-\\
	$=$	\>	\>$\{$	\>\+\+\+domains: (\ref{ldom.and.rdom}) and Theorem \ref{domains}(b), and $J$ is difunctional  (i.e.\ $J\ms{2}{=}\ms{2}J\ms{1}{\MPcomp}\ms{1}J^{\MPrev}\ms{1}{\MPcomp}\ms{1}J$)\-\-$~~~ \}$\pop\\
	$R{\MPperldom{}}\ms{1}{\MPcomp}\ms{1}J\ms{1}{\MPcomp}\ms{1}R{\MPperrdom{}}$\push\-\\
	$=$	\>	\>$\{$	\>\+\+\+\ref{gen.index}(b)\-\-$~~~ \}$\pop\\
	$R~~.$
\end{mpdisplay}
\end{shortproof}

{} 

The property that $R$ is a difunction is equivalent to  $R{\MPperldom{}}\ms{2}{=}\ms{2}R\ms{1}{\MPcomp}\ms{1}R^{\MPrev}$  (and symmetrically to  $R{\MPperrdom{}}\ms{2}{=}\ms{2}R^{\MPrev}\ms{1}{\MPcomp}\ms{1}R$).
Also, since $R\ms{2}{=}\ms{2}R\ms{1}{\MPcomp}\ms{1}R^{\MPrev}\ms{1}{\MPcomp}\ms{1}R$,  the right side of Lemma \ref{index.simp} simplifies to $R$.  In this way, the
definition of an index of a difunction can be restated as follows.
\begin{Definition}[Difunction Index]\label{difunction.index}{\rm \ \ \ An index  of a difunction $R$ is a relation $J$ that  has the following properties:
\begin{description}
\item[(a)]$J\ms{1}{\subseteq}\ms{1}R~~,$ 
\item[(b)]$R\ms{1}{\MPcomp}\ms{1}J^{\MPrev}\ms{1}{\MPcomp}\ms{1}R\ms{3}{=}\ms{3}R~~.$ 
\item[(c)]$J{\MPldom{}}\ms{1}{\MPcomp}\ms{1}R\ms{1}{\MPcomp}\ms{1}R^{\MPrev}\ms{1}{\MPcomp}\ms{1}J{\MPldom{}}\ms{3}{=}\ms{3}J{\MPldom{}}~~,$ 
\item[(d)]$J{\MPrdom{}}\ms{1}{\MPcomp}\ms{1}R^{\MPrev}\ms{1}{\MPcomp}\ms{1}R\ms{1}{\MPcomp}\ms{1}J{\MPrdom{}}\ms{3}{=}\ms{3}J{\MPrdom{}}~~.$ \QED
\end{description}
}
\end{Definition}

\begin{Lemma}\label{difunction.bijection}{\rm \ \ \ An index  $J$  of a difunction  $R$  is a bijection between $J{\MPldom{}}$ and $J{\MPrdom{}}$.
}%
\end{Lemma}%
\begin{shortproof}
\begin{mpdisplay}{0.15em}{6.5mm}{0mm}{2}
	$J{\MPldom{}}$\push\-\\
	$=$	\>	\>$\{$	\>\+\+\+\ref{difunction.index}(c)\-\-$~~~ \}$\pop\\
	$J{\MPldom{}}\ms{1}{\MPcomp}\ms{1}R^{\MPrev}\ms{1}{\MPcomp}\ms{1}R\ms{1}{\MPcomp}\ms{1}J{\MPldom{}}$\push\-\\
	$\supseteq$	\>	\>$\{$	\>\+\+\+\ref{difunction.index}(a)\-\-$~~~ \}$\pop\\
	$J{\MPldom{}}\ms{1}{\MPcomp}\ms{1}J^{\MPrev}\ms{1}{\MPcomp}\ms{1}J\ms{1}{\MPcomp}\ms{1}J{\MPldom{}}$\push\-\\
	$=$	\>	\>$\{$	\>\+\+\+domains: (\ref{ldom.and.rdom}) and Theorem \ref{domains}(b)\-\-$~~~ \}$\pop\\
	$J^{\MPrev}\ms{1}{\MPcomp}\ms{1}J$\push\-\\
	$\supseteq$	\>	\>$\{$	\>\+\+\+domains: Definition \ref{lr.squares}\-\-$~~~ \}$\pop\\
	$J{\MPldom{}}~~.$
\end{mpdisplay}
Thus, by anti-symmetry,  \begin{displaymath}J{\MPldom{}}\ms{4}{=}\ms{4}J^{\MPrev}\ms{1}{\MPcomp}\ms{1}J~~.\end{displaymath}Symmetrically,  $J{\MPrdom{}}\ms{4}{=}\ms{4}J\ms{1}{\MPcomp}\ms{1}J^{\MPrev}$.   That is, $J$ is a bijection.
\end{shortproof}

Corollary \ref{difunction.difunction} formulates a method to  determine whether a relation is a
difunction:  compute an index of the relation and then determine whether it is a difunction.   By
\ref{gen.index}(a), the second step in this process will be no less efficient than determining difunctionality
directly and, in many cases, may be substantially more efficient.  (There is, however, no guarantee of
improved efficiency since the inequality in \ref{gen.index}(a) may be an equality.)
\begin{Corollary}\label{difunction.difunction}{\rm \ \ \ Suppose $J$ is an index of relation $R$.  Then  $R$ is a difunction iff $J$ is a
difunction.
}%
\end{Corollary}%
\begin{shortproof}
Lemma \ref{difun.index.implies.difun} establishes ``if''.  Lemma \ref{difunction.bijection} establishes ``only
if'' (since a bijection is a difunction).%
\end{shortproof}

\subsection{Indexes of Pers}\label{Indices and Pers}

That every difunction has an index  is a desirable property but it is not provable in standard
axiomatic formulations of relation algebra.  Rather than postulate its truth, we shall postulate that all
pers have an index, and then show that a consequence of the postulate is that all difunctions have an
index.

A relation $R$ is a per iff $R\ms{1}{=}\ms{1}R{\MPperldom{}}\ms{1}{=}\ms{1}R{\MPperrdom{}}$.  Using this property, the definition of index can be simplified for pers.
Specifically, an index $J$ of per $R$ has the following properties. (Cf.  definition \ref{gen.index}.)
\begin{description}
\item[(a)]$J\ms{1}{\subseteq}\ms{1}R~~,$ 
\item[(b)]$R{\MPcomp}J{\MPcomp}R\ms{2}{=}\ms{2}R~~,$ 
\item[(c)]$J{\MPldom{}}\ms{1}{\MPcomp}\ms{1}R\ms{1}{\MPcomp}\ms{1}J{\MPldom{}}\ms{3}{=}\ms{3}J{\MPldom{}}~~,$ 
\item[(d)]$J{\MPrdom{}}\ms{1}{\MPcomp}\ms{1}R\ms{1}{\MPcomp}\ms{1}J{\MPrdom{}}\ms{3}{=}\ms{3}J{\MPrdom{}}~~.$ 
 
\end{description}

Lemmas \ref{index.per.corefl} and \ref{index.per} prepare the way for Definition \ref{per.index}.
\begin{Lemma}\label{index.per.corefl}{\rm \ \ \ If a per has an index, then it has an index that is a coreflexive.
}%
\end{Lemma}%
\begin{proof}
Suppose $R$ is a per and $J$ is an index of $R$. 
The lemma is proved if we show that $J{\MPldom{}}$ is an index of $R$.   We thus have to prove that 
\begin{description}
\item[(e)]$J{\MPldom{}}\ms{1}{\subseteq}\ms{1}R~~,$ 
\item[(f)]$R\ms{1}{\MPcomp}\ms{1}J{\MPldom{}}\ms{1}{\MPcomp}\ms{1}R\ms{3}{=}\ms{3}R~~,$ 
\item[(g)]$(J{\MPldom{}}){\MPldom{}}\ms{1}{\MPcomp}\ms{1}R\ms{1}{\MPcomp}\ms{1}(J{\MPldom{}}){\MPldom{}}\ms{3}{=}\ms{3}(J{\MPldom{}}){\MPldom{}}~~,$ 
\item[(h)]$(J{\MPrdom{}}){\MPrdom{}}\ms{1}{\MPcomp}\ms{1}R\ms{1}{\MPcomp}\ms{1}(J{\MPrdom{}}){\MPrdom{}}\ms{3}{=}\ms{3}(J{\MPrdom{}}){\MPrdom{}}~~,$ 
 
\end{description}
assuming the properties (a), (b), (c) and (d) above.

Of the four properties,  only (f) is non-trivial.  (Properties (g) and (h) follow because $J{\MPldom{}}\ms{1}{=}\ms{1}(J{\MPldom{}}){\MPldom{}}$ and 
$J{\MPrdom{}}\ms{1}{=}\ms{1}(J{\MPrdom{}}){\MPrdom{}}$.  Property (e) follows because, since $R$ is a per,  $R{\MPldom{}}\ms{1}{\subseteq}\ms{1}R$.)

Property (f) is proved as follows.
\begin{mpdisplay}{0.15em}{6.5mm}{0mm}{2}
	$R\ms{1}{\MPcomp}\ms{1}J{\MPldom{}}\ms{1}{\MPcomp}\ms{1}R$\push\-\\
	$=$	\>	\>$\{$	\>\+\+\+by Lemma \ref{difunction.bijection},  $J\ms{1}{\MPcomp}\ms{1}J^{\MPrev}\ms{2}{=}\ms{2}J{\MPldom{}}$\-\-$~~~ \}$\pop\\
	$R\ms{1}{\MPcomp}\ms{1}J\ms{1}{\MPcomp}\ms{1}J^{\MPrev}\ms{1}{\MPcomp}\ms{1}R$\push\-\\
	$=$	\>	\>$\{$	\>\+\+\+domains: (\ref{ldom.and.rdom})\-\-$~~~ \}$\pop\\
	$R\ms{1}{\MPcomp}\ms{1}J\ms{1}{\MPcomp}\ms{1}J{\MPrdom{}}\ms{1}{\MPcomp}\ms{1}J^{\MPrev}\ms{1}{\MPcomp}\ms{1}R$\push\-\\
	$=$	\>	\>$\{$	\>\+\+\+(d)\-\-$~~~ \}$\pop\\
	$R\ms{1}{\MPcomp}\ms{1}J\ms{1}{\MPcomp}\ms{1}J{\MPrdom{}}\ms{1}{\MPcomp}\ms{1}R\ms{1}{\MPcomp}\ms{1}J{\MPrdom{}}\ms{1}{\MPcomp}\ms{1}J^{\MPrev}\ms{1}{\MPcomp}\ms{1}R$\push\-\\
	$=$	\>	\>$\{$	\>\+\+\+domains: (\ref{ldom.and.rdom})\-\-$~~~ \}$\pop\\
	$R\ms{1}{\MPcomp}\ms{1}J\ms{1}{\MPcomp}\ms{1}R\ms{1}{\MPcomp}\ms{1}J^{\MPrev}\ms{1}{\MPcomp}\ms{1}R$\push\-\\
	$=$	\>	\>$\{$	\>\+\+\+(b)\-\-$~~~ \}$\pop\\
	$R\ms{1}{\MPcomp}\ms{1}J^{\MPrev}\ms{1}{\MPcomp}\ms{1}R$\push\-\\
	$=$	\>	\>$\{$	\>\+\+\+$R$ is a per, so $R\ms{1}{=}\ms{1}R^{\MPrev}$; converse\-\-$~~~ \}$\pop\\
	$(R{\MPcomp}J{\MPcomp}R)^{\MPrev}$\push\-\\
	$=$	\>	\>$\{$	\>\+\+\+$R$ is a per, so $R\ms{1}{=}\ms{1}R^{\MPrev}$;  (b) and converse\-\-$~~~ \}$\pop\\
	$R~~.$
\end{mpdisplay}
\end{proof}

\begin{Lemma}\label{index.per}{\rm \ \ \ For all pers $R$,   if $R$ has an index then there is a relation $J$ such that  
\begin{description}
\item[(a)]$J\ms{1}{\subseteq}\ms{1}R{\MPldom{}}~~,$ 
\item[(b)]$J{\MPcomp}R{\MPcomp}J\ms{2}{=}\ms{2}J~~,$ 
\item[(c)]$R{\MPcomp}J{\MPcomp}R\ms{2}{=}\ms{2}R~~.$ 
 
\end{description}
Conversely,  for all pers $R$,  if relation $J$ satisfies the properties (a), (b) and (c) above, then $J$ is an index of
$R$. 
}%
\end{Lemma}%
\begin{proof}
First, suppose $R$ is a per that has an index.  By Lemma \ref{index.per.corefl},  $R$ has a coreflexive index.
 Let  $J$ be such  a coreflexive  index of $R$.  We must show that properties (a), (b) and (c) hold.
We have
\begin{mpdisplay}{0.15em}{6.5mm}{0mm}{2}
	$J\ms{1}{\subseteq}\ms{1}R{\MPldom{}}$\push\-\\
	$\Leftarrow$	\>	\>$\{$	\>\+\+\+\ref{gen.index}(a) and monotonicity\-\-$~~~ \}$\pop\\
	$J\ms{1}{=}\ms{1}J{\MPldom{}}$\push\-\\
	$=$	\>	\>$\{$	\>\+\+\+$J$ is a coreflexive\-\-$~~~ \}$\pop\\
	$\mathsf{true}~~.$
\end{mpdisplay}
This proves (a). Now for (b):
\begin{mpdisplay}{0.15em}{6.5mm}{0mm}{2}
	$J{\MPcomp}R{\MPcomp}J$\push\-\\
	$=$	\>	\>$\{$	\>\+\+\+$J$ is a coreflexive, so $J\ms{1}{=}\ms{1}J{\MPldom{}}$,\\
	$R$ is a per, so $R\ms{1}{=}\ms{1}R{\MPperldom{}}$\-\-$~~~ \}$\pop\\
	$J{\MPldom{}}\ms{1}{\MPcomp}\ms{1}R{\MPperldom{}}\ms{1}{\MPcomp}\ms{1}J{\MPldom{}}$\push\-\\
	$=$	\>	\>$\{$	\>\+\+\+\ref{gen.index}(c)\-\-$~~~ \}$\pop\\
	$J{\MPldom{}}$\push\-\\
	$=$	\>	\>$\{$	\>\+\+\+$J$ is a coreflexive, so $J\ms{1}{=}\ms{1}J{\MPldom{}}$\-\-$~~~ \}$\pop\\
	$J~~.$
\end{mpdisplay}
Finally, (c):
\begin{mpdisplay}{0.15em}{6.5mm}{0mm}{2}
	$R{\MPcomp}J{\MPcomp}R$\push\-\\
	$=$	\>	\>$\{$	\>\+\+\+$R$ is a per, so $R\ms{1}{=}\ms{1}R{\MPperldom{}}$\-\-$~~~ \}$\pop\\
	$R{\MPperldom{}}\ms{1}{\MPcomp}\ms{1}J\ms{1}{\MPcomp}\ms{1}R{\MPperldom{}}$\push\-\\
	$=$	\>	\>$\{$	\>\+\+\+\ref{gen.index}(b)\-\-$~~~ \}$\pop\\
	$R~~.$
\end{mpdisplay}
For the converse, assume $R$ is a per and $J$ satisifies the properties (a), (b) and (c) above.  We have to check
the four properties listed in Definition \ref{gen.index}.  First, \ref{gen.index}(a):
\begin{mpdisplay}{0.15em}{6.5mm}{0mm}{2}
	$J$\push\-\\
	$\subseteq$	\>	\>$\{$	\>\+\+\+assumption: (a) above\-\-$~~~ \}$\pop\\
	$R{\MPldom{}}$\push\-\\
	$\subseteq$	\>	\>$\{$	\>\+\+\+$R$ is per\-\-$~~~ \}$\pop\\
	$R~~.$
\end{mpdisplay}
The properties \ref{gen.index}(b), (c) and (d) follow because $J\ms{1}{=}\ms{1}J{\MPldom{}}\ms{1}{=}\ms{1}J{\MPrdom{}}$ and $R\ms{1}{=}\ms{1}R{\MPperldom{}}\ms{1}{=}\ms{1}R{\MPperrdom{}}$.
\end{proof}

As a consequence of Lemma \ref{index.per}, we postulate the following definition of an index of a per.

\begin{Definition}[Index of a Per]\label{per.index}{\rm \ \ \ Suppose $P$ is a per.  Then a (\emph{coreflexive}) 
 \emph{index} of $P$ is a relation $J$ such  that 
\begin{description}
\item[(a)]$J\ms{1}{\subseteq}\ms{1}P{\MPldom{}}~~,$ 
\item[(b)]$J{\MPcomp}P{\MPcomp}J\ms{2}{=}\ms{2}J~~,$ 
\item[(c)]$P{\MPcomp}J{\MPcomp}P\ms{2}{=}\ms{2}P~~.$ \QED 
\end{description}
}
\end{Definition}

We also postulate that every per has a coreflexive index.  We call this the \emph{axiom of choice}.

\begin{Axiom}[Axiom of Choice]\label{Axiom of Choice}{\rm \ \ \ Every per has a coreflexive  index.
}
\QED\end{Axiom}

\subsection{From Pers To Relations}\label{From Pers To Relations}

It is a desirable property that every relation has an index.  However, as mentioned earlier, this can't be
proved in standard relation algebra.  It can be proved if we assume that every per has an index.  The
construction is suggested by Theorem \ref{index-perdoms}.

\begin{Theorem}\label{per.to.relation}{\rm \ \ \ Suppose  $J$ and $K$ are (coreflexive)  indices of $R{\MPperldom{}}$ and
$R{\MPperrdom{}}$, respectively.  Then $J{\MPcomp}R{\MPcomp}K$ is an index of $R$.
}
\end{Theorem}
\begin{proof}
For convenience, we list the properties of $J$ and $K.$  These are obtained by instantiating Definition 
\ref{per.index} with $J{,}R\ms{2}{:=}\ms{2}J\ms{1}{,}\ms{1}R{\MPperldom{}}$ and $J{,}R\ms{2}{:=}\ms{2}K\ms{1}{,}\ms{1}R{\MPperrdom{}}$.  (Domain properties have been used to simplify
(a) and (d).)

\begin{description}
\item[(a)]$J\ms{1}{\subseteq}\ms{1}R{\MPldom{}}~~,$ 
\item[(b)]$J\ms{1}{\MPcomp}\ms{1}R{\MPperldom{}}\ms{1}{\MPcomp}\ms{1}J\ms{3}{=}\ms{3}J~~,$ 
\item[(c)]$R{\MPperldom{}}{\MPcomp}\ms{1}J\ms{1}{\MPcomp}\ms{1}R{\MPperldom{}}\ms{3}{=}\ms{3}R{\MPperldom{}}~~,$ 
\item[(d)]$K\ms{1}{\subseteq}\ms{1}R{\MPrdom{}}~~,$ 
\item[(e)]$K\ms{1}{\MPcomp}\ms{1}R{\MPperrdom{}}\ms{1}{\MPcomp}\ms{1}K\ms{3}{=}\ms{3}K~~,$ 
\item[(f)]$R{\MPperrdom{}}{\MPcomp}\ms{1}K\ms{1}{\MPcomp}\ms{1}R{\MPperrdom{}}\ms{3}{=}\ms{3}R{\MPperrdom{}}~~.$ 
 
\end{description}

We have to prove the four  properties \ref{gen.index}(a)-(d) with the instantiation $J{,}R\ms{2}{:=}\ms{2}J{\MPcomp}R{\MPcomp}K\ms{1}{,}\ms{1}R$.  
By (a),  $J\ms{1}{=}\ms{1}J^{\MPrev}\ms{1}{=}\ms{1}J{\MPldom{}}\ms{1}{=}\ms{1}J{\MPrdom{}}$.  Similarly for $K$.  The proof obligations  are thus:
\begin{description}
\item[(g)]$J{\MPcomp}R{\MPcomp}K\ms{2}{\subseteq}\ms{2}R~~,$ 
\item[(h)]$R{\MPperldom{}}\ms{1}{\MPcomp}\ms{1}J\ms{1}{\MPcomp}\ms{1}R\ms{1}{\MPcomp}\ms{1}K\ms{1}{\MPcomp}\ms{1}R{\MPperrdom{}}\ms{3}{=}\ms{3}R~~.$ 
\item[(i)]$(J{\MPcomp}R{\MPcomp}K){\MPldom{}}\ms{1}{\MPcomp}\ms{1}R{\MPperldom{}}\ms{1}{\MPcomp}\ms{1}(J{\MPcomp}R{\MPcomp}K){\MPldom{}}\ms{3}{=}\ms{3}(J{\MPcomp}R{\MPcomp}K){\MPldom{}}~~,$ 
\item[(j)]$(J{\MPcomp}R{\MPcomp}K){\MPrdom{}}\ms{1}{\MPcomp}\ms{1}R{\MPperrdom{}}\ms{1}{\MPcomp}\ms{1}(J{\MPcomp}R{\MPcomp}K){\MPrdom{}}\ms{3}{=}\ms{3}(J{\MPcomp}R{\MPcomp}K){\MPrdom{}}~~,$ 
 
\end{description}
Property (g) is an easy combination of (a) and (d).  For  (h) we have:
\begin{mpdisplay}{0.15em}{6.5mm}{0mm}{2}
	$R{\MPperldom{}}\ms{1}{\MPcomp}\ms{1}J\ms{1}{\MPcomp}\ms{1}R\ms{1}{\MPcomp}\ms{1}K\ms{1}{\MPcomp}\ms{1}R{\MPperrdom{}}$\push\-\\
	$=$	\>	\>$\{$	\>\+\+\+per domains: (\ref{per.leftandrightdomain.eq})\-\-$~~~ \}$\pop\\
	$R{\MPperldom{}}\ms{1}{\MPcomp}\ms{1}J\ms{1}{\MPcomp}\ms{1}R{\MPperldom{}}\ms{1}{\MPcomp}\ms{1}R\ms{1}{\MPcomp}\ms{1}R{\MPperrdom{}}\ms{1}{\MPcomp}\ms{1}K\ms{1}{\MPcomp}\ms{1}R{\MPperrdom{}}$\push\-\\
	$=$	\>	\>$\{$	\>\+\+\+(b) and (f)\-\-$~~~ \}$\pop\\
	$R{\MPperldom{}}\ms{1}{\MPcomp}\ms{1}R\ms{1}{\MPcomp}\ms{1}R{\MPperrdom{}}$\push\-\\
	$=$	\>	\>$\{$	\>\+\+\+per domains: (\ref{per.leftandrightdomain.eq})\-\-$~~~ \}$\pop\\
	$R~~.$
\end{mpdisplay}
 For (i), we have
\begin{mpdisplay}{0.15em}{6.5mm}{0mm}{2}
	$(J{\MPcomp}R{\MPcomp}K){\MPrdom{}}\ms{1}{\MPcomp}\ms{1}R{\MPperrdom{}}\ms{1}{\MPcomp}\ms{1}(J{\MPcomp}R{\MPcomp}K){\MPrdom{}}$\push\-\\
	$=$	\>	\>$\{$	\>\+\+\+ $(J{\MPcomp}R{\MPcomp}K){\MPrdom{}}\ms{3}{\subseteq}\ms{3}K{\MPrdom{}}\ms{3}{=}\ms{3}K$,\\
	composition of coreflexives is intersection\-\-$~~~ \}$\pop\\
	$(J{\MPcomp}R{\MPcomp}K){\MPrdom{}}\ms{1}{\MPcomp}\ms{1}K\ms{1}{\MPcomp}\ms{1}R{\MPperrdom{}}\ms{1}{\MPcomp}\ms{1}K\ms{1}{\MPcomp}\ms{1}(J{\MPcomp}R{\MPcomp}K){\MPrdom{}}$\push\-\\
	$=$	\>	\>$\{$	\>\+\+\+(e)\-\-$~~~ \}$\pop\\
	$(J{\MPcomp}R{\MPcomp}K){\MPrdom{}}\ms{1}{\MPcomp}\ms{1}K\ms{1}{\MPcomp}\ms{1}(J{\MPcomp}R{\MPcomp}K){\MPrdom{}}$\push\-\\
	$=$	\>	\>$\{$	\>\+\+\+$(J{\MPcomp}R{\MPcomp}K){\MPrdom{}}\ms{3}{\subseteq}\ms{3}K{\MPrdom{}}\ms{3}{=}\ms{3}K$\\
	composition of coreflexives is intersection\-\-$~~~ \}$\pop\\
	$(J{\MPcomp}R{\MPcomp}K){\MPrdom{}}~~.$
\end{mpdisplay} 
The proof is (j) is symmetrical.
\end{proof}

Theorem \ref{per.to.relation} shows how to construct an index of a  relation $R$  from indexes $J$ and $K$ of its left
and right per  domains.  In combination with Lemma \ref{index.dom.def} and Corollary \ref{index.doms}, the
construction is unique.  Specifically, the steps are, first to choose from each equivalence class of $R{\MPperldom{}}$ and
each equivalence class of $R{\MPperrdom{}}$ a single representative.  The collection of such representatives defines the
coreflexives  $J$  and $K$.  Then the index  is defined to be $J{\MPcomp}R{\MPcomp}K$.

\section{Characterisations of Pers and Difunctions}\label{per.difun.char}

This section is about characterising pers and difunctions in terms of functional relations.   Although the
characterisations are well known,  they are not derivable in point-free relation algebra.  We show that
they are derivable using our axiom of choice.

\subsection{Characterisation of Pers}\label{per.char}

A well-known property  is that a relation $R$ is a per iff\begin{equation}\label{fwokf.char}
{\left\langle\exists{}f\ms{3}{:}\ms{3}f\ms{1}{\MPcomp}\ms{1}f^{\MPrev}\ms{3}{=}\ms{3}f{\MPldom{}}\ms{3}{:}\ms{3}R\ms{2}{=}\ms{2}f^{\MPrev}\ms{1}{\MPcomp}\ms{1}f\right\rangle}~~.
\end{equation}This property is said to be a \emph{characteristic} property of pers.  Perhaps surprisingly, it is  \emph{not} derivable in
systems that axiomatise point-free relation algebra.   Freyd and \v{S}\v{c}edrov  \cite[1.281]{FRSC90}  call
the functional $f$ witnessing the existential quantification a ``splitting\footnote{~Freyd and
\v{S}\v{c}edrov define a ``splitting'' in the more general context of a category rather than an allegory; 
the notion is applicable to ``idempotents'' which are also more general than pers.}'' of  $R.$   Typically, the
existence of ``splittings'' is either postulated  as an axiom (eg.\ Winter \cite{Winter2004}) or by adding
axioms formulating relations as a so-called ``power allegory''  \cite[2.422]{FRSC90}, or by adding the
so-called ``all-or-nothing'' axiom \cite{Glueck17}.  (See Section  \ref{Pairs of Points} for discussion of ``all or
nothing''.)  See \cite{BO23} for a comparison of the techniques used to establish (\ref{fwokf.char}) using
these different axiom systems.  Here we show that the existence of ``splittings'' is a consequence of our  
axiom of choice:

\begin{Theorem}\label{split.existence}{\rm \ \ \ If per $P$  has a coreflexive index $J$ , then\begin{displaymath}P\ms{4}{=}\ms{4}(J{\MPcomp}P)^{\MPrev}\ms{1}{\MPcomp}\ms{1}(J{\MPcomp}P)\ms{8}{\wedge}\ms{8}J\ms{4}{=}\ms{4}(J{\MPcomp}P)\ms{1}{\MPcomp}\ms{1}(J{\MPcomp}P)^{\MPrev}~~.\end{displaymath}Thus, assuming the  axiom of choice,   for all relations $R$,\begin{displaymath}\mathsf{per}{.}R\ms{5}{\equiv}\ms{5}{\left\langle\exists{}f\ms{3}{:}\ms{3}f\ms{1}{\MPcomp}\ms{1}f^{\MPrev}\ms{3}{=}\ms{3}f{\MPldom{}}\ms{3}{:}\ms{3}R\ms{2}{=}\ms{2}f^{\MPrev}\ms{1}{\MPcomp}\ms{1}f\right\rangle}~~.\end{displaymath}
}
\end{Theorem}
\begin{shortproof}
The proof is very straightforward.  We have
\begin{mpdisplay}{0.15em}{6.5mm}{0mm}{2}
	$(J{\MPcomp}P)^{\MPrev}\ms{1}{\MPcomp}\ms{1}(J{\MPcomp}P)$\push\-\\
	$=$	\>	\>$\{$	\>\+\+\+distributivity\-\-$~~~ \}$\pop\\
	$P^{\MPrev}\ms{1}{\MPcomp}\ms{1}J\ms{1}{\MPcomp}\ms{1}J\ms{1}{\MPcomp}\ms{1}P$\push\-\\
	$=$	\>	\>$\{$	\>\+\+\+$J$ is coreflexive, so $J{\MPcomp}J\ms{1}{=}\ms{1}J$;   $P\ms{1}{=}\ms{1}P^{\MPrev}$\-\-$~~~ \}$\pop\\
	$P{\MPcomp}J{\MPcomp}P$\push\-\\
	$=$	\>	\>$\{$	\>\+\+\+$J$ is an index of $P$,  Definition \ref{per.index}(c)\-\-$~~~ \}$\pop\\
	$P$
\end{mpdisplay}
and 
\begin{mpdisplay}{0.15em}{6.5mm}{0mm}{2}
	$(J{\MPcomp}P)\ms{1}{\MPcomp}\ms{1}(J{\MPcomp}P)^{\MPrev}$\push\-\\
	$=$	\>	\>$\{$	\>\+\+\+distributivity\-\-$~~~ \}$\pop\\
	$J\ms{1}{\MPcomp}\ms{1}P\ms{1}{\MPcomp}\ms{1}P^{\MPrev}\ms{1}{\MPcomp}\ms{1}J$\push\-\\
	$=$	\>	\>$\{$	\>\+\+\+$P$ is a per,   so  by Lemma \ref{per.perdom}(ii),  $P\ms{2}{=}\ms{2}P^{\MPrev}\ms{1}{\MPcomp}\ms{1}P$\-\-$~~~ \}$\pop\\
	$J{\MPcomp}P{\MPcomp}J$\push\-\\
	$=$	\>	\>$\{$	\>\+\+\+$J$ is an index of $P$,  Definition \ref{per.index}(b)\-\-$~~~ \}$\pop\\
	$J~~.$
\end{mpdisplay}
This proves the first property.  It also establishes that (assuming the axiom of choice), for all $R$, \begin{displaymath}\mathsf{per}{.}R\ms{5}{\Rightarrow}\ms{5}{\left\langle\exists{}f\ms{3}{:}\ms{3}f\ms{1}{\MPcomp}\ms{1}f^{\MPrev}\ms{3}{=}\ms{3}f{\MPldom{}}\ms{3}{:}\ms{3}R\ms{2}{=}\ms{2}f^{\MPrev}\ms{1}{\MPcomp}\ms{1}f\right\rangle}~~.\end{displaymath}(The witness is $J{\MPcomp}R$.)  The converse is obvious: see \cite{BO23} for details. 
The equivalence follows by mutual implication.
\end{shortproof}

A second so-called ``characteristic''  property is that  a relation $R$ is a difunctional  iff\begin{displaymath}{\left\langle\exists\ms{1}{}f{,}g\ms{6}{:}\ms{6}f\ms{1}{\MPcomp}\ms{1}f^{\MPrev}\ms{3}{=}\ms{3}f{\MPldom{}}\ms{3}{=}\ms{3}g\ms{1}{\MPcomp}\ms{1}g^{\MPrev}\ms{3}{=}\ms{3}g{\MPldom{}}\ms{6}{:}\ms{6}R\ms{2}{=}\ms{2}f^{\MPrev}\ms{1}{\MPcomp}\ms{1}g\right\rangle}~~.\end{displaymath}Like the characteristic property of pers, it is   not derivable in systems that axiomatise point-free relation 
algebra.   It  is, however,  a  corollary of the existence of ``splittings'' (and thus of Theorem \ref{split.existence}), 
as shown by  Winter \cite{Winter2004}.

\subsection{Unicity of Characterisations}\label{per.unicity}

The characterisation of a per  in the form $f^{\MPrev}\ms{1}{\MPcomp}\ms{1}f$  where $f$ is a functional relation  is not unique.   (There are
typically many representatives one can choose for each equivalence class; so there are very many
distinct indexes of a per.) The characterisation  is sometimes described as being  ``essentially'' unique or 
sometimes as unique ``up to isomorphism''.  See our working document \cite{VB2022} for full details.  




\section{Enabling Pointwise Reasoning}\label{Formulations of Points}

In this section, our goal is to capture the notion that a relation is a set with elements pairs of points. 

In traditional pointwise reasoning about relations, a basic assumption is that a type is a  set  that forms  a
complete, universally distributive   lattice  under the subset ordering; the type of a (binary) relation is a set 
of pairs.  The set of relations of a given type thus forms a powerset of a set of pairs. 

In  Section  \ref{Indices:Powersets},  we recall a general theorem  on the structure of powersets.  Briefly,
Theorem \ref{saturated.is.powerset} states that a set is isomorphic to the powerset of its ``atoms'' iff it is
``saturated''.   The section defines these concepts;  the concepts then  form the backbone of later sections
where we specialise the theorem  to relations.  

One (of several) mechanisms for introducing pointwise reasoning   within the framework of point-free
relation algebra involves the introduction of the so-called ``all-or-nothing rule'' which was postulated as
an axiom by Gl\"uck \cite{Glueck17}.  This rule is combined with completeness and  ``extensionality'' axioms 
which state  that, for each type $A$,  the coreflexives of type $A$ form a complete, saturated lattice.   This was the
approach taken in  \cite{BACKHOUSE2022100730}  where  pointwise  reasoning was used to
formulate and prove  properties of  graphs.   Theorem \ref{all.or.nothing.rule} establishes that the
all-or-nothing rule is a consequence of our axiom of choice (Axiom \ref{Axiom of Choice}: 
every per has an index).  Together with the ``extensionality'' axiom,  this enables the application of
Theorem \ref{saturated.is.powerset} to establish that the type $A{\sim}B$ is isomorphic to the powerset $2^{A{\MPtimes}B}$ (the set
of subsets of the cartesian product $A{\MPtimes}B$).  See Theorems \ref{all.or.nothing.rule} and \ref{all.or.nothing.sat} in Section
\ref{Pairs of Points}.

Section \ref{Indices:The Axiom System Points} introduces ``points'' and states the extensionality axiom
that we assume.  A number of sections are then necessary in order to establish Theorem \ref{all.or.nothing.sat}.
Section \ref{PSSS} introduces ``particles'' and ``pairs''; it is then  shown that particles  are points whilst  section 
\ref{Points are Singleton Squares}  shows that ---assuming the axiom of choice--- points are particles.   (For
this reason, the  terminology ``particle'' is  temporary.)   Section~\ref{ProperAtomsarePairs} shows that proper
atoms (of a given type) are ``pairs''.  These are the ingredients for deriving the ``all-or-nothing'' rule  in
Section \ref{Pairs of Points}.  Section \ref{Pairs of Points} also shows that the point-free  definition of a ``pair'' in 
Section \ref{PSSS}  does correspond to what one normally understands to be a pair of points.  The section
concludes with Theorem \ref{all.or.nothing.sat}.

\subsection{Powersets}\label{Indices:Powersets}
 
As mentioned above, this section defines ``atoms'' and ``saturated'' in the context of a partially ordered
set.  We then state a fundamental theorem relating these concepts to powersets. 

The definition of an atom is the following.
\begin{Definition}[Atom and Atomicity]\label{atom}{\rm \ \ \ Suppose $\mathcal{A}$ is a set partially ordered by the relation 
${\sqsubseteq}$.  Then, the element $p$ is an \emph{atom}\mpindexnew{atom}{\mpT }{\mpT }{\mpT }{\mpT }{\mpT }{\mpT }{} iff\begin{displaymath}{\left\langle\forall{}q\ms{4}{:}{:}\ms{4}q\ms{1}{\sqsubseteq}\ms{1}p\ms{5}{\equiv}\ms{5}q\ms{1}{=}\ms{1}p\ms{3}{\vee}\ms{3}q\ms{1}{=}\ms{1}{\MPplatbottom}\right\rangle}~~.\end{displaymath}Note that  ${\MPplatbottom}$ is an atom according to this definition.  If $p$ is an atom that is different from ${\MPplatbottom}$
we say that it is a \emph{proper} atom\mpindexnew{atom}{proper}{\mpT }{\mpT }{\mpT }{\mpT }{\mpT }{}\mpindexnew{proper}{atom}{\mpT }{\mpT }{\mpT }{\mpT }{\mpT }{}.   A lattice is said to be \emph{atomic}\mpindexnew{atomic lattice}{\mpT }{\mpT }{\mpT }{\mpT }{\mpT }{\mpT }{} if\begin{displaymath}{\left\langle\forall{}q\ms{4}{:}{:}\ms{4}q\ms{1}{\neq}\ms{1}{\MPplatbottom}\ms{4}{\equiv}\ms{4}{\left\langle\exists{}a\ms{2}{:}\ms{2}\mathsf{atom}{.}a\ms{1}{\wedge}\ms{1}a\ms{1}{\neq}\ms{1}{\MPplatbottom}\ms{2}{:}\ms{2}a\ms{1}{\sqsubseteq}\ms{1}q\right\rangle}\right\rangle}~~.\end{displaymath}In words, a lattice is atomic if every proper  element includes a proper  atom.
}
\QED
\end{Definition}

The definition of saturated is as follows.
\begin{Definition}[Saturated]\label{saturated}{\rm \ \ \ A  complete lattice (ordered by ${\sqsubseteq}$)   is \emph{saturated}\mpindexnew{saturated lattice}{\mpT }{\mpT }{\mpT }{\mpT }{\mpT }{\mpT }{}   iff \begin{displaymath}{\left\langle\forall{}p\ms{3}{:}{:}\ms{3}p\ms{4}{=}\ms{4}{\left\langle\sqcup{}a\ms{2}{:}\ms{2}\mathsf{atom}{.}a\ms{2}{\wedge}\ms{2}a\ms{1}{\sqsubseteq}\ms{1}p\ms{2}{:}\ms{2}a\right\rangle}\right\rangle}~~.\end{displaymath}
}
\QED
\end{Definition}

The set of subsets of a type is a
powerset iff the lattice is saturated, as formulated in the following theorem.
\begin{Theorem}\label{saturated.is.powerset}{\rm \ \ \ Suppose $\mathcal{A}$ is a complete, universally distributive lattice. Then the
following statements are equivalent.
\begin{description}
\item[(a)]$\mathcal{A}$ is  saturated\mpindexnew{saturated lattice}{\mpT }{\mpT }{\mpT }{\mpT }{\mpT }{\mpT }{}, 
\item[(b)]$\mathcal{A}$ is atomic\mpindexnew{atomic lattice}{\mpT }{\mpT }{\mpT }{\mpT }{\mpT }{\mpT }{} and complemented, 
\item[(c)]$\mathcal{A}$ is isomorphic to the powerset\mpindexnew{powerset}{\mpT }{\mpT }{\mpT }{\mpT }{\mpT }{\mpT }{} of its atoms. 
\QED
\end{description}
}
\end{Theorem}

(See \cite[Theorem 6.43]{REL92a} for the  proof of Theorem \ref{saturated.is.powerset}.)

We use Theorem \ref{saturated.is.powerset} in two ways.  Firstly, for all types $A$, we simply postulate that the
set of coreflexives of type $A$  is  isomorphic to a powerset under the ${\subseteq}$ ordering:  the atoms are the ``points''
introduced in Section \ref{Indices:The Axiom System Points}.  Second, we use this postulate together with our
axiom of choice to show that, for all types $A$ and $B$,  the type $A{\sim}B$ of (heterogeneous) relations is also 
isomorphic to a powerset under the ${\subseteq}$ ordering:   the atoms are ``pairs'' introduced in Section \ref{PSSS}.  The
proof that ``pairs'' are indeed atoms is the subject of Section \ref{ProperAtomsarePairs}.  A prelude to this is
Theorem \ref{point=singleton.square}, proved in Sections \ref{PSSS} and  \ref{Points are Singleton Squares}, which 
asserts that ``points'' are a special case of ``pairs''.

\subsection{Points}\label{Indices:The Axiom System Points}

We begin by postulating that each type $A$ is a set of ``points''.  We also postulate that 
the set of coreflexives of type $A$ forms a complete, universally distributive lattice under the subset
ordering.  Finally,  we postulate that the lattice is saturated.  
With Theorem \ref{saturated.is.powerset} in mind, we define ``points'' to be the proper atoms of the lattice:
\begin{Definition}[Point\mpindexnew{point}{\mpT }{\mpT }{\mpT }{\mpT }{\mpT }{\mpT }{}]\label{point}{\rm \ \ \ A homogeneous relation $a$ of type $A$  is a \emph{point} iff it has the following 
three properties.
\begin{description}
\item[(a)]$a\ms{1}{\neq}\ms{1}{\MPplatbottom}$~~, 
\item[(b)]$a\ms{1}{\subseteq}\ms{1}\mathbb{I}$~~, and 
\item[(c)]${\left\langle\forall{}b\ms{2}{:}\ms{2}b\ms{1}{\neq}\ms{1}{\MPplatbottom}\ms{2}{\wedge}\ms{2}b\ms{1}{\subseteq}\ms{1}a\ms{2}{:}\ms{2}b\ms{1}{=}\ms{1}a\right\rangle}$~~. 
 
\end{description}
In words, a point is a proper, coreflexive atom.\QED
}
\end{Definition}

If $A$ is a type,  we use $a$, $a'$ etc.  to denote points of type $A$.   Similarly for points of type $B$.  Points 
represent elements of the appropriate type.  

For points $a$ and $a'$ of the same type,\begin{equation}\label{point.neq}
a\ms{1}{=}\ms{1}a'\ms{4}{\vee}\ms{4}a{\MPcomp}a'\ms{1}{=}\ms{1}{\MPplatbottom}~~.
\end{equation}The proof is straightforward.  Suppose $a$ and $a'$ are points.  Then
\begin{mpdisplay}{0.15em}{6.5mm}{0mm}{2}
	$a\ms{1}{=}\ms{1}a{\MPcomp}a'$\push\-\\
	$\Leftarrow$	\>	\>$\{$	\>\+\+\+$a$ is an atom, Definition \ref{atom}\-\-$~~~ \}$\pop\\
	$a{\MPcomp}a'\ms{1}{\neq}\ms{1}{\MPplatbottom}\ms{3}{\wedge}\ms{3}a{\MPcomp}a'\ms{1}{\subseteq}\ms{1}a$\push\-\\
	$\Leftarrow$	\>	\>$\{$	\>\+\+\+$a'\ms{1}{\subseteq}\ms{1}\mathbb{I}$\-\-$~~~ \}$\pop\\
	$a{\MPcomp}a'\ms{1}{\neq}\ms{1}{\MPplatbottom}~~.$
\end{mpdisplay}
Interchanging $a$ and $a'$,   \begin{displaymath}a'\ms{1}{=}\ms{1}a{\MPcomp}a'\ms{2}{\Leftarrow}\ms{2}a'{\MPcomp}a\ms{1}{\neq}\ms{1}{\MPplatbottom}~~.\end{displaymath}But, since composition of coreflexives is symmetric,  $a{\MPcomp}a'\ms{1}{=}\ms{1}a'{\MPcomp}a$.   We conclude that \begin{displaymath}a\ms{1}{=}\ms{1}a{\MPcomp}a'\ms{1}{=}\ms{1}a'\ms{5}{\Leftarrow}\ms{5}a{\MPcomp}a'\ms{1}{\neq}\ms{1}{\MPplatbottom}~~.\end{displaymath}This is equivalent to (\ref{point.neq}).

In point-free relation algebra, subsets of a type
are modelled by coreflexives of that type.  In order to model the property that the coreflexives of a given
type form a lattice that is isomorphic to the set of subsets of the type we need to add to our 
axiom system a   \emph{saturation} property, viz.:
\begin{Definition}[Saturation]\label{extensional}{\rm \ \ \ Suppose $A$ is a type.  
The lattice of coreflexives of type $A$  is said to be \emph{saturated} iff \begin{equation}\label{point.sat}
{\left\langle\forall{}p\ms{4}{:}{:}\ms{4}p\ms{1}{\subseteq}\ms{1}\mathbb{I}_{A}\ms{5}{\equiv}\ms{5}p\ms{2}{=}\ms{2}{\left\langle\cup{}a\ms{3}{:}\ms{3}\mathsf{point}{.}a\ms{2}{\wedge}\ms{2}a\ms{1}{\subseteq}\ms{1}p\ms{3}{:}\ms{3}a\right\rangle}\right\rangle}~~.
\end{equation}
}
\QED
\end{Definition}

The axiom that we call ``extensionality'' is then:
\begin{Axiom}[Extensionality]\label{extensional.axiom}{\rm \ \ \ For each type $A$, the points of type $A$ form a 
complete, universally  distributive, saturated  lattice under the  subset ordering.
}\QED
\end{Axiom}

Applying Theorem \ref{saturated.is.powerset},  a consequence of Axiom 
\ref{extensional.axiom} is that the coreflexives of type $A$ form a lattice  that is
isomorphic to the powerset $2^{A}$.  In this sense, the coreflexives in point-free relation algebra represent sets
of points in traditional pointwise formulations of relation algebra.

We now want to show how to formulate the property that the set of relations of type $A{\sim}B$  is isomorphic to
the powerset $2^{A{\MPtimes}B}$, i.e.\ relations in point-free relation algebra represent pairs $(a,b)$ of points $a$ and $b$ of
type $A$ and $B$, respectively.  

\subsection{Pairs  and Particles}\label{PSSS}

We now turn our attention to the lattice  of relations of a given type.
We begin with a point-free definition of a ``pair''.  In Subsection \ref{Pairs of Points},  we show 
that Definition  \ref{singleton} does indeed capture the notion of a ``pair of points'' whereby the points are the
``particles'' also introduced in the definition.
\begin{Definition}[Pair]\label{singleton}{\rm \ \ \ A relation $Z$ is a \emph{pair} iff it has the following properties:
\begin{description}
\item[(a)]$Z\ms{1}{\neq}\ms{1}{\MPplatbottom}~~,$ 
\item[(b)]$Z\ms{2}{=}\ms{2}Z{\MPcomp}{\MPplattop}{\MPcomp}Z~~,$ 
\item[(c)]$Z{\MPldom{}}\ms{3}{=}\ms{3}Z\ms{1}{\MPcomp}\ms{1}Z^{\MPrev}~~,$ 
\item[(d)]$Z{\MPrdom{}}\ms{3}{=}\ms{3}Z^{\MPrev}\ms{1}{\MPcomp}\ms{1}Z~~.$ 
 
\end{description}
We call a relation a \emph{particle} if it is a pair and it is symmetric.
}
\QED
\end{Definition}

In words, a pair $Z$  is a non-empty rectangle  (properties \ref{singleton}(a) and \ref{singleton}(b))
that is a bijection  on its  left domain and right domains  (properties \ref{singleton}(c) and \ref{singleton}(d)). 

(Definition \ref{singleton} was introduced in \cite{Vo99}  but using the terminology ``singleton'' instead of ``pair'',
and ``singleton square'' instead of ``particle''.)

Our goal is to prove that the points are exactly the particles.  This section is about showing that a
particle is a point.  See Corollary \ref{singletonsquare.is.point}.

One task is to show that particles are atoms.  The more general property, which we need in later sections,
is that pairs are atoms. 
\begin{Lemma}\label{atomic.singleton}{\rm \ \ \ A pair is an atom.
}%
\end{Lemma}%
\begin{proof}
Suppose $Z$ is a pair and suppose $Y$ is such that $Y\ms{1}{\subseteq}\ms{1}Z$.  By the definition of atom, definition \ref{atom},
 we must
show that $Y\ms{1}{=}\ms{1}{\MPplatbottom}\ms{2}{\vee}\ms{2}Y\ms{1}{=}\ms{1}Z$.   Equivalently, assuming $Y\ms{1}{\neq}\ms{1}{\MPplatbottom}$, we must show that $Y\ms{1}{=}\ms{1}Z$.  This is done as follows.
\begin{mpdisplay}{0.15em}{6.5mm}{0mm}{2}
	$Y$\push\-\\
	$=$	\>	\>$\{$	\>\+\+\+assumption:  $Y\ms{1}{\subseteq}\ms{1}Z$.  So, $Y{\MPldom{}}\ms{1}{\subseteq}\ms{1}Z{\MPldom{}}$ and $Y{\MPrdom{}}\ms{1}{\subseteq}\ms{1}Z{\MPrdom{}}$;  domains: (\ref{monos10.10})\-\-$~~~ \}$\pop\\
	$Z{\MPldom{}}\ms{1}{\MPcomp}\ms{1}Y\ms{1}{\MPcomp}\ms{1}Z{\MPrdom{}}$\push\-\\
	$=$	\>	\>$\{$	\>\+\+\+$Z$ is a pair, so $Z{\MPldom{}}\ms{2}{=}\ms{2}Z\ms{1}{\MPcomp}\ms{1}Z^{\MPrev}\ms{2}{=}\ms{2}(Z{\MPcomp}{\MPplattop}{\MPcomp}Z)\ms{1}{\MPcomp}\ms{1}Z^{\MPrev}$;  similarly for $Z{\MPrdom{}}$\-\-$~~~ \}$\pop\\
	$Z\ms{1}{\MPcomp}\ms{1}{\MPplattop}\ms{1}{\MPcomp}\ms{1}Z\ms{1}{\MPcomp}\ms{1}Z^{\MPrev}\ms{1}{\MPcomp}\ms{1}Y\ms{1}{\MPcomp}\ms{1}Z^{\MPrev}\ms{1}{\MPcomp}\ms{1}Z\ms{1}{\MPcomp}\ms{1}{\MPplattop}\ms{1}{\MPcomp}\ms{1}Z$\push\-\\
	$=$	\>	\>$\{$	\>\+\+\+domains: Theorem \ref{domains}(a) and Theorem \ref{domains}(b)\-\-$~~~ \}$\pop\\
	$Z\ms{1}{\MPcomp}\ms{1}{\MPplattop}\ms{1}{\MPcomp}\ms{1}Z{\MPldom{}}\ms{1}{\MPcomp}\ms{1}Y\ms{1}{\MPcomp}\ms{1}Z{\MPrdom{}}\ms{1}{\MPcomp}\ms{1}{\MPplattop}\ms{1}{\MPcomp}\ms{1}Z$\push\-\\
	$=$	\>	\>$\{$	\>\+\+\+$Z{\MPldom{}}\ms{1}{\MPcomp}\ms{1}Y\ms{1}{\MPcomp}\ms{1}Z{\MPrdom{}}\ms{2}{=}\ms{2}Y$  (see first step above)\-\-$~~~ \}$\pop\\
	$Z{\MPcomp}{\MPplattop}{\MPcomp}Y{\MPcomp}{\MPplattop}{\MPcomp}Z$\push\-\\
	$=$	\>	\>$\{$	\>\+\+\+assumption: $Y\ms{1}{\neq}\ms{1}{\MPplatbottom}$,  cone rule\-\-$~~~ \}$\pop\\
	$Z{\MPcomp}{\MPplattop}{\MPcomp}Z$\push\-\\
	$=$	\>	\>$\{$	\>\+\+\+$Z$ is a pair\-\-$~~~ \}$\pop\\
	$Z~~.$
\end{mpdisplay}
\end{proof}

Since a particle is, by definition, a pair, we have:
\begin{Corollary}\label{singleton.square.is.atom}{\rm \ \ \ A particle is an atom.
}%
\QED
\end{Corollary}%
\begin{Lemma}\label{singletonsquare.is.coreflexive}{\rm \ \ \ A particle is coreflexive.
}%
\end{Lemma}%
\begin{shortproof}
Suppose $Z$ is square and a pair.  Then
\begin{mpdisplay}{0.15em}{6.5mm}{0mm}{2}
	$Z$\push\-\\
	$=$	\>	\>$\{$	\>\+\+\+assumption: $Z$ is a pair, so $Z\ms{1}{=}\ms{1}Z{\MPcomp}{\MPplattop}{\MPcomp}Z$;  \\
	 $\left[\ms{3}{\MPplattop}{\MPcomp}Z\ms{3}{=}\ms{3}{\MPplattop}\ms{1}{\MPcomp}\ms{1}Z{\MPldom{}}\ms{1}{\MPcomp}\ms{1}Z\ms{3}{=}\ms{3}{\MPplattop}\ms{1}{\MPcomp}\ms{1}Z^{\MPrev}\ms{1}{\MPcomp}\ms{1}Z\ms{3}\right]$\-\-$~~~ \}$\pop\\
	$Z\ms{1}{\MPcomp}\ms{1}{\MPplattop}\ms{1}{\MPcomp}\ms{1}Z^{\MPrev}\ms{1}{\MPcomp}\ms{1}Z$\push\-\\
	$=$	\>	\>$\{$	\>\+\+\+assumption:  $Z$ is a square, so $Z\ms{2}{=}\ms{2}Z\ms{1}{\MPcomp}\ms{1}{\MPplattop}\ms{1}{\MPcomp}\ms{1}Z^{\MPrev}\ms{2}{=}\ms{2}Z^{\MPrev}$ \-\-$~~~ \}$\pop\\
	$Z^{\MPrev}\ms{1}{\MPcomp}\ms{1}Z$\push\-\\
	$=$	\>	\>$\{$	\>\+\+\+assumption: $Z$ is a pair, so $Z{\MPrdom{}}\ms{2}{=}\ms{2}Z^{\MPrev}\ms{1}{\MPcomp}\ms{1}Z$\-\-$~~~ \}$\pop\\
	$Z{\MPrdom{}}~~.$
\end{mpdisplay}
That is, $Z$ equals  $Z{\MPrdom{}}$ which is coreflexive.
\end{shortproof}

\begin{Corollary}[Particle]\label{singleton.square}{\rm \ \ \ A  relation $Z$   is a \emph{particle} iff it has the following  three 
properties.
\begin{description}
\item[(a)]$Z\ms{1}{\neq}\ms{1}{\MPplatbottom}$~~, 
\item[(b)]$Z\ms{1}{\subseteq}\ms{1}\mathbb{I}$~~, and 
\item[(c)]$Z\ms{2}{=}\ms{2}Z{\MPcomp}{\MPplattop}{\MPcomp}Z$~~. 
 
\end{description}
In words, a particle  is a proper, coreflexive rectangle.
}%
\end{Corollary}%
\begin{shortproof}
``Only-if'' is the combination of the definition of a particle and Lemma
 \ref{singletonsquare.is.coreflexive}.  ``If'' is a straightforward consequence of the properties of domains and
coreflexives.%
\end{shortproof}
\begin{Corollary}\label{singletonsquare.is.point}{\rm \ \ \ A particle is a proper, coreflexive atom. 
That is,  a particle is a point.
}%
\end{Corollary}%
\begin{shortproof}
This is a combination of Lemmas \ref{atomic.singleton} and \ref{singletonsquare.is.coreflexive}.%
\end{shortproof}

\subsection{Points are Particles}\label{Points are Singleton Squares}

We now prove the converse of Corollary \ref{singletonsquare.is.point}.  We use the assumption that every per
has a coreflexive  index:  the axiom of choice (Axiom \ref{Axiom of Choice}).
\begin{Lemma}\label{point.is.singletonsquare}{\rm \ \ \ Assuming Axiom \ref{Axiom of Choice}, a  point  is a particle.
}%
\end{Lemma}%
\begin{proof}
Suppose that $a$ is a point.  Comparing the definition of a point, Definition \ref{point},   with the defining
properties  of a particle, Corollary  \ref{singletonsquare.is.point},   it suffices to prove that  $a\ms{1}{=}\ms{1}a{\MPcomp}{\MPplattop}{\MPcomp}a$. 
Clearly $a{\MPcomp}{\MPplattop}{\MPcomp}a$ is a per.  (The simple proof uses the fact that $a\ms{1}{=}\ms{1}a^{\MPrev}$, because $a$ is coreflexive,   and
${\MPplattop}{\MPcomp}a{\MPcomp}{\MPplattop}\ms{1}{=}\ms{1}{\MPplattop}$ because $a\ms{1}{\neq}\ms{1}{\MPplatbottom}$.)  So, by the axiom of choice,  $a{\MPcomp}{\MPplattop}{\MPcomp}a$ has an index $J$, say.  We show that $J$ is a
particle and $J\ms{1}{=}\ms{1}a$.

To show that $J$ is a particle, we must establish the three properties listed  in 
Corollary  \ref{singleton.square} with the  instantiation $Z\ms{1}{:=}\ms{1}J$.    Part (a) is proved as follows.
\begin{mpdisplay}{0.15em}{6.5mm}{0mm}{2}
	$J\ms{1}{=}\ms{1}{\MPplatbottom}$\push\-\\
	$\Rightarrow$	\>	\>$\{$	\>\+\+\+${\MPplatbottom}$ is zero of composition\-\-$~~~ \}$\pop\\
	$a{\MPcomp}{\MPplattop}{\MPcomp}a{\MPcomp}J{\MPcomp}a{\MPcomp}{\MPplattop}{\MPcomp}a\ms{2}{=}\ms{2}{\MPplatbottom}$\push\-\\
	$=$	\>	\>$\{$	\>\+\+\+$J$ is an index of per $a{\MPcomp}{\MPplattop}{\MPcomp}a$, Definition  \ref{per.index}(c) \-\-$~~~ \}$\pop\\
	$a{\MPcomp}{\MPplattop}{\MPcomp}a\ms{2}{=}\ms{2}{\MPplatbottom}$\push\-\\
	$\Rightarrow$	\>	\>$\{$	\>\+\+\+$a{\MPcomp}a{\MPcomp}a\ms{1}{\subseteq}\ms{1}a{\MPcomp}{\MPplattop}{\MPcomp}a$ and $a{\MPcomp}a{\MPcomp}a\ms{1}{=}\ms{1}a$ (because $a\ms{1}{\subseteq}\ms{1}\mathbb{I}$)\-\-$~~~ \}$\pop\\
	$a\ms{1}{\subseteq}\ms{1}{\MPplatbottom}$\push\-\\
	$=$	\>	\>$\{$	\>\+\+\+$\left[\ms{2}R\ms{1}{\subseteq}\ms{1}{\MPplatbottom}\ms{4}{\equiv}\ms{4}R\ms{1}{=}\ms{1}{\MPplatbottom}\ms{2}\right]$ with $R\ms{1}{:=}\ms{1}a$\-\-$~~~ \}$\pop\\
	$a\ms{1}{=}\ms{1}{\MPplatbottom}$\push\-\\
	$=$	\>	\>$\{$	\>\+\+\+assumption:   $a$ is proper, i.e.\ $a\ms{1}{\neq}\ms{1}{\MPplatbottom}$\-\-$~~~ \}$\pop\\
	$\mathsf{false}~~.$
\end{mpdisplay}
We conclude that $J\ms{1}{\neq}\ms{1}{\MPplatbottom}$.   The next step is to show that $J\ms{1}{=}\ms{1}a$.  
\begin{mpdisplay}{0.15em}{6.5mm}{0mm}{2}
	$J\ms{1}{=}\ms{1}a$\push\-\\
	$\Leftarrow$	\>	\>$\{$	\>\+\+\+assumption: $a$ is an atom\-\-$~~~ \}$\pop\\
	$J\ms{1}{=}\ms{1}{\MPplatbottom}\ms{3}{\vee}\ms{3}J\ms{1}{\subseteq}\ms{1}a$\push\-\\
	$=$	\>	\>$\{$	\>\+\+\+$J\ms{1}{\neq}\ms{1}{\MPplatbottom}$ (see above)\-\-$~~~ \}$\pop\\
	$J\ms{1}{\subseteq}\ms{1}a$\push\-\\
	$=$	\>	\>$\{$	\>\+\+\+assumption: $a\ms{1}{\subseteq}\ms{1}\mathbb{I}$,   so $a\ms{1}{=}\ms{1}(a{\MPcomp}{\MPplattop}{\MPcomp}a){\MPldom{}}$\-\-$~~~ \}$\pop\\
	$J\ms{2}{\subseteq}\ms{2}(a{\MPcomp}{\MPplattop}{\MPcomp}a){\MPldom{}}$\push\-\\
	$=$	\>	\>$\{$	\>\+\+\+assumption:  $J$ is an index of $a{\MPcomp}{\MPplattop}{\MPcomp}a$\\
	Definition \ref{per.index}(a)\-\-$~~~ \}$\pop\\
	$\mathsf{true}~~.$
\end{mpdisplay}
Property  (b) of Corollary \ref{singleton.square} immediately follows because $a$ is coreflexive.  
We now show that $J\ms{1}{=}\ms{1}J{\MPcomp}{\MPplattop}{\MPcomp}J$.  
\begin{mpdisplay}{0.15em}{6.5mm}{0mm}{2}
	$J{\MPcomp}{\MPplattop}{\MPcomp}J$\push\-\\
	$=$	\>	\>$\{$	\>\+\+\+$J\ms{1}{=}\ms{1}a$ (proved above) and $a\ms{1}{\subseteq}\ms{1}\mathbb{I}$\-\-$~~~ \}$\pop\\
	$J{\MPcomp}a{\MPcomp}{\MPplattop}{\MPcomp}a{\MPcomp}J$\push\-\\
	$=$	\>	\>$\{$	\>\+\+\+assumption:  $J$ is an index of $a{\MPcomp}{\MPplattop}{\MPcomp}a$\\
	Definition \ref{per.index}(c) with $P\ms{1}{:=}\ms{1}a{\MPcomp}{\MPplattop}{\MPcomp}a$\-\-$~~~ \}$\pop\\
	$J~~.$
\end{mpdisplay}
We conclude that  $J\ms{1}{=}\ms{1}a\ms{1}{=}\ms{1}J{\MPcomp}{\MPplattop}{\MPcomp}J$.  Thus $a\ms{1}{=}\ms{1}a{\MPcomp}{\MPplattop}{\MPcomp}a$ as required.
\end{proof}

Combining Corollary \ref{singletonsquare.is.point} with Lemma \ref{point.is.singletonsquare}, we conclude:
\begin{Theorem}\label{point=singleton.square}{\rm \ \ \ Assuming Axiom \ref{Axiom of Choice}, a  relation is a point iff it is a
particle.
}
\QED
\end{Theorem}

\subsection{Proper Atoms are Pairs}\label{ProperAtomsarePairs}

The goal of this section is to show that a proper atom is a pair.  Aiming to exploit the equivalence of points
and particles, we begin with lemmas on the left and right domains of a proper atom.
\begin{Lemma}\label{atom.domain.is.atomic}{\rm \ \ \ Suppose $R$ is a proper  atom.  Then $R{\MPldom{}}$ and $R{\MPrdom{}}$ are proper 
atoms\footnote{~Note:  strictly we should detail the lattice under consideration here.  However,  it is easy to 
show that a coreflexive being an atom in the lattice of coreflexives is equivalent to its
being an atom in the lattice of relations.  This justifies the omission.}.
}%
\end{Lemma}%
\begin{proof}
First,  that $R{\MPldom{}}$ and $R{\MPrdom{}}$ are both proper is immediate from (\ref{monos10.14'}).

To show that $R{\MPldom{}}$ is an atom we have to show that, for all $p$, \begin{displaymath}p\ms{1}{\subseteq}\ms{1}R{\MPldom{}}\ms{3}{\wedge}\ms{3}p\ms{1}{\neq}\ms{1}{\MPplatbottom}\ms{5}{\equiv}\ms{5}p\ms{1}{=}\ms{1}R{\MPldom{}}~~.\end{displaymath}We do this by mutual implication.  First, the follows-from:
\begin{mpdisplay}{0.15em}{6.5mm}{0mm}{2}
	$p\ms{1}{\subseteq}\ms{1}R{\MPldom{}}\ms{3}{\wedge}\ms{3}p\ms{1}{\neq}\ms{1}{\MPplatbottom}\ms{5}{\Leftarrow}\ms{5}p\ms{1}{=}\ms{1}R{\MPldom{}}$\push\-\\
	$=$	\>	\>$\{$	\>\+\+\+predicate calculus\-\-$~~~ \}$\pop\\
	$(p\ms{1}{\subseteq}\ms{1}R{\MPldom{}}\ms{4}{\Leftarrow}\ms{4}p\ms{1}{=}\ms{1}R{\MPldom{}})\ms{4}{\wedge}\ms{4}(p\ms{1}{\neq}\ms{1}{\MPplatbottom}\ms{4}{\Leftarrow}\ms{4}p\ms{1}{=}\ms{1}R{\MPldom{}})$\push\-\\
	$\Leftarrow$	\>	\>$\{$	\>\+\+\+left conjunct: anti-symmetry, right conjunct: Leibniz\-\-$~~~ \}$\pop\\
	$\mathsf{true}\ms{5}{\wedge}\ms{5}R{\MPldom{}}\ms{2}{\neq}\ms{2}{\MPplatbottom}$\push\-\\
	$\Leftarrow$	\>	\>$\{$	\>\+\+\+$R{\MPldom{}}$ is proper (see above)\-\-$~~~ \}$\pop\\
	$\mathsf{true}~~.$
\end{mpdisplay}
Now we prove the converse.   Assume $p\ms{1}{\subseteq}\ms{1}R{\MPldom{}}$ and  $p\ms{1}{\neq}\ms{1}{\MPplatbottom}$.  Then
\begin{mpdisplay}{0.15em}{6.5mm}{0mm}{2}
	$p\ms{1}{=}\ms{1}R{\MPldom{}}$\push\-\\
	$=$	\>	\>$\{$	\>\+\+\+anti-symmetry and assumption:  $p\ms{1}{\subseteq}\ms{1}R{\MPldom{}}$\-\-$~~~ \}$\pop\\
	$R{\MPldom{}}\ms{2}{\subseteq}\ms{2}p$\push\-\\
	$\Leftarrow$	\>	\>$\{$	\>\+\+\+assumption:  $p\ms{1}{\subseteq}\ms{1}R{\MPldom{}}$ and $R{\MPldom{}}\ms{1}{\subseteq}\ms{1}\mathbb{I}$~,  so  $p\ms{1}{=}\ms{1}p{\MPldom{}}$;  $(p{\MPcomp}R){\MPldom{}}\ms{1}{\subseteq}\ms{1}p{\MPldom{}}$\-\-$~~~ \}$\pop\\
	$R{\MPldom{}}\ms{3}{=}\ms{3}(p{\MPcomp}R){\MPldom{}}$\push\-\\
	$\Leftarrow$	\>	\>$\{$	\>\+\+\+Leibniz\-\-$~~~ \}$\pop\\
	$R\ms{2}{=}\ms{2}p{\MPcomp}R$\push\-\\
	$=$	\>	\>$\{$	\>\+\+\+$p{\MPcomp}R\ms{2}{\neq}\ms{2}{\MPplatbottom}$  (see below for proof)\\
	$R$ is an atom, Definition \ref{atom} (appropriately instantiated)\-\-$~~~ \}$\pop\\
	$p{\MPcomp}R\ms{2}{\subseteq}\ms{2}R$\push\-\\
	$=$	\>	\>$\{$	\>\+\+\+assumption:  $p\ms{1}{\subseteq}\ms{1}R{\MPldom{}}$ and $R{\MPldom{}}\ms{1}{\subseteq}\ms{1}\mathbb{I}$~, monotonicity\-\-$~~~ \}$\pop\\
	$\mathsf{true}~~.$
\end{mpdisplay}
In order to verify the penultimate step  in the above calculation,  we  show that $p{\MPcomp}R\ms{1}{=}\ms{1}{\MPplatbottom}\ms{2}{\Rightarrow}\ms{2}\mathsf{false}$
 under the assumption that  $p\ms{1}{\subseteq}\ms{1}R{\MPldom{}}$ and  $p\ms{1}{\neq}\ms{1}{\MPplatbottom}$.
\begin{mpdisplay}{0.15em}{6.5mm}{0mm}{2}
	$p{\MPcomp}R\ms{1}{=}\ms{1}{\MPplatbottom}$\push\-\\
	$=$	\>	\>$\{$	\>\+\+\+cone rule: (\ref{cone.rule})\-\-$~~~ \}$\pop\\
	${\MPplattop}{\MPcomp}p{\MPcomp}R{\MPcomp}{\MPplattop}\ms{3}{=}\ms{3}{\MPplatbottom}$\push\-\\
	$=$	\>	\>$\{$	\>\+\+\+domains: Theorem \ref{domains}(a)\-\-$~~~ \}$\pop\\
	${\MPplattop}\ms{1}{\MPcomp}\ms{1}p\ms{1}{\MPcomp}\ms{1}R{\MPldom{}}\ms{1}{\MPcomp}\ms{1}{\MPplattop}\ms{4}{=}\ms{4}{\MPplatbottom}$\push\-\\
	$\Rightarrow$	\>	\>$\{$	\>\+\+\+assumption:  $p\ms{1}{\subseteq}\ms{1}R{\MPldom{}}$, composition of coreflexives is intersection\-\-$~~~ \}$\pop\\
	${\MPplattop}{\MPcomp}p{\MPcomp}{\MPplattop}\ms{3}{=}\ms{3}{\MPplatbottom}$\push\-\\
	$=$	\>	\>$\{$	\>\+\+\+assumption:  $p\ms{1}{\neq}\ms{1}{\MPplatbottom}$, cone rule:  (\ref{cone.rule})\-\-$~~~ \}$\pop\\
	$\mathsf{false}~~.$
\end{mpdisplay}
\end{proof}

\begin{Corollary}\label{proper.atom.dom.particles}{\rm \ \ \ If $R$ is a proper atom,  $R{\MPldom{}}$ and $R{\MPrdom{}}$ are particles.
}%
\end{Corollary}%
\begin{shortproof}
By Lemma \ref{atom.domain.is.atomic} and  Definition \ref{point}  of a point, if $R$ is a proper atom, $R{\MPldom{}}$ and $R{\MPrdom{}}$
are points.  Thus, by Lemma \ref{point.is.singletonsquare},  $R{\MPldom{}}$ and $R{\MPrdom{}}$ are particles.%
\end{shortproof}

We now aim to verify properties \ref{singleton}(b), (c) and (d)  of a pair,  with $Z$ instantiated to proper atom
$R$.   Property \ref{singleton}(b) is the following lemma.   
\begin{Lemma}\label{proper.atom.is.pair}{\rm \ \ \ A proper atom is a rectangle.
}%
\end{Lemma}%
\begin{shortproof}
Suppose $R$ is a proper atom.  Then
\begin{mpdisplay}{0.15em}{6.5mm}{0mm}{2}
	$R\ms{1}{\MPcomp}\ms{1}{\MPplattop}\ms{1}{\MPcomp}\ms{1}R$\push\-\\
	$=$	\>	\>$\{$	\>\+\+\+domains: Theorem \ref{domains}(a)\-\-$~~~ \}$\pop\\
	$R{\MPldom{}}\ms{2}{\MPcomp}\ms{2}{\MPplattop}\ms{2}{\MPcomp}\ms{2}R{\MPrdom{}}$\push\-\\
	$=$	\>	\>$\{$	\>\+\+\+$R\ms{1}{\neq}\ms{1}{\MPplatbottom}$, cone rule: (\ref{cone.rule})\-\-$~~~ \}$\pop\\
	$R{\MPldom{}}\ms{2}{\MPcomp}\ms{2}{\MPplattop}\ms{2}{\MPcomp}\ms{2}R\ms{2}{\MPcomp}\ms{2}{\MPplattop}\ms{2}{\MPcomp}\ms{2}R{\MPrdom{}}$\push\-\\
	$=$	\>	\>$\{$	\>\+\+\+domains: (\ref{ldom.and.rdom})\-\-$~~~ \}$\pop\\
	$R{\MPldom{}}\ms{2}{\MPcomp}\ms{2}{\MPplattop}\ms{2}{\MPcomp}\ms{2}R{\MPldom{}}\ms{2}{\MPcomp}\ms{2}R\ms{2}{\MPcomp}\ms{2}R{\MPrdom{}}\ms{2}{\MPcomp}\ms{2}{\MPplattop}\ms{2}{\MPcomp}\ms{2}R{\MPrdom{}}$\push\-\\
	$=$	\>	\>$\{$	\>\+\+\+by Corollary \ref{proper.atom.dom.particles},  $R{\MPldom{}}$ and $R{\MPrdom{}}$ are particles;\\
	Corollary  \ref{singleton.square}(c) with $Z\ms{1}{:=}\ms{1}R{\MPldom{}}$ and $Z\ms{1}{:=}\ms{1}R{\MPrdom{}}$\-\-$~~~ \}$\pop\\
	$R{\MPldom{}}\ms{2}{\MPcomp}\ms{2}R\ms{2}{\MPcomp}\ms{2}R{\MPrdom{}}$\push\-\\
	$=$	\>	\>$\{$	\>\+\+\+domains: (\ref{ldom.and.rdom})\-\-$~~~ \}$\pop\\
	$R~~.$
\end{mpdisplay}
That is,  $R\ms{1}{\MPcomp}\ms{1}{\MPplattop}\ms{1}{\MPcomp}\ms{1}R\ms{2}{=}\ms{2}R$.  Thus, by definition, $R$ is a rectangle.
\end{shortproof}

We now have all the ingredients for our goal.
\begin{Lemma}\label{atoms.are.pairs}{\rm \ \ \ Suppose $R$ is a proper atom.  Then, assuming Axiom \ref{Axiom of Choice},
$R$ is a pair.
}%
\end{Lemma}%
\begin{proof}
Suppose $R$ is a proper atom.  
We have to verify properties \ref{singleton}(b), (c) and (d) (with $Z\ms{1}{:=}\ms{1}R$)  of a pair.  

Property \ref{singleton}(b) is Lemma \ref{proper.atom.is.pair}.  
Properties \ref{singleton}(c) and (d) assert that $R$ is a bijection.  To prove this, 
let $J$ be an index of $R$.  (This is where Axiom \ref{Axiom of Choice} is assumed.)  Then
\begin{mpdisplay}{0.15em}{6.5mm}{0mm}{2}
	$J\ms{1}{=}\ms{1}R$\push\-\\
	$=$	\>	\>$\{$	\>\+\+\+$R$ is an atom\-\-$~~~ \}$\pop\\
	$J\ms{1}{\neq}\ms{1}{\MPplatbottom}\ms{3}{\wedge}\ms{3}J\ms{1}{\subseteq}\ms{1}R$\push\-\\
	$=$	\>	\>$\{$	\>\+\+\+$J$ is an index of $R$, Definition \ref{gen.index} \-\-$~~~ \}$\pop\\
	$\mathsf{true}~~.$
\end{mpdisplay}
That is, $J\ms{1}{=}\ms{1}R$.  But $R$ is a rectangle and thus a difunction.  So, applying
Lemma \ref{difunction.bijection},  $J$ ---and thus $R$--- is a bijection, as required.
\end{proof}

To  conclude this section and Sections \ref{PSSS} and   \ref{Points are Singleton Squares}, we have:
\begin{Theorem}\label{atom.equals.pair}{\rm \ \ \ Assuming  Axiom  \ref{Axiom of Choice}, for all types $A$ and $B$, and all 
relations $R$ of type $A{\sim}B$,   $R$ is a proper atom iff  $R$ is a pair.
}
\end{Theorem}
\begin{shortproof}
This is a combination of Lemmas \ref{atomic.singleton} and \ref{atoms.are.pairs}.
\end{shortproof}

\subsection{Pairs of Points and the All-or-Nothing Rule}\label{Pairs of Points}

The final step is to show that we can derive the ``all-or-nothing'' rule.  Throughout this section we
assume Axiom  \ref{Axiom of Choice}. 
\begin{Lemma}\label{singleton.atomic.domain}{\rm \ \ \ If $Z$ is a pair then $Z{\MPldom{}}$ and $Z{\MPrdom{}}$ are particles.
}%
\end{Lemma}%
\begin{proof}
Suppose $Z$ is a pair.  We begin by showing 
 that its left and right domains are also pairs.  

Properties \ref{singleton}(a), (c) and (d) ---with $Z\ms{1}{:=}\ms{1}Z{\MPldom{}}$ and $Z\ms{1}{:=}\ms{1}Z{\MPrdom{}}$---  are properties of the domain operators. 
This leaves \ref{singleton}(b).  For the instance $Z\ms{1}{:=}\ms{1}Z{\MPldom{}}$, we have:
\begin{mpdisplay}{0.15em}{6.5mm}{0mm}{2}
	$Z{\MPldom{}}\ms{1}{\MPcomp}\ms{1}{\MPplattop}\ms{1}{\MPcomp}\ms{1}Z{\MPldom{}}$\push\-\\
	$=$	\>	\>$\{$	\>\+\+\+domains: Theorem \ref{domains}(a) and (b) \-\-$~~~ \}$\pop\\
	$Z\ms{1}{\MPcomp}\ms{1}{\MPplattop}\ms{1}{\MPcomp}\ms{1}Z\ms{1}{\MPcomp}\ms{1}Z^{\MPrev}$\push\-\\
	$=$	\>	\>$\{$	\>\+\+\+assumption:  $Z$ is a pair, so $Z{\MPcomp}{\MPplattop}{\MPcomp}Z\ms{1}{=}\ms{1}Z$\-\-$~~~ \}$\pop\\
	$Z\ms{1}{\MPcomp}\ms{1}Z^{\MPrev}$\push\-\\
	$=$	\>	\>$\{$	\>\+\+\+assumption:  $Z$ is a pair, so $Z\ms{1}{\MPcomp}\ms{1}Z^{\MPrev}\ms{2}{=}\ms{2}Z{\MPldom{}}$\-\-$~~~ \}$\pop\\
	$Z{\MPldom{}}~~.$
\end{mpdisplay}
The proof that $Z{\MPrdom{}}$ is a pair is symmetrical.

It now follows immediately that $Z{\MPldom{}}$ and $Z{\MPrdom{}}$ are squares:  a square is a symmetric rectangle, and both are
rectangles (see above); also,  both are coreflexives, and coreflexives are symmetric.
\end{proof}

The following theorem is \cite[Lemma 4.41(d)]{Vo99}.
\begin{Theorem}\label{singleton.is.pair}{\rm \ \ \ For all $Z$,  \begin{displaymath}\mathsf{pair}{.}Z\ms{4}{\equiv}\ms{4}{\left\langle\exists\ms{1}{}a{,}b\ms{2}{:}\ms{2}\mathsf{point}{.}a\ms{1}{\wedge}\ms{1}\mathsf{point}{.}b\ms{2}{:}\ms{2}Z\ms{1}{=}\ms{1}a{\MPcomp}{\MPplattop}{\MPcomp}b\right\rangle}~~.\end{displaymath}
}
\end{Theorem}
\begin{shortproof}
By mutual implication.  First,  
\begin{mpdisplay}{0.15em}{6.5mm}{0mm}{2}
	$\mathsf{pair}{.}Z$\push\-\\
	$\Rightarrow$	\>	\>$\{$	\>\+\+\+Lemma \ref{singleton.atomic.domain};\\
	Definition \ref{singleton}(b) and  $\left[\ms{3}Z{\MPcomp}{\MPplattop}{\MPcomp}Z\ms{3}{=}\ms{3}Z{\MPldom{}}\ms{1}{\MPcomp}\ms{1}{\MPplattop}\ms{1}{\MPcomp}\ms{1}Z{\MPrdom{}}\ms{3}\right]$\-\-$~~~ \}$\pop\\
	$\mathsf{particle}\ms{1}{.}\ms{1}Z{\MPldom{}}\ms{4}{\wedge}\ms{4}\mathsf{particle}\ms{1}{.}\ms{1}Z{\MPrdom{}}\ms{4}{\wedge}\ms{4}Z\ms{2}{=}\ms{2}Z{\MPldom{}}\ms{1}{\MPcomp}\ms{1}{\MPplattop}\ms{1}{\MPcomp}\ms{1}Z{\MPrdom{}}$\push\-\\
	$\Rightarrow$	\>	\>$\{$	\>\+\+\+Corollary \ref{singletonsquare.is.point}\-\-$~~~ \}$\pop\\
	$\mathsf{point}\ms{1}{.}\ms{1}Z{\MPldom{}}\ms{4}{\wedge}\ms{4}\mathsf{point}\ms{1}{.}\ms{1}Z{\MPrdom{}}\ms{4}{\wedge}\ms{4}Z\ms{2}{=}\ms{2}Z{\MPldom{}}\ms{1}{\MPcomp}\ms{1}{\MPplattop}\ms{1}{\MPcomp}\ms{1}Z{\MPrdom{}}$\push\-\\
	$\Rightarrow$	\>	\>$\{$	\>\+\+\+$a{,}b\ms{2}{:=}\ms{2}Z{\MPldom{}}\ms{1}{,}\ms{1}Z{\MPrdom{}}$\-\-$~~~ \}$\pop\\
	${\left\langle\exists\ms{1}{}a{,}b\ms{2}{:}\ms{2}\mathsf{point}{.}a\ms{1}{\wedge}\ms{1}\mathsf{point}{.}b\ms{2}{:}\ms{2}Z\ms{1}{=}\ms{1}a{\MPcomp}{\MPplattop}{\MPcomp}b\right\rangle}~~.$
\end{mpdisplay}
For the converse, assume that $a$ and $b$ are points.  We have to prove that $a{\MPcomp}{\MPplattop}{\MPcomp}b$ is a pair.  Applying
Definition \ref{singleton}, this means checking four properties:
\begin{description}
\item[(a)]$a{\MPcomp}{\MPplattop}{\MPcomp}b\ms{2}{\neq}\ms{2}{\MPplatbottom}~~,$ 
\item[(b)]$a{\MPcomp}{\MPplattop}{\MPcomp}b\ms{2}{=}\ms{2}a{\MPcomp}{\MPplattop}{\MPcomp}b{\MPcomp}{\MPplattop}{\MPcomp}a{\MPcomp}{\MPplattop}{\MPcomp}b~~,$ 
\item[(c)]$(a{\MPcomp}{\MPplattop}{\MPcomp}b){\MPldom{}}\ms{3}{=}\ms{3}(a{\MPcomp}{\MPplattop}{\MPcomp}b)\ms{1}{\MPcomp}\ms{1}(a{\MPcomp}{\MPplattop}{\MPcomp}b)^{\MPrev}~~,$ 
\item[(d)]$(a{\MPcomp}{\MPplattop}{\MPcomp}b){\MPrdom{}}\ms{3}{=}\ms{3}(a{\MPcomp}{\MPplattop}{\MPcomp}b)^{\MPrev}\ms{1}{\MPcomp}\ms{1}(a{\MPcomp}{\MPplattop}{\MPcomp}b)~~.$ 
\end{description}
Properties (a) and (b) are instances of the cone rule together with the assumption that $a$ and $b$ are
proper.  We prove (c) as follows.
\begin{mpdisplay}{0.15em}{6.5mm}{0mm}{2}
	$(a{\MPcomp}{\MPplattop}{\MPcomp}b)\ms{1}{\MPcomp}\ms{1}(a{\MPcomp}{\MPplattop}{\MPcomp}b)^{\MPrev}$\push\-\\
	$=$	\>	\>$\{$	\>\+\+\+converse\-\-$~~~ \}$\pop\\
	$a\ms{1}{\MPcomp}\ms{1}{\MPplattop}\ms{1}{\MPcomp}\ms{1}b\ms{1}{\MPcomp}\ms{1}b^{\MPrev}\ms{1}{\MPcomp}\ms{1}{\MPplattop}\ms{1}{\MPcomp}\ms{1}a$\push\-\\
	$=$	\>	\>$\{$	\>\+\+\+assumption: $b$ is a point, cone rule: (\ref{cone.rule})\-\-$~~~ \}$\pop\\
	$a{\MPcomp}{\MPplattop}{\MPcomp}a$\push\-\\
	$=$	\>	\>$\{$	\>\+\+\+assumption:  $a$ is a point; so, by Corollary \ref{point.is.singletonsquare}, $a$ is a pair;\\
	Definition \ref{singleton}(b) with $Z\ms{1}{:=}\ms{1}a$\-\-$~~~ \}$\pop\\
	$a$\push\-\\
	$=$	\>	\>$\{$	\>\+\+\+$a{\MPcomp}{\MPplattop}{\MPcomp}b$ is a non-empty rectangle\-\-$~~~ \}$\pop\\
	$(a{\MPcomp}{\MPplattop}{\MPcomp}b){\MPldom{}}~~.$
\end{mpdisplay}
Property (d) is proved symmetrically.
\end{shortproof}

We conclude with the theorem that Gl\"uck's ``all-or-nothing'' axiom \cite{Glueck17}  is a consequence of
our axiom of choice.
\begin{Theorem}[All or Nothing]\label{all.or.nothing.rule}{\rm \ \ \ \begin{displaymath}{\left\langle\forall\ms{1}{}a{,}b{,}R\ms{3}{:}\ms{3}\mathsf{point}{.}a\ms{1}{\wedge}\ms{1}\mathsf{point}{.}b\ms{3}{:}\ms{3}a{\MPcomp}R{\MPcomp}b\ms{1}{=}\ms{1}{\MPplatbottom}\ms{2}{\vee}\ms{2}a{\MPcomp}R{\MPcomp}b\ms{1}{=}\ms{1}a{\MPcomp}{\MPplattop}{\MPcomp}b\right\rangle}~~.\end{displaymath}
}
\end{Theorem}
\begin{proof}
Suppose  $a$ and $b$ are points.  By Theorem \ref{singleton.is.pair},  $a{\MPcomp}{\MPplattop}{\MPcomp}b$ is a pair. 
So, by Lemma \ref{atomic.singleton},  $a{\MPcomp}{\MPplattop}{\MPcomp}b$ is an atom.  Applying the definition of atomic, we have, for all  $R$,
\begin{mpdisplay}{0.15em}{6.5mm}{0mm}{2}
	$\mathsf{true}$\push\-\\
	$=$	\>	\>$\{$	\>\+\+\+monotonicity,  $R\ms{1}{\subseteq}\ms{1}{\MPplattop}$\-\-$~~~ \}$\pop\\
	$a{\MPcomp}R{\MPcomp}b\ms{2}{\subseteq}\ms{2}a{\MPcomp}{\MPplattop}{\MPcomp}b$\push\-\\
	$=$	\>	\>$\{$	\>\+\+\+$a{\MPcomp}{\MPplattop}{\MPcomp}b$ is an atom, Definition \ref{atom}\-\-$~~~ \}$\pop\\
	$a{\MPcomp}R{\MPcomp}b\ms{1}{=}\ms{1}{\MPplatbottom}\ms{2}{\vee}\ms{2}a{\MPcomp}R{\MPcomp}b\ms{1}{=}\ms{1}a{\MPcomp}{\MPplattop}{\MPcomp}b~~.$
\end{mpdisplay}
\end{proof}

The significance of the all-or-nothing rule is that, together with  Theorem \ref{saturated.is.powerset},
 it follows that the lattice  of relations of type $A{\sim}B$ is isomorphic to the  powerset $2^{A{\MPtimes}B}$.  
\begin{Theorem}\label{all.or.nothing.sat}{\rm \ \ \ Suppose, for  types  $A$ and $B$,  the lattices   of coreflexives\mpindexnew{coreflexive}{lattice of}{\mpT }{\mpT }{\mpT }{\mpT }{\mpT }{} of types $A$ and $B$
are both  extensional (i.e.\ complete,  universally distributive and saturated).   Then the lattice  of  relations  of type  
$A{\sim}B$ is saturated; the atoms are elements of the form $a{\MPcomp}{\MPplattop}{\MPcomp}b$ where $a$ and $b$ are atoms of
the poset of coreflexives (of types $A$ and $B$, respectively).   It follows that, if  the lattice of relations of type
$A{\sim}B$  is complete and universally distributive, it  is isomorphic to the powerset\mpindexnew{powerset}{\mpT }{\mpT }{\mpT }{\mpT }{\mpT }{\mpT }{} of the set of elements of
the form $a{\MPcomp}{\MPplattop}{\MPcomp}b$ where $a$ and $b$ are points of types $A$ and $B$, respectively.
}
\end{Theorem}
\begin{proof} By Theorems  \ref{singleton.is.pair} and \ref{atom.equals.pair}, 
$a{\MPcomp}{\MPplattop}{\MPcomp}b$ is an atom.   It suffices  to prove that the lattice  of relations of type $A{\sim}B$  is saturated.  
This  is easy: for all $R$ of type $A{\sim}B$,
\begin{mpdisplay}{0.15em}{6.5mm}{0mm}{2}
	$R$\push\-\\
	$=$	\>	\>$\{$	\>\+\+\+$\mathbb{I}$ is unit of composition,  \\
	lattices of coreflexives of types $A$ and $B$  are extensional\-\-$~~~ \}$\pop\\
	${\left\langle{\cup}a\ms{1}{:}\ms{1}\mathsf{point}{.}a\ms{1}{:}\ms{1}a\right\rangle}\ms{1}{\MPcomp}\ms{1}R\ms{1}{\MPcomp}\ms{1}{\left\langle{\cup}b\ms{1}{:}\ms{1}\mathsf{point}{.}b\ms{1}{:}\ms{1}b\right\rangle}$\push\-\\
	$=$	\>	\>$\{$	\>\+\+\+distributivity of composition over ${\cup}$\-\-$~~~ \}$\pop\\
	${\left\langle{\cup}\ms{1}a{,}b\ms{2}{:}\ms{2}\mathsf{point}{.}a\ms{1}{\wedge}\ms{1}\mathsf{point}{.}b\ms{2}{:}\ms{2}a{\MPcomp}R{\MPcomp}b\right\rangle}$\push\-\\
	$=$	\>	\>$\{$	\>\+\+\+all-or-nothing rule:  Theorem \ref{all.or.nothing.rule}, ${\MPplatbottom}$ is zero of supremum\-\-$~~~ \}$\pop\\
	${\left\langle{\cup}\ms{1}a{,}b\ms{3}{:}\ms{3}\mathsf{point}{.}a\ms{2}{\wedge}\ms{2}\mathsf{point}{.}b\ms{2}{\wedge}\ms{2}a{\MPcomp}R{\MPcomp}b\ms{1}{\neq}\ms{1}{\MPplatbottom}\ms{3}{:}\ms{3}a{\MPcomp}{\MPplattop}{\MPcomp}b\right\rangle}~~.$
\end{mpdisplay}
That the lattice of relations is a powerset follows from Theorem \ref{saturated.is.powerset}.   By Theorem 
\ref{singleton.is.pair}, every pair is a relation of the form $a{\MPcomp}{\MPplattop}{\MPcomp}b$; also,  by Lemma \ref{atomic.singleton}, 
$a{\MPcomp}{\MPplattop}{\MPcomp}b$ is an atom.
\end{proof}

Summarising  Theorem \ref{all.or.nothing.sat},  the \emph{saturation} property is that \begin{equation}\label{saturation}
{\left\langle\forall{}R\ms{4}{:}{:}\ms{4}R\ms{2}{=}\ms{2}{\left\langle\cup\ms{1}{}a{,}b\ms{2}{:}\ms{2}a{\MPcomp}{\MPplattop}{\MPcomp}b\ms{1}{\subseteq}\ms{1}R\ms{2}{:}\ms{2}a{\MPcomp}{\MPplattop}{\MPcomp}b\right\rangle}\right\rangle}~~.
\end{equation}
Combining Theorem \ref{all.or.nothing.sat} with Theorem \ref{saturated.is.powerset}, we get the   \emph{irreducibility}
property:   if $\mathcal{R}$ is a function with    range relations of type $A{\sim}B$ and source $K$, then, for all points $a$ and $b$ of
appropriate type,   \begin{equation}\label{irreducible}
a{\MPcomp}{\MPplattop}{\MPcomp}b\ms{3}{\subseteq}\ms{3}{\cup}\mathcal{R}\ms{5}{\equiv}\ms{5}{\left\langle\exists{}k\ms{2}{:}\ms{2}k{\in}K\ms{2}{:}\ms{2}a{\MPcomp}{\MPplattop}{\MPcomp}b\ms{1}{\subseteq}\ms{1}\mathcal{R}{.}k\right\rangle}~~.
\end{equation}Property (\ref{saturation})  formalises the interpretation of  the property $a{\MPcomp}{\MPplattop}{\MPcomp}b\ms{1}{\subseteq}\ms{1}R$ as  the
property $(a,b){\in}R$ in standard set-theoretic accounts of relation algebra.

Theorem \ref{all.or.nothing.sat} assumes that the lattices of coreflexives (of appropriate type) are extensional.  
Conversely,  if we assume that, for all types $A$ and $B$, the lattice of relations of type $A{\sim}B$ is extensional then
so is the lattice of coreflexives of type $A$, for all $A$.  This is Theorem \ref{converse.all.or.nothing.sat}.  
(The proof of Theorem \ref{converse.all.or.nothing.sat} can be found in the companion document \cite{VB2022}.)


\begin{Theorem}\label{converse.all.or.nothing.sat}{\rm \ \ \ Suppose, for  all types  $A$ and $B$,  the lattice  of  relations  of type  
$A{\sim}B$ is extensional, whereby the  atoms are elements of the form $a{\MPcomp}{\MPplattop}{\MPcomp}b$ where $a$ and $b$ are atoms of
the poset of coreflexives (of types $A$ and $B$, respectively).   Then, for all $A$, the lattice of coreflexives of type $A$
is extensional.
}
\QED
\end{Theorem}

Combining Theorems \ref{all.or.nothing.sat} and \ref{converse.all.or.nothing.sat}, we get:
\begin{Corollary}\label{extensional.equiv}{\rm \ \ \ Suppose, for all types $A$ and $B$, the lattice of relations of type $A{\sim}B$ is
complete and universally distributive.  
Then for all types  $A$ and $B$, the lattice of relations of type $A{\sim}B$ is extensional
iff for all types $A$, the lattice of coreflexives of type $A$ is extensional.
}%
\QED
\end{Corollary}%

Although the saturation property  allows us to identify atoms of the form $a{\MPcomp}{\MPplattop}{\MPcomp}b$ with
elements $(a,b)$ of the set $A{\MPtimes}B$, it does not establish that the operators of relation algebra (converse,
composition etc.) correspond to their standard set-theoretic interpretations.   This is straightforward.  
For example, for  composition we have, for all $R$ and $S$,   
\begin{mpdisplay}{0.15em}{6.5mm}{0mm}{2}
	$R{\MPcomp}S$\push\-\\
	$=$	\>	\>$\{$	\>\+\+\+saturation: (\ref{saturation})\-\-$~~~ \}$\pop\\
	${\left\langle\cup\ms{1}{}a{,}b\ms{2}{:}\ms{2}a{\MPcomp}{\MPplattop}{\MPcomp}b\ms{1}{\subseteq}\ms{1}R\ms{2}{:}\ms{2}a{\MPcomp}{\MPplattop}{\MPcomp}b\right\rangle}\ms{2}{\MPcomp}\ms{2}{\left\langle\cup\ms{1}{}b'{,}c\ms{2}{:}\ms{2}b'{\MPcomp}{\MPplattop}{\MPcomp}c\ms{1}{\subseteq}\ms{1}S\ms{2}{:}\ms{2}b'{\MPcomp}{\MPplattop}{\MPcomp}c\right\rangle}$\push\-\\
	$=$	\>	\>$\{$	\>\+\+\+distributivity\-\-$~~~ \}$\pop\\
	${\left\langle\cup\ms{1}{}a{,}b{,}b'{,}c\ms{3}{:}\ms{3}a{\MPcomp}{\MPplattop}{\MPcomp}b\ms{1}{\subseteq}\ms{1}R\ms{2}{\wedge}\ms{2}b'{\MPcomp}{\MPplattop}{\MPcomp}c\ms{1}{\subseteq}\ms{1}S\ms{3}{:}\ms{3}a{\MPcomp}{\MPplattop}{\MPcomp}b{\MPcomp}b'{\MPcomp}{\MPplattop}{\MPcomp}c\right\rangle}$\push\-\\
	$=$	\>	\>$\{$	\>\+\+\+$b$ and $b'$ are points, so $b{\MPcomp}b'\ms{1}{\neq}\ms{1}{\MPplatbottom}\ms{2}{\equiv}\ms{2}b'\ms{1}{=}\ms{1}b$\\
	case analysis on $b'\ms{1}{=}\ms{1}b\ms{3}{\vee}\ms{3}b'\ms{1}{\neq}\ms{1}b$, one-point rule\-\-$~~~ \}$\pop\\
	${\left\langle\cup\ms{1}{}a{,}b{,}c\ms{4}{:}\ms{4}a{\MPcomp}{\MPplattop}{\MPcomp}b\ms{1}{\subseteq}\ms{1}R\ms{3}{\wedge}\ms{3}b{\MPcomp}{\MPplattop}{\MPcomp}c\ms{2}{\subseteq}\ms{2}S\ms{4}{:}\ms{4}a{\MPcomp}{\MPplattop}{\MPcomp}b{\MPcomp}b{\MPcomp}{\MPplattop}{\MPcomp}c\right\rangle}$\push\-\\
	$=$	\>	\>$\{$	\>\+\+\+$b$ ranges over points,  so $b{\MPcomp}b\ms{1}{=}\ms{1}b\ms{1}{\neq}\ms{1}{\MPplatbottom}$,  cone rule: (\ref{cone.rule})\-\-$~~~ \}$\pop\\
	${\left\langle\cup\ms{1}{}a{,}b{,}c\ms{4}{:}\ms{4}a{\MPcomp}{\MPplattop}{\MPcomp}b\ms{1}{\subseteq}\ms{1}R\ms{3}{\wedge}\ms{3}b{\MPcomp}{\MPplattop}{\MPcomp}c\ms{2}{\subseteq}\ms{2}S\ms{4}{:}\ms{4}a{\MPcomp}{\MPplattop}{\MPcomp}c\right\rangle}$\push\-\\
	$=$	\>	\>$\{$	\>\+\+\+range disjunction\-\-$~~~ \}$\pop\\
	${\left\langle\cup\ms{1}{}a{,}c\ms{4}{:}\ms{4}{\left\langle\exists{}b\ms{3}{:}{:}\ms{3}a{\MPcomp}{\MPplattop}{\MPcomp}b\ms{1}{\subseteq}\ms{1}R\ms{2}{\wedge}\ms{2}b{\MPcomp}{\MPplattop}{\MPcomp}c\ms{1}{\subseteq}\ms{1}S\right\rangle}\ms{4}{:}\ms{4}a{\MPcomp}{\MPplattop}{\MPcomp}c\right\rangle}~~.$
\end{mpdisplay}
Comparing the first and last lines of this calculation (and interpreting $a{\MPcomp}{\MPplattop}{\MPcomp}b\ms{1}{\subseteq}\ms{1}R$ as $(a,b){\in}R$ and 
$b{\MPcomp}{\MPplattop}{\MPcomp}c\ms{1}{\subseteq}\ms{1}S$ as $(b,c){\in}S$) we recognise the standard set-theoretic definition of $R{\MPcomp}S$.  

The important step to note in the above calculation is the use of the distributivity of composition
over union.  The validity of such universal distributivity --- both from the left and from the right---
 is a consequence of the Galois connections (\ref{under}) and (\ref{over}) defining factors.  A similar step needed
in the   calculation for converse relies on the fact that converse is the upper and lower adjoint of itself.

We conclude this section with a brief comparison of extensionality as formulated here with the
notion of extensionality formulated by Voermans \cite{Vo99}.   

Voermans \cite[Section 4.5]{Vo99} postulated that the lattice of binary relations of  a given type is saturated
by relations of the form $X{\MPcomp}{\MPplattop}{\MPcomp}Y$ where $X$ and $Y$ are \emph{particles}.  Relations of this form are then shown to
model pairs $(x,y)$ in standard set-theoretic presentations of relation algebra. Here, we have postulated
that each type $A$ forms a lattice that is saturated by \emph{points}: see Axiom \ref{extensional.axiom}; this postulate
is  combined with our axiom of choice: all pers have an index.  Then  pairs in standard set-theoretic 
presentations of relation algebra are modelled by relations of the form  $a{\MPcomp}{\MPplattop}{\MPcomp}b$, where $a$ and $b$ are points. 
Because particles are points (Corollary \ref{singletonsquare.is.point}), the saturation property postulated by
Voermans is formally stronger than Axiom \ref{extensional.axiom}.  As a consequence, it becomes slightly
harder to establish that, for example,  the composition of two relations does indeed correspond to the
set-theoretic notion of composition.  (See \cite[Section 4.5]{Vo99} for details of what is involved.)  More
importantly, the
combination of Axioms \ref{Axiom of Choice} and \ref{extensional.axiom} facilitates a better separation of
concerns:  Axiom \ref{Axiom of Choice} provides a powerful extension of point-free reasoning, whilst Axiom
\ref{extensional.axiom} fills the gap where pointwise reasoning is unavoidable.

\section{Independence of the Axioms}\label{IndependenceOfTheAxioms}

In this paper, we have considered various additions to the basic axioms of point-free relation algebra: the
cone rule,  our axiom of choice, extensionality and the all-or-nothing rule.  The question arises to what
extent these additional rules are independent of one another.

Elementary examples of   point-free relation algebra give some insight.   In all the examples, there is just
one,  unnamed,  type.  The 1-element algebra
has just one element:  in this algebra ${\MPplatbottom}\ms{1}{=}\ms{1}\mathbb{I}\ms{1}{=}\ms{1}{\MPplattop}$.  The 2-element algebra has two elements: 
 in this algebra ${\MPplatbottom}\ms{1}{\neq}\ms{1}\mathbb{I}\ms{1}{=}\ms{1}{\MPplattop}$.   There is a  3-element algebra that 
has distinct   elements ${\MPplatbottom}$, $\mathbb{I}$ and ${\MPplattop}$.    (There is also a 3-element algebra in which $\mathbb{I}\ms{1}{=}\ms{1}{\MPplattop}$.)

The cone rule is independent of the other rules since, given any two   algebras,  one can define their
product in such a way that the cone rule  is not satisfied but the validity of other rules is preserved.  See
\cite[section 3.4.3]{Vo99} for more details. 

The 1-element algebra is the algebra of concrete relations on the empty set and 
the 2-element algebra is the algebra of concrete relations on a set with exactly one element.  Both these 
algebras satisfy  all four additional axioms.  

The 1-element and 2-element algebras  have the property that  $\mathbb{I}\ms{1}{=}\ms{1}{\MPplattop}$.  So, for any $R$,  
$${\MPplattop}{\MPcomp}R{\MPcomp}{\MPplattop}\ms{1}{=}\ms{1}\mathbb{I}{\MPcomp}R{\MPcomp}\mathbb{I}\ms{1}{=}\ms{1}R.$$  In such algebras, our axiom of choice is  valid (because every element is an index of
itself); extensionality is also valid but, if there 
are more than two elements,  the cone rule is not valid.  (If the algebra is complemented, it is a Boolean 
algebra.)

The 3-element algebra does not satisfy our axiom of choice since ${\MPplattop}$ does not have an index.  Nor  does 
it satisfy the  all-or-nothing rule.  It does, however, satisfy the cone rule  as well as being extensional.   

A slightly more complicated example is needed to show that the all-or-nothing rule does not imply
our axiom of choice.  Consider the algebra with a single point $a$ in addition to the three constants 
${\MPplatbottom}$,  $\mathbb{I}$ and ${\MPplattop}$.  That is, ${\MPplatbottom}\ms{1}{\subseteq}\ms{1}a\ms{1}{\subseteq}\ms{1}\mathbb{I}\ms{1}{\subseteq}\ms{1}{\MPplattop}$.  Add the requirement that $a{\MPcomp}{\MPplattop}\ms{1}{=}\ms{1}a\ms{1}{=}\ms{1}{\MPplattop}{\MPcomp}a$.    Then $a\ms{2}{=}\ms{2}a{\MPcomp}{\MPplattop}{\MPcomp}a$ and the
all-or-nothing rule is valid.  However, ${\MPplattop}$ does not have an index:  the only two possibilities are $a$ and $\mathbb{I}$, and
it is easily checked that neither satisfies the requirements.  

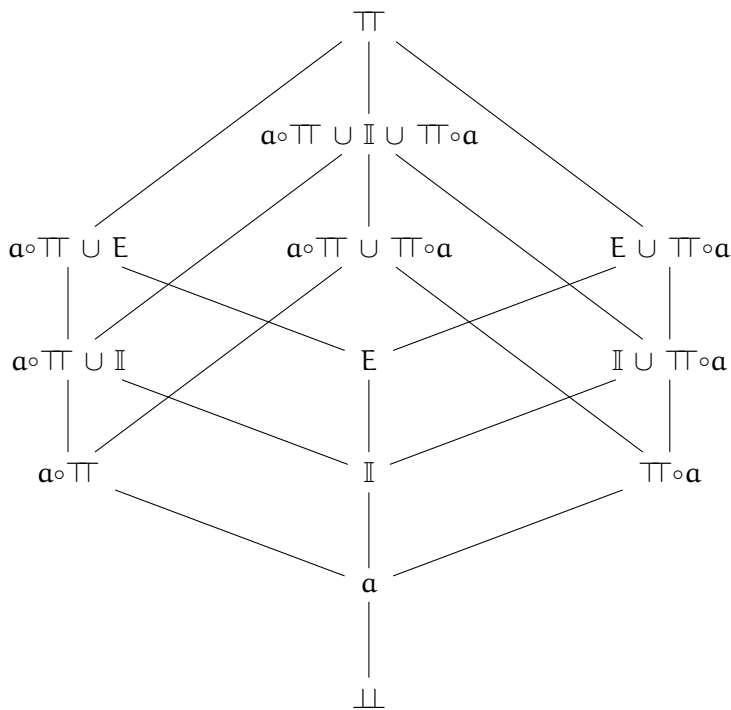
\begin{figure}[h]
\centering \omittik{\begin{center}
\begin{tikzpicture}[>=stealth,initial text=]
  \node (1) at (0,9) {$\MPplattop$};
  \node (10) at (0,7.5){$a{\MPcomp}{\MPplattop}\ms{1}\cup\ms{1}{\mathbb{I}}\ms{1}\cup\ms{1}{\MPplattop}{\MPcomp}a$};
\node (11) at (-4,6) {$a{\MPcomp}{\MPplattop}\ms{1}\cup\ms{1}E$};
\node (3) at (0,6) {$a{\MPcomp}{\MPplattop}\ms{1}\cup\ms{1}{\MPplattop}{\MPcomp}a$};
\node (12) at (4,6) {$E\ms{1}\cup\ms{1}{\MPplattop}{\MPcomp}a$};
\node (2) at (-4,4.5){$a{\MPcomp}{\MPplattop}\ms{1}\cup\ms{1}{\mathbb{I}}$};
\node (13) at (0,4.5){$E$};
\node (4) at (4,4.5) {${\mathbb{I}}\ms{1}\cup\ms{1}{\MPplattop}{\MPcomp}a$};
\node (5) at (-4,3) {$a{\MPcomp}{\MPplattop}$};
\node (6) at (0,3) {$\mathbb{I}$};
\node (7) at (4,3) {${\MPplattop}{\MPcomp}a$};
\node (8) at (0,1.5) {$a$};
\node (9) at (0,0) {$\MPplatbottom$};
\path[-]
(1) edge node {} (11)
(1) edge node {} (10)
(1) edge node {} (12)
(10) edge node {} (3)
(10) edge node {} (2)
(10) edge node {} (4)
(11) edge node {} (2)
(11) edge node {} (13)
(12) edge node {} (13)
(12) edge node {} (4)
(13) edge node {} (6)
(2) edge node {} (5)
(2) edge node {} (6)
(3) edge node {} (5)
(3) edge node {} (7)
(4) edge node {} (6)
(4) edge node {} (7)
(5) edge node {} (8)
(6) edge node {} (8)
(7) edge node {} (8)
(8) edge node {} (9)
;
\end{tikzpicture}
\end{center}}

\caption{All-or-Nothing and Cone Rule but not Axiom of Choice}\label{fig:NonChoiceExample}
\end{figure}

The above example does not satisfy extensionality (because there is exactly one point different from $\mathbb{I}$); 
nor does it satisfy  the cone rule (because ${\MPplattop}{\MPcomp}a{\MPcomp}{\MPplattop}\ms{2}{=}\ms{2}a$).   Adding  the cone rule as a requirement 
has the implication that $a{\MPcomp}{\MPplattop}$,  $a$ and ${\MPplattop}{\MPcomp}a$ must be different  --- with the knock-on effect that several
additional elements must also be added to the algebra.   The resulting 
algebra,  which was  constructed by  Jules Desharnais [private
communication] using \textsf{Mace4} \cite{Prover9Mace4},   is the minimal algebra with elements $a$ and $E$,  in addition to the
constants ${\MPplatbottom}$,  $\mathbb{I}$ and ${\MPplattop}$,  such that \begin{displaymath}{\MPplatbottom}\ms{1}{\subseteq}\ms{1}a\ms{1}{\subseteq}\ms{1}\mathbb{I}\ms{1}{\subseteq}\ms{1}E\ms{5}{\wedge}\ms{5}a{\MPcomp}E\ms{2}{=}\ms{2}a\ms{2}{=}\ms{2}E{\MPcomp}a\mbox{~~,}\end{displaymath}and the cone rule and the all-or-nothing rule are both satisfied.   The lattice structure of the algebra is
shown in Fig.\ \ref{fig:NonChoiceExample}.

Note that the original 4-element algebra forms the ``stem'' of the algebra --- ${\MPplattop}$  has been
renamed $E$.  Adding the cone rule causes the algebra to  ``blossom out''  into a much larger structure.  
The  algebra  does not satisfy our axiom of choice,   because  $E$ does not have an index.  

The combination of the all-or-nothing rule,  the cone rule and extensionality implies  that the lattice
of relations is isomorphic to the  powerset of the set of elements of the form $a{\MPcomp}{\MPplattop}{\MPcomp}b$ where $a$ and $b$ are
points  \cite[Theorem 57]{BACKHOUSE2022100730}.  (We have already seen two examples of the theorem:
the 1-element and the 2-element algebras.)
In turn, this implies that, in standard set theory,  our
axiom of choice is valid.  In constructive set theory, the axiom might be deemed to be inadmissible --- in the
same way that the universal axiom of choice might be  deemed to be inadmissible.   (Inadmissible is
different from invalid.  The law of the excluded middle is inadmissible in constructive logic but is not
invalid;  in constructive logic, particular instances of excluded middle must be proven by exhibiting
witnesses, but current wisdom is  that a counterexample to excluded middle will never  be found.)

The final question we   ask is whether, in point-free relation algebra,  
 the universal axiom of choice is implied by our axiom of choice.  Formally, the universal axiom of choice is that, for any relation $R$, there is a relation $f$ such that
$f\ms{1}{\subseteq}\ms{1}R$,  $f\ms{1}{\MPcomp}\ms{1}f^{\MPrev}\ms{2}{\subseteq}\ms{2}\mathbb{I}$ and  $f{\MPrdom{}}\ms{1}{=}\ms{1}R{\MPrdom{}}$.  (In words, $f$ is a functional approximation to $R$ with the same right domain
as $R$; for given point $b$,  $f{\MPcomp}b$  ``chooses'' a point $a$ in the left domain of $R$ such that $a$ and $b$ are related by $R$.)

The 3-element algebra satisfies the universal axiom of choice but not our axiom of
choice.    So the universal axiom of choice does not imply our axiom of choice. It is an open question
whether or not our axiom of choice implies the universal axiom of choice. 

\section{Conclusion}\label{IndicesAndCoresConclusion}

Point-free relation algebra has been developed over many, many years (beginning in the 19th century) 
and is generally regarded as a much better basis for the development of the theory of relations than
pointwise reasoning.  However, for practical applications,   pointwise reasoning is at times unavoidable. 
For example, path-finding algorithms on graphs   must ultimately  be expressed in terms of the nodes
and edges of the graph (the points and elements  of the relation defined by the graph).  Good practice is to
develop such algorithms in a stepwise fashion, beginning with point-free reasoning (typically using
regular algebra)  and delaying the introduction of points until absolutely necessary.

It is common practice to represent an equivalence relation by choosing a specific  element of each
equivalence class.  For example,  the class of  integers modulo $3$ is commonly represented by the set of
three elements $0$, $1$ and $2$.  The characterisation of an equivalence relation by a representative function is
not derivable in point-free relation algebra since there is a constructive element in the choice of
representatives.  Extensions to point-free relation algebra, such as the postulate that relations form a 
so-called ``power allegory''  \cite[2.4]{FRSC90}, are intended to enable pointwise reasoning but nevertheless
fail to properly capture the use of representatives. Our axiom of choice (Axiom \ref{Axiom of Choice})
together with our point-free formulation of the notion of an index of a relation does capture the use of
representatives.  The
strength of the axiom together with the fact that an index of a relation has the same type as the relation
makes the notion of an index ---in our view--- very attractive and useful.  
Moreover, its  combination with the extensionality axiom (Axiom \ref{extensional.axiom})
permits the derivation of Gl\"uck's ``all-or-nothing'' axiom \cite{Glueck17}.  In this way,  point-free
reasoning has been strengthened whilst also facilitating pointwise reasoning when unavoidable.

It might be argued that our axiom of choice is too strong.  On the contrary, we would argue that it
corresponds much better to standard  practice.    For example,   the computation of the strongly
connected components of a graph involves computing a representative node for each component. 
(Tarjan   \cite{Tarjan1972} and Sharir \cite{Sharir81} call the representative of a strongly  connected
component  of a graph  the ``root'' of the component;  Cormen, Leiserson and Rivest \cite[p.490]{CLR90} call
it  the  ``forefather'' of the component.) A suggestion for future work is to exploit our notion of an index in
order to reformulate ---much more succinctly--- the properties of depth-first search that underlie its
effectiveness in such computations.

Our focus in this paper has been on documenting the properties of indexes and the consequences for
axiom systems enabling pointwise reasoning.   The original motivation for this work was, however, quite
different.   Seventy years ago, in a series of publications \cite{Riguet48,Riguet50,Riguet51},   Jacques Riguet
introduced the notions of a ``relation difonctionelle'',   the ``diff\a'{e}rence'' of a relation and  ``relations de
Ferrers''.   In view of possible practical implications, particularly in respect of relational databases,  our
goal was to bring Riguet's work up to date, making it more accessible to modern audiences.   In the
process, we began to realise that substantial improvements could be made by introducing the notions of 
``core'' and ``index'' of a relation, drawing inspiration from Voerman's \cite{Vo99} notion of the (left- and
right-) per domains of a relation.  Results on cores and indexes relevant to Riguet's work, in particular
practical applications of his notion of the ``diff\a'{e}rence'' of a relation (which we rename the ``diagonal'' of a
relation),  are documented in \cite{VB2023b}.   Further insight into the nature of cores and indices is
documented in   \cite{VJB2024} where we introduce the ``\textsf{thins}'' ordering on relations; among the theorems
we prove is that an index of a relation is minimal with respect to the \textsf{thins} ordering.   Proofs we have
omitted here are documented in the companion working  document \cite{VB2022}.    

\paragraph{Appendix. Notational Conventions}\label{IDCAppendix}

The predence rules we assume are that unary operators take precedence over all binary operators;  
composition takes precedence over union and intersection, 
which have equal precedence; next comes  the subset ordering and finally the logical operators, which
follow the usual precedence rules except that disjunction and conjuction have the same precedence.    We
endeavour to add spacing to suggest the precedence.  

Our calculations consist of a sequence of terms, successive terms being 
separated by a relation symbol and a   ``hint'' indicating  why the relation holds between the terms.   
The relations between terms are always  ordering relations  and, typically but not always,  the conclusion
of the calculation is the relation between the initial and final terms (which is implicitly deduced by
transitivity properties of the connecting relations).  We
endeavour to use equality steps whenever possible, even in the case that the final conclusion is an
ordering.  (This is because an equality step is safe in the sense that no loss of information is incurred when
making the step.  Calculations that reach a dead end often do so because an ordering is introduced that is
too strong.)    Sometimes an equality is established by exploiting the
anti-symmetry of an ordering relation.  In such cases, the initial and final terms are identical and the 
conclusion is that all terms in the calculation are equal.  Sometimes we announce the intention to exploit
anti-symmetry by the use of the word ``mutual'' (as in, for example, ``mutual implication'' or ``mutual
inclusion'').  

The first rule of logic is the rule sometimes known as ``substitution of equals for equals'':  two things are
equal exactly when one can be substituted for the other in any context.   The rule is
generally attributed to Gottfried Wilhelm Leibniz.   The use of the rule is often implicit in our calculations
but occasionally we refer to it by the hint ``Leibniz''.  The centrality of the rule is reinforced by the
ubiquitous overloading of Robert Recorde's ``${=}$'' symbol to denote equality.   So the symbol is used for
equality of numbers, sets, functions, etc.  In some circumstances,
however, the overloading can be confusing.    A simple example is that the boolean $x\ms{1}{=}\ms{1}0$ is equal to the
boolean $x{+}1\ms{1}{=}\ms{1}1$, whatever the value of $x$.  However,  it could be confusing  to overload Robert Recorde's equality
symbol to express their equality  as in \begin{displaymath}(x\ms{1}{=}\ms{1}0)\ms{3}{=}\ms{3}(x{+}1\ms{1}{=}\ms{1}1)~~.\end{displaymath}For this reason, it is convenient to introduce a special symbol ``${\equiv}$''  which we sometimes use to denote
the equality of booleans; this symbol is given a lower precedence than Robert Recorde's symbol, as
exemplified by \begin{displaymath}x\ms{1}{=}\ms{1}0\ms{3}{\equiv}\ms{3}x{+}1\ms{1}{=}\ms{1}1~~.\end{displaymath}A more important reason for introducing an additional  symbol is that boolean equality is associative.  In a
continued equality $a\ms{1}{=}\ms{1}b\ms{1}{=}\ms{1}c$  (whatever the type of $a$, $b$ and $c$), the transitivity of equality is implicitly 
assumed.  For booleans $a$, $b$ and $c$,  $a\ms{1}{\equiv}\ms{1}b\ms{1}{\equiv}\ms{1}c$  has a different meaning:  its value is determined by
evaluating  $(a\ms{1}{=}\ms{1}b)\ms{1}{=}\ms{1}c$ or  $a\ms{1}{=}\ms{1}(b\ms{1}{=}\ms{1}c)$; both yield the same result.   We don't exploit the associativity of
boolean equality in this paper.  We do exploit its transitivity as well as Leibniz's rule frequently in our
calculations.  Consequently,  we most commonly use Robert Recorde's symbol to denote the equality of
boolean expressions (as well as expressions of other types).

\paragraph{Acknowledgement}\label{difun:Acknowledgements}

Many thanks to Jules Desharnais for supplying models of the different axiom systems.  We take full responsibility for any errors we may have introduced.  Thanks also to the referees for their helpful comments.


\bibliographystyle{alpha}
\bibliography{bibliogr}

\end{document}